\documentclass[aps,prb,twocolumn,amsmath,amssymb]{revtex4-1}
\usepackage{amsmath,amssymb,amsfonts,bm}
\usepackage{graphicx}
\usepackage{color}
\usepackage{bbold}
\usepackage{graphicx}
\usepackage{dcolumn}
\usepackage{epstopdf}
\usepackage{color}
\usepackage{epsfig}
\usepackage{url}
\usepackage{hyperref}
\usepackage{verbatim}

\newcommand{\be}{\begin{equation} }
\newcommand{\ee}{\end{equation} }
\newcommand{\ba}{\begin{eqnarray} }
\newcommand{\ea}{\end{eqnarray} }
\newcommand{\n}{\nonumber \\ }

\newcommand{\bit}{\begin{itemize}}
\newcommand{\eit}{\end{itemize}}

\begin{document}

\title{Classical spin liquids in stacked triangular lattice Ising antiferromagnets}
\author{D. T. Liu$^1$, F. J. Burnell$^{1,2}$, L. D. C. Jaubert$^3$ and J. T. Chalker$^1$}
\affiliation{$^1$Theoretical Physics, Oxford University, 1 Keble Road, Oxford OX1 3NP, United Kingdom}
\affiliation{$^2$School of Physics and Astronomy, University of Minnesota, Minneapolis, MN 55455, 
USA
}
\affiliation{$^3$Okinawa Institute of Science and Technology Graduate University, Onna-son, Okinawa 904-0395, Japan}
\date{submitted: 18 August 2016, revised: 8 November 2016}

\begin{abstract}
We study Ising antiferromagnets that have nearest-neighbour interactions on multilayer triangular lattices with frustrated ($abc$ and $abab$) stacking, and make comparisons with the unfrustrated ($aaa$) stacking. If interlayer couplings are much weaker than in-plane ones, the paramagnetic phase of models with frustrated stackings has a classical spin-liquid regime at low temperature, in which correlations are strong both within and between planes, but there is no long-range order. We investigate this regime using Monte Carlo simulations and by mapping the spin models to coupled height models, which are treated using renormalisation group methods and an analysis of the effects of vortex excitations. The classical spin-liquid regime is parametrically wide at small interlayer coupling in models with frustrated stackings. By contrast, for the unfrustrated stacking there is no extended regime in which interlayer correlations are strong without three-dimensional order.
\end{abstract}


\maketitle

\section{Introduction}
The triangular lattice Ising antiferromagnet is arguably the simplest model of a highly frustrated magnet and was probably the earliest such system to be studied in detail \cite{tlafm}. At low temperatures it is both highly fluctuating and strongly correlated; indeed, it remains disordered down to zero temperature and has a macroscopically degenerate ground state. The combination of fluctuations with correlations is typical more generally of highly frustrated magnets, which in this regime have been termed cooperative paramagnets or classical spin-liquids \cite{review}. 

In this paper we consider three-dimensional (3D) Ising antiferromagnets built from triangular layers that are stacked in such a way that nearest-neighbour interlayer interactions are frustrated, and make comparisons with the unfrustrated stacking. We focus on low-temperature behaviour in systems with weak interlayer coupling, where correlations within each layer are necessarily strong but correlations between layers are controlled by a competition between fluctuations and interactions. Using a combination of perturbative and non-perturbative analytical techniques and Monte Carlo simulations, we show that this competition leads to a classical spin liquid regime, in which strong correlations exist without long range order.

Models for frustrated magnets can be classified at the mean-field level according to the properties of the matrix of exchange interactions. 
In this approach, the eigenvectors associated with the minimum eigenvalues of the interaction matrix provide candidate ordering patterns. These minimum eigenvalues appear at isolated points in reciprocal space for unfrustrated systems, but may be highly degenerate for frustrated systems. For example, for nearest neighbour interactions on the kagome and pyrochlore lattices, the subspace of minimum eigenvalues forms a flat band that spans the entire Brillouin zone \cite{review,kagome,pyrochlore}. Other cases display intermediate behaviour: on the diamond lattice with nearest and next-nearest neighbour interactions, the minimum eigenvalues form a two-dimensional surface in the 3D Brillouin zone \cite{bergman}. The systems we discuss here are distinctive in having minimum eigenvalues that lie on \emph{lines} in the 3D Brillouin zone \cite{rastelli}. One of our central findings is that these systems have a cooperative paramagnetic regime in which they develop strong correlations that are centred near these reciprocal-space lines.

\begin{figure}
	\includegraphics[height=0.25\textwidth]{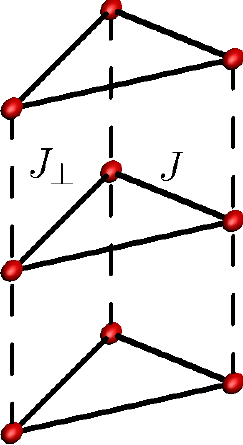}
	\includegraphics[height=0.25\textwidth]{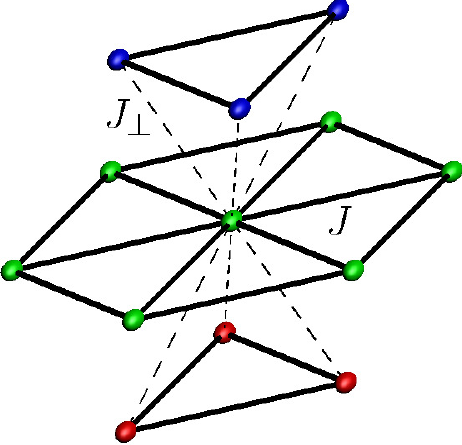}
	\includegraphics[height=0.25\textwidth]{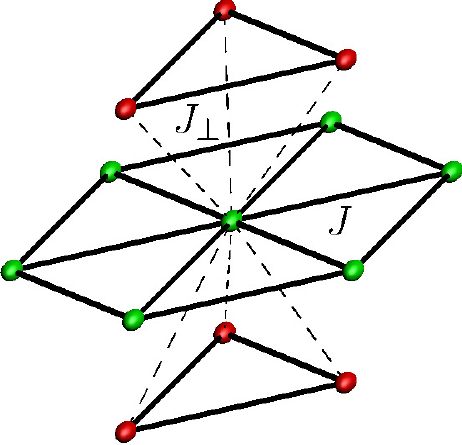}
	\caption{The three different ways of stacking triangular lattices that are considered in this paper: $aaa$ (top left), $abc$ (top right), $abab$ stacking (bottom). In-plane interactions $J$ and interlayer interactions $J_\perp$ are indicated with full and dashed lines, respectively.
}
	\label{fig:stacking_interactions}
\end{figure}

The three different ways of stacking triangular layers that we compare in this work are indicated in standard notation by
$aaa$, $abc$, and $abab$:  
see Fig.~\ref{fig:stacking_interactions}. Of these, the first provides a reference model without interlayer frustration, while the $abc$ stacking yields minimum eigenvalues along helices in the Brillouin zone, and the $abab$ stacking gives minimum eigenvalues on a ring around the Brillouin zone corner. The $abc$ stacking with equal in-plane and interlayer interactions is equivalent to a nearest-neighbour model on the face-centered-cubic (fcc) lattice, while the $abab$ stacking forms the hexagonal-close-packed (hcp) lattice.

Moving beyond a mean-field classification, the theoretical understanding of stacked triangular lattice Ising antiferromagnets (TLIAFMs) that we develop here is based on the height model description of low-temperature states for a single layer \cite{heightmodel,zeng}. This long-established model represents ground states of a layer in terms of an emergent height field, with a simple effective Hamiltonian that captures the entropy of fluctuations. A spin-flip excitation fractionalises into an unbound vortex-antivortex excitation pair in the height field, and the vortex separation sets the correlation length at finite temperature in the single-layer model. In the following we derive and study height models for weakly coupled multilayer systems, showing how the interplay of interlayer coupling and vortex excitations allows strong correlations to develop between layers, without long-range order. We also use the results of extensive Monte Carlo simulations to test these conclusions and to examine behaviour when interlayer coupling is not weak.

Our study is motivated in part by observations \cite{yamada, radaelli} of charge ordering in the materials LuFe$_2$O$_4$ and YbFe$_2$O$_4$. The charge states of $\text{Fe}^{2+}$ and $\text{Fe}^{3+}$ ions in these systems can be represented using Ising pseudospins, with antiferromagnetic coupling between pseudospins arising from screened Coulomb interactions \cite{yamada,harris}. The pseudospins occupy the sites of an $abc$-stacked triangular lattice, though with an alternating layer spacing that is not included in the models we study. Experimental studies \cite{yamada,radaelli,fe-review}, in particular of YbFe$_2$O$_4$ \cite{radaelli}, find helices of scattering intensity in a temperature range above a three-dimensional charge-ordering transition. These helices mirror in their reciprocal space location the positions of minimum eigenvalues of the interaction matrix discussed above. While an accurate description of these materials would require treating additional (magnetic) degrees of freedom \cite{fe-review}, the results we present in this paper demonstrate how strong interlayer correlations can arise over an extended temperature range without long-range order.

Past theoretical work on charge ordering in these materials has included quite detailed mean-field treatments  \cite{yamada,harris} and Monte Carlo simulations of a bilayer model \cite{nagano}, but has not made use of the understanding of single-layer TLIAFMs provided by height models, or used simulations to study correlations in the paramagnetic phase with the detail we present here.

TLIAFMs with other stackings have been examined previously in a variety of contexts. Treatments of the $abab$ case include mean-field theory, a low temperature expansion, and Monte Carlo simulations \cite{domany,kallin, diep}. That work has probed the ordering transition, but without examining the limit of weakly coupled layers or correlations in the paramagnetic phase.  TLIAFMs with unfrustrated ($aaa$) stacking have been of long-standing interest \cite{kallin-aaa}.  They display a continuous phase transition that, strikingly, is in the 3D XY universality class despite the absence of a microscopic continuous symmetry \cite{coppersmith,ma}. The two components of the order parameter represent ordering at the two inequivalent Brillouin zone corners, and the XY symmetry is broken in the ordered phase by dangerously irrelevant six-fold anisotropies. This model and transition are also important as an imaginary time representation of the quantum dimer model on the hexagonal lattice \cite{moessner2001}. 

The remainder of the paper is organised as follows. We introduce the models studied and give an overview of their physical behaviour in Sec.~\ref{overview}. We describe Monte Carlo results in Sec.~\ref{MC}. We introduce height models in Sec.~\ref{height} and analyse their behaviour in Secs.~\ref{behaviour} and \ref{beyond}. Results from our different approaches are compared in Sec.~\ref{discussion}. Some technical details are described in a series of appendices. An outline of some of the results has been presented previously in Ref.~\onlinecite{previous}.

\section{Models and overview}\label{overview}

The starting point for our investigation is the nearest neighbour Ising antiferromagnet on stacked triangular layers with anisotropic couplings. Each spin is coupled to its six in-plane neighbours with an exchange constant $J>0$ and to the closest spins in the layers above and below with an exchange constant $J_{\perp}$ (see Fig. \ref{fig:stacking_interactions}). The Hamiltonian is
\begin{align} \label{TheHamiltonian}
H &= J  \sum_{\langle ij\rangle, z} \sigma_{i,z} \sigma_{j,z} + J_\perp \sum_{\{ ij\}, z}  \sigma_{i, z} \sigma_{j,z+1} + H^{(1)}
\end{align}
where $H^{(1)}$ indicates further-neighbour interactions, which may be present in the bare Hamiltonian or may represent terms generated under renormalisation. Here $\sigma_{i,z} = \pm1$, the notation $\langle i,j \rangle$ denotes nearest neighbour pairs of sites from the same layer, and $\{i,j\}$ nearest neighbour pairs from adjacent layers. 
The sign of $J_\perp$ may be taken positive without loss of generality, since it can be reversed by the transformation: $\sigma_{i,z} \to \sigma^\prime_{i,z} = (-1)^z\sigma_{i,z}$.

\begin{figure}
	\includegraphics[height=0.25\textwidth]{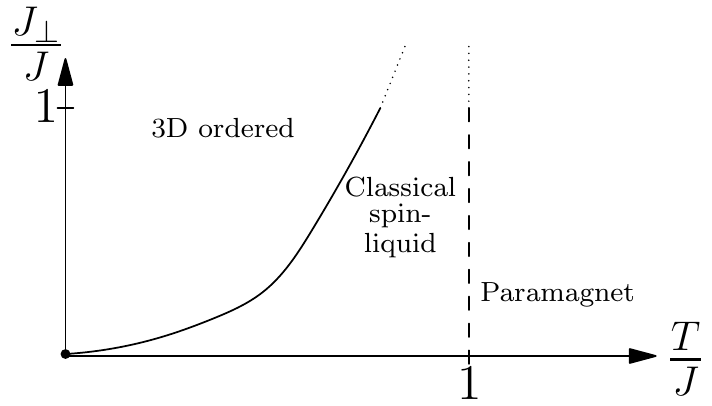}
	\caption{Schematic phase diagram for stacked triangular lattice Ising antiferromagnets. The full line represents the phase boundary, and the dashed line indicates a smooth crossover.}
	\label{fig:schematic_phase_diag}
\end{figure}

We are concerned with the statistical mechanics of these models as a function of temperature $T$ and the interaction strength ratio $J_\perp/J$. At $J_\perp/J=1$, one expects ordering below a temperature $T_{\rm c} \sim J$, while for $J_\perp/J=0$ the system of uncoupled layers remains disordered at all temperatures. A schematic phase diagram obtained by interpolating between these limits has the form shown in Fig.~\ref{fig:schematic_phase_diag}. For $J_\perp/J \ll 1$ the paramagnetic phase extends to temperatures $T\ll J$. In this regime, spins are highly correlated within each layer. Our objectives are to understand interlayer correlations and the form of the phase boundary for small $T/J$ and $J_\perp/J$, in each of the three stackings. For the two frustrated stackings we find that at small $J_\perp/J$ there is a low-temperature regime in which the correlation lengths, both in-layer and inter-layer, are much larger than the lattice spacing. A system in this regime is termed a {\it cooperative paramagnet} or {\it classical spin liquid}. This regime is smoothly connected to the conventional paramagnetic state at $T\gg J$ but distinguished from it by strong correlations.

For orientation it is useful to have a simple approach that gives an initial indication of likely behaviour. Mean field theory can often be employed in this way but fails here, wrongly predicting an ordering temperature set by $J$, even for small $J_\perp$. An alternative that has been widely applied in geometrically frustrated magnets is the self-consistent Gaussian approximation (SCGA) \cite{SCGA}. It is well-controlled only for $n$-component spins at large $n$, but is known in some instances to be quite accurate even for Ising systems \cite{isakov}. In the SCGA, correlations are given in terms of the interaction matrix ${\bf J}$ and the inverse temperature $\beta$ by
\begin{align}
\langle \sigma_{i} \sigma_{j}\rangle = \left[\left(\beta \mathbf{J} + \lambda \mathbf {I} \right)^{-1}\right]_{ij}\,.\label{scga_correlator}
\end{align}
Here, $\lambda$ is a parameter fixed by the consistency condition $\langle \lvert \sigma_i \rvert^2 \rangle = 1$, which can be satisfied throughout the paramagnetic phase. Using a spectral decomposition of $\bf J$ in terms of its eigenvalues $\epsilon_{\mathbf{q}}^l$ and eigenvectors $u^{l}_{\mathbf{q}}\left(\alpha\right)$, where $\alpha$ labels sites within a unit cell and $l$ labels the bands of $\bf J$, the SCGA expression for the structure factor is
\begin{align}\label{SCGAS(q)}
S(\mathbf{q}) 	&=\frac{1}{N}\sum_{i,j} \left[\left(\beta \mathbf{J} + \lambda \mathbf {I} \right)^{-1}\right]_{ij}e^{i\mathbf{q}\cdot(\mathbf{r}_i-\mathbf{r}_j)}\nonumber\\
&= \sum_{l,\alpha,\alpha'} \frac{u^{l*}_{\mathbf{q}}\left(\alpha\right) u^{l}_{\mathbf{q}}\left(\alpha'\right)}{\beta\epsilon_{\mathbf{q}}^l+\lambda}
\end{align}
From this it is apparent [barring cancellations in the sum $\sum_{\alpha,\alpha'}u^{l*}_{\mathbf{q}}\left(\alpha\right) u^{l}_{\mathbf{q}}\left(\alpha'\right)$] that maxima in $S({\bf q})$ arise from minima in $\epsilon_{\mathbf{q}}^l$. 

Applying the SCGA to stacked triangular lattice antiferromagnets, the paramagnetic phase extends to temperatures $T\ll J$ if $J_\perp \ll J$, and in this regime the maxima in $S({\bf q})$ are sharply defined. To find the location of these maxima in reciprocal space, we examine the minima of $\epsilon_{\mathbf{q}}^l$. We take axes with $\hat{z}$ perpendicular to the triangular layers, unit spacing between neighbouring layers for the $aaa$ and $abc$ stackings, and unit spacing between neighbouring $a$-layers in the $abab$ stacking, which has two sites per primitive unit cell. We choose in-plane lattice vectors 
\be\label{latticevectors}
{\bf a}_1=(1,0,0) \qquad {\rm and}\qquad {\bf a}_2=(1/2,\sqrt{3}/2,0),
\ee
The corresponding in-plane reciprocal lattice vectors are ${\bf A}_1 = 2\pi(1,-1/\sqrt{3},0)$ and ${\bf A}_2 = 2\pi(0,2/\sqrt{3},0)$. We use $\bm{\delta}$ to denote the separation vector between neighbouring sites in adjacent layers. Hence $\bm{\delta} = (0,0,1)$, $(1/2,1/(2\sqrt{3}),1)$ and $(1/2,1/(2\sqrt{3}),1/2)$ for the $aaa$, $abc$ and $abab$ stackings, respectively. 

The contribution to $\epsilon_{\mathbf{q}}^l$ from in-plane interactions  has a minimum at the $K$-points of the triangular lattice Brillouin zone: 
\be\label{Kpoints}
\mathbf{K} = (\frac{ 4 \pi}{3},0) \qquad {\rm and}\qquad \mathbf{K}^\prime = (\frac{ 2 \pi}{3},\frac{ 2 \pi}{\sqrt{3} }). 
\ee
Upon inclusion of  small $J_\perp$, these minima evolve in different ways for each of the stackings we consider. For the $aaa$ stacking, they lie at isolated points, undisplaced in-plane and at $q_z =\pi$. For the frustrated stackings, their locations can be specified in terms of the wavevector-dependent complex scalar $\zeta = 1+e^{i\mathbf{q}\cdot\mathbf{a}_1}+e^{i\mathbf{q}\cdot\mathbf{a}_2}$. In the $abc$ case they lie on the curve
\begin{equation}\label{Qabc}
\zeta = -\frac{J_\perp}{J}e^{i\mathbf{q}\cdot\bm{\delta}}
\end{equation}
and in the $abab$ case they lie on
\begin{equation}\label{Qabab}
\lvert \zeta\rvert =  \frac{J_\perp}{J},\quad q_z=0\,.
\end{equation}
These conditions respectively define helices and rings centred on the zone corners, as shown in Fig.~\ref{fig:scga_plots}. Further discussion of the interaction matrix eigenvalues is given in Appendix~\ref{scga_app}.
\begin{figure}
	\includegraphics[height=0.3\textwidth]{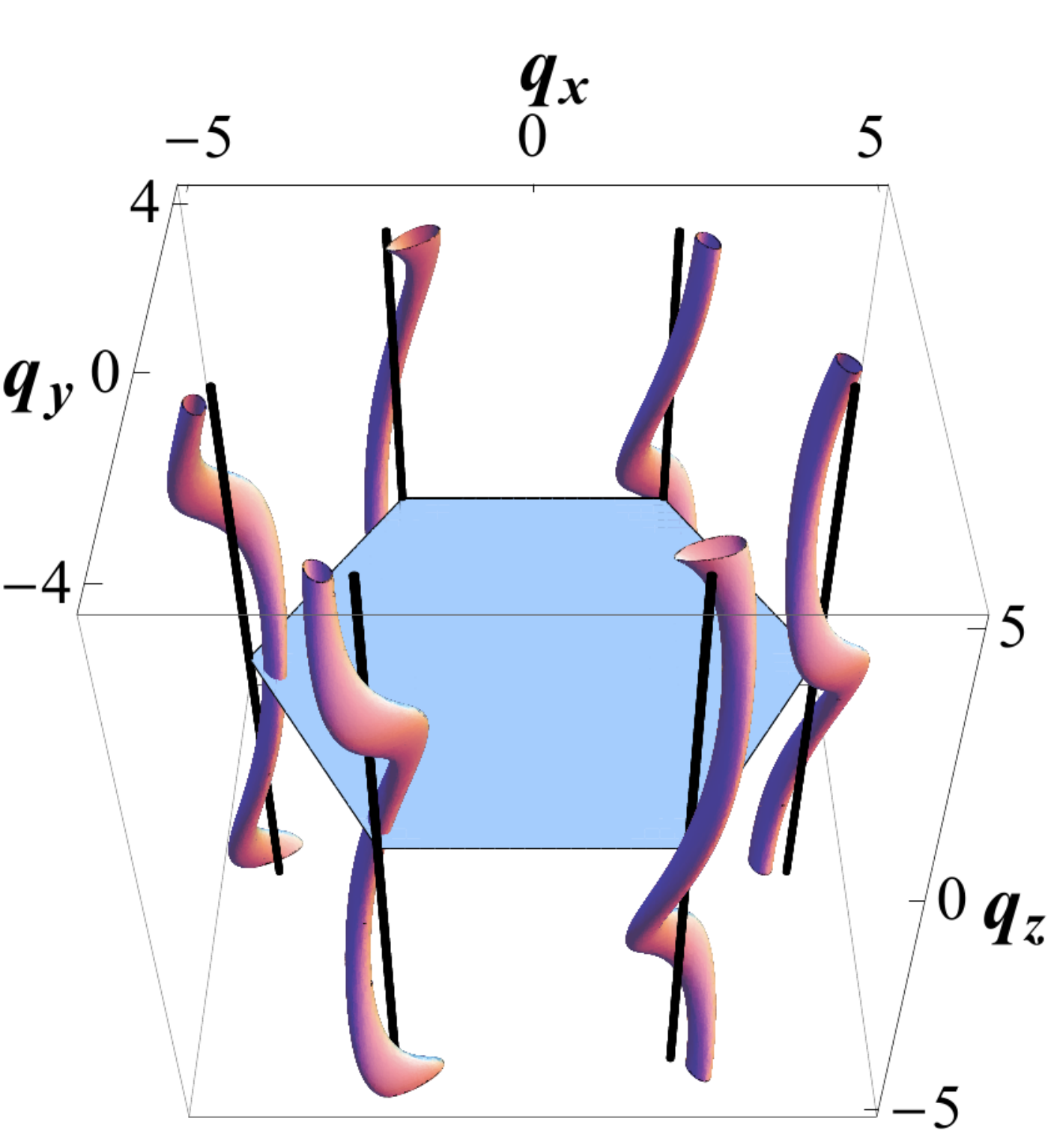}
	\includegraphics[width=0.37\textwidth]{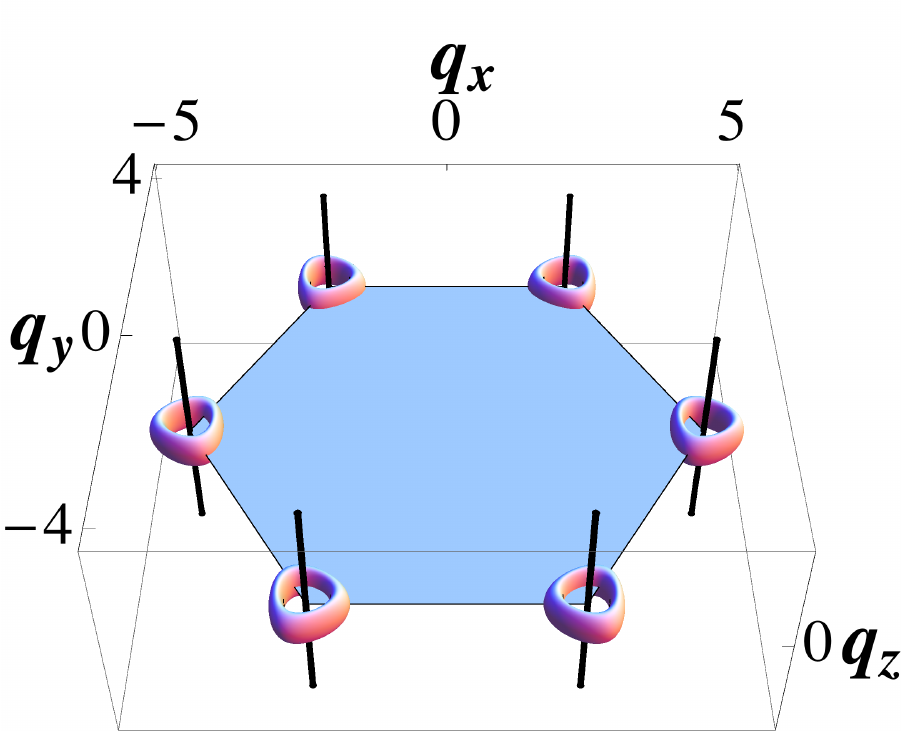}
	\caption{Location of surfaces on which eigenvalues of the interaction matrix are constant and close to the minimum, for (top) the $abc$ stacking, and (bottom) the $abab$ stacking, at $J_\perp/ J = 0.2$.
		} 
	\label{fig:scga_plots}
\end{figure}

\section{Monte Carlo Simulations}\label{MC}

We use extensive Monte Carlo simulations to find the ordering temperature for all three models and to study correlations in the paramagnetic phase of models on the $abc$ and $abab$ stacked lattices.
The primary observables computed are the energy $E$, specific heat $C$, and the structure factor $S(\mathbf{q})$, 
which is obtained from the Fourier transform of magnetisation 
\begin{equation}
 \tilde{\sigma}(\mathbf{q}) = \sum_{i} e^{i\mathbf{q}\cdot\mathbf{r}_i}\sigma_i\label{fourier_spins}
\end{equation}
as
\begin{equation}
S\left(\mathbf{q}\right) = \frac{1}{L^2L_z}\langle \lvert\tilde{\sigma}(\mathbf{q})\rvert^2 \rangle.\label{struct_fact}
\end{equation}

Because of the complex energy landscape arising from geometrical frustration, we employ a parallel tempering algorithm with single-spin-flip Metropolis dynamics\cite{Swendsen1986,Parisi1992}. Specifically, we simulate $N_\text{r}$ replicas (taking $ N_\text{r} \sim 100$) at geometrically spaced temperatures, with the highest temperature  $\sim 5J$. A Monte Carlo sweep involves one single-spin-flip attempt per site, followed by one parallel tempering swap attempt between replicas at adjacent temperatures. A system  consists of $L_z$ rhombic layers, each of size $L \times L$ lattice constants, with periodic boundary conditions in all directions. A typical simulation treats $\approx 10^5$ sites $\left( L = 72-200, L_z = 12-48 \right)$ using $10^5$ sweeps. We measure $E$ and $C$ each sweep, and $S(\mathbf{q})$ every $N_\text{r}$ sweeps.  Further details of the data analysis are presented in Appendix \ref{num_app}.

\subsection{Ordering Transition}

\begin{figure}
	\includegraphics[width=0.45\textwidth]{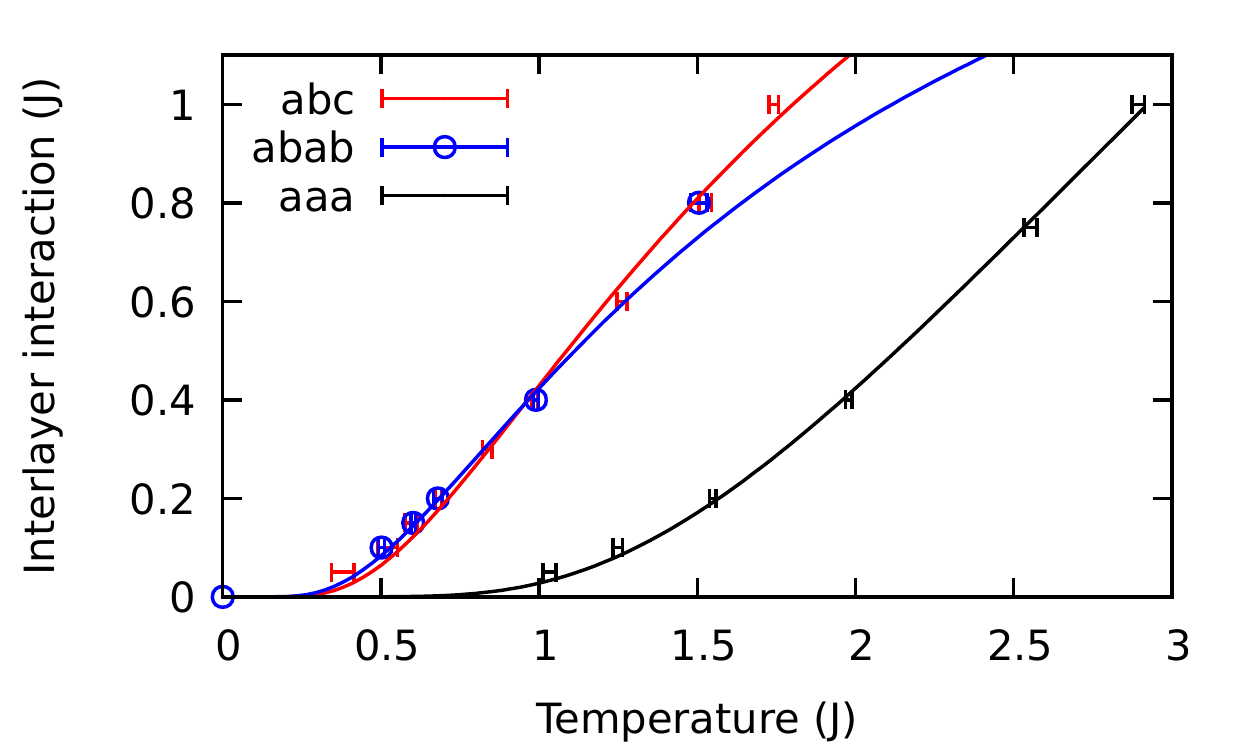}
	\caption{Phase boundaries for the unfrustrated ($aaa$) and frustrated ($abc$ and $abab$) stackings. Points: data from Monte Carlo simulations. Lines: fits to theory of Sec.~\ref{phasediagram}; see discussion in Sec.~\ref{discussion}.
	}
	\label{fig:phase_diagram}
\end{figure}

Phase diagrams as a function of $T$ and $J_\perp$ are shown in Fig.~\ref{fig:phase_diagram}  for both the unfrustrated ($aaa$) and the frustrated ($abc$ and $abab$) stackings. For a given strength of interlayer coupling, the ordering temperature (determined from the maximum of the heat capacity) is much lower in the systems with frustrated stackings compared with the unfrustrated one. In addition, over most of the range of $J_\perp/J$ studied, the transitions in the systems with frustrated stackings are strongly first order:  the probability distribution of the energy is strongly bimodal at the transition unless $J_\perp/J\ll 1$. The discontinuity in the energy at the transition decreases with decreasing $J_\perp$, and for $J_\perp \lesssim 0.05J$ the order of the transition is not discernible from the simulations. Differences in transition temperature between the two frustrated stackings are very small for $J_\perp/J\leq 1$. Our results for the $abc$ stacking at $J_\perp=J$ can be compared with earlier work on the fcc lattice, and are in good agreement with the transition temperature of $T_c \approx 1.72J$ found in Refs.~\onlinecite{FCC1980,FCC2006}. 

\begin{figure}
	\includegraphics[width=0.48\textwidth]{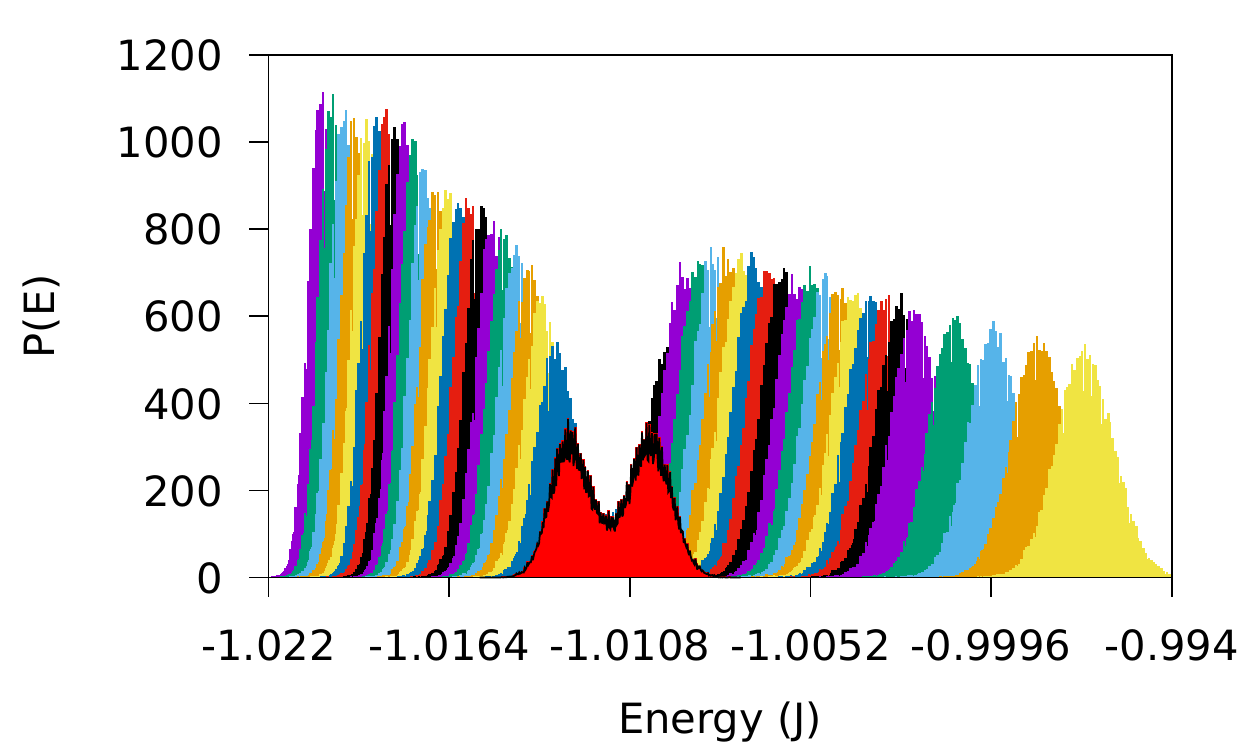}\\
	\includegraphics[width=0.48\textwidth]{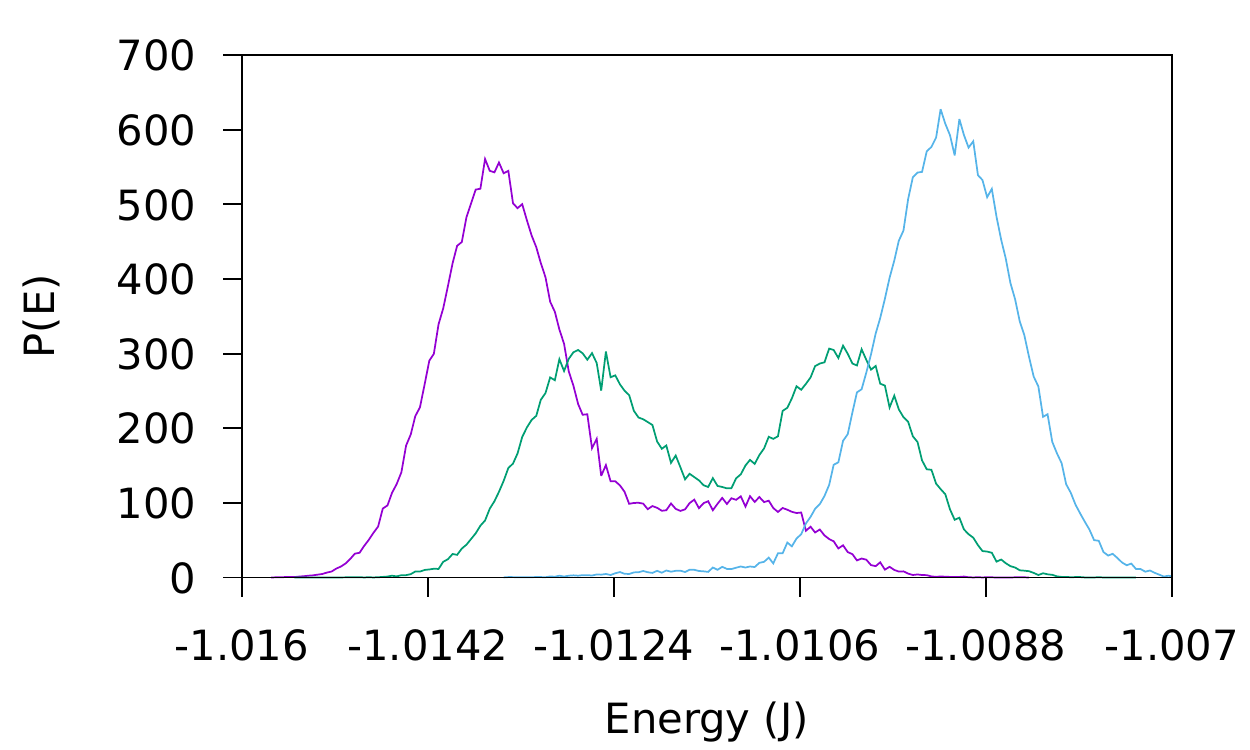}
	\includegraphics[width=0.48\textwidth]{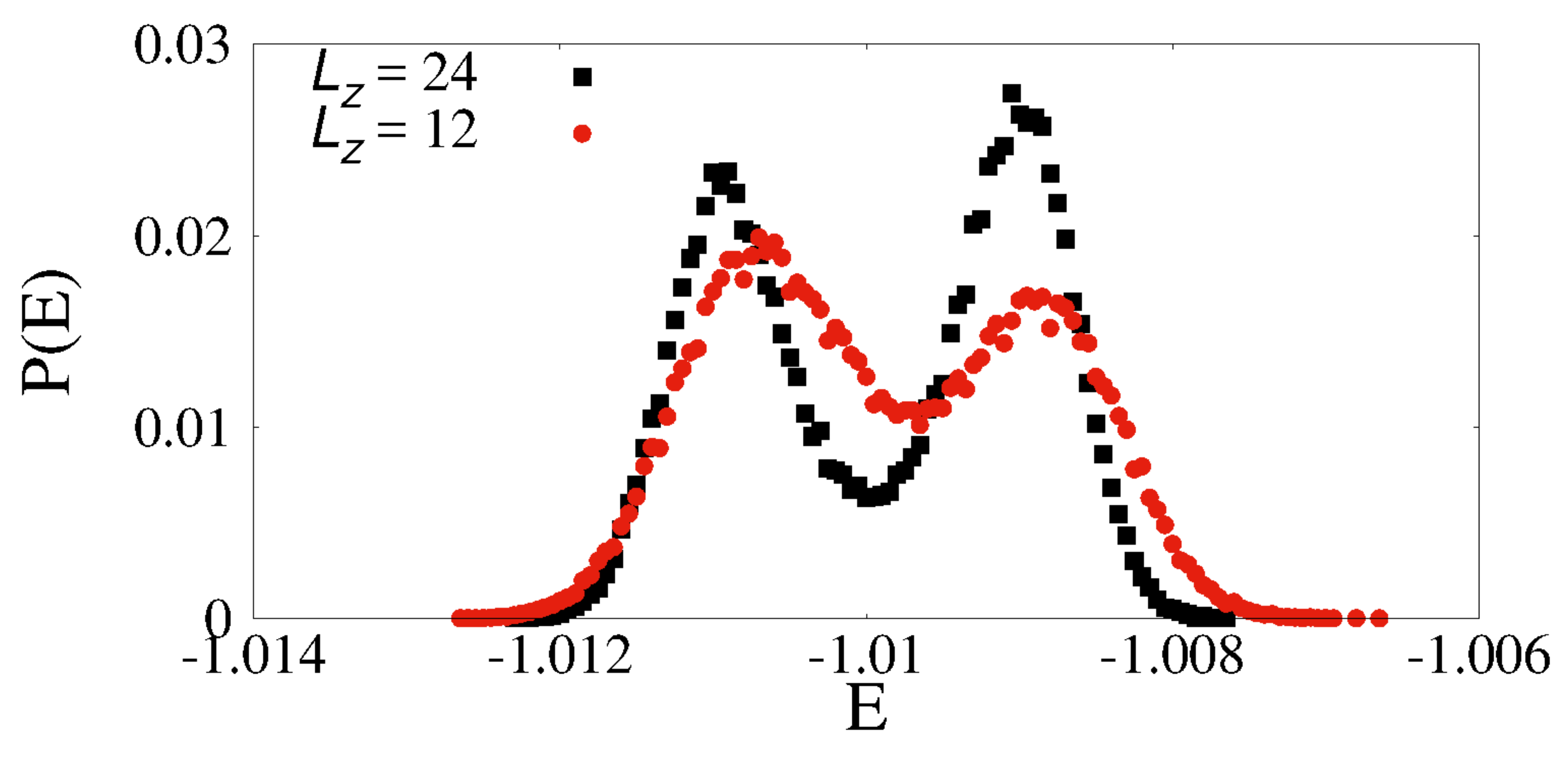}
	\caption{Distributions $P(E)$ of energy $E$ for the $abc$ stacking with $J_\perp = 0.1J$. Top: temperatures in the range $0.4J\leq T \leq 0.65J$. Middle: temperatures $T = 0.50J$, $0.51J$ and $0.52J$ close to the transition. System size $L=72$, $L_z = 12$. The distribution closest to the transition is solid red outlined in black, centered around E = -1.011 in the top panel, and is the middle temperature in the middle panel. Its bimodal form indicates a first-order transition. Bottom: finite size effects, illustrated for $L=96$ and $T=0.52J$.}
	\label{fig:output206_equilibration}\label{fig:output206_transition}
\end{figure}

Examples of the energy distribution at different temperatures are shown in Fig.~\ref{fig:output206_equilibration}. We monitor the overlap of distributions at adjacent temperatures in the parallel tempering scheme, as substantial overlap is a requirement for effective exchange of replicas. The top panel demonstrates that this is the case in our simulations. At a first-order transition, the energy distribution is bimodal. The middle panel illustrates this. Finite size shifts in our estimates of the transition temperature are a few percent, as indicated by a comparison of the middle and lower panels.

\subsection{Correlation functions}

A characteristic feature of classical spin liquids is the presence of strong correlations and a large correlation length, without long-range order or proximity to a critical point. In this subsection we present correlation functions and correlation lengths for TLIAFMs with frustrated stackings, determined from Monte Carlo simulations.

\subsubsection{The $abc$ stacking}\label{abc}

\begin{figure*}
	\includegraphics[width=0.99\textwidth]{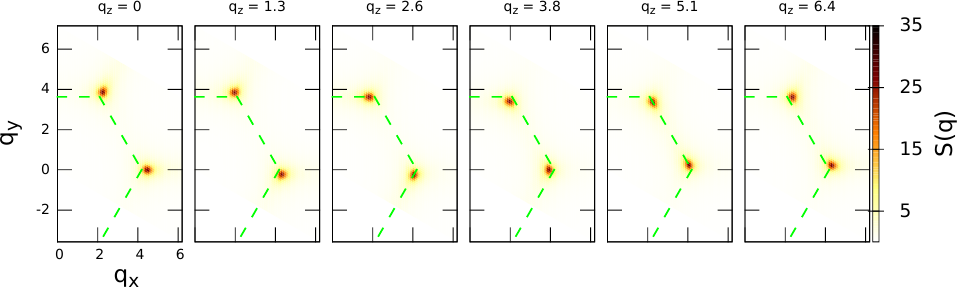}
	\caption{Cross-sections of structure factor at constant $q_z$ in a system with $abc$ stacking. For each $q_z$, sharp maxima in $S({\bf q})$ occur near the Brillouin zone boundary, which is shown as a green dashed line. As $q_z$ increases, the maxima precess around the zone corners without significant change in intensity, indicating that they form helices in the three-dimensional reciprocal space. Parameter values are $J_\perp = 0.2J$, $T = 0.8J$, $L=72, L_z = 12$; for this value of $J_\perp$, $T_c = (0.68 \pm 0.01)J$.}
	\label{fig:structure_factor_slices}
\end{figure*}

The behaviour of the structure factor for a system with $abc$ stacking in the classical spin-liquid regime is illustrated in Fig.~\ref{fig:structure_factor_slices}. Combining information from the series of slices in reciprocal space that are shown in this figure, it is apparent that maxima in $S({\bf q})$ lie on helices in reciprocal space. The axes of these helices pass through corners of the triangular-lattice Brillouin zone. 

To analyse this behaviour quantitatively, we extract a reciprocal-space radius $Q$ for the helix and a correlation length $\xi_\perp$ by fitting data for $S({\bf q})$ near the maxima to a sum of in-plane Lorentzians 
\begin{equation}\label{abcS}
S\left(\mathbf{q}\right) = \frac{I}{\xi_\perp^{2}\left(\mathbf{q}_\perp-\mathbf{q}^0_\perp(q_z)\right)^2 + 1}
\end{equation}
from each helix. Provided any dependence of $|{\bf q}^0(q_z)|$ on $q_z$ is weak, we can make the identification 
$Q=|{\bf q}^0(q_z)|$. (See Appendix~\ref{num_app} for further discussion.)

\begin{figure}
	\includegraphics[width=0.48\textwidth,page=1]{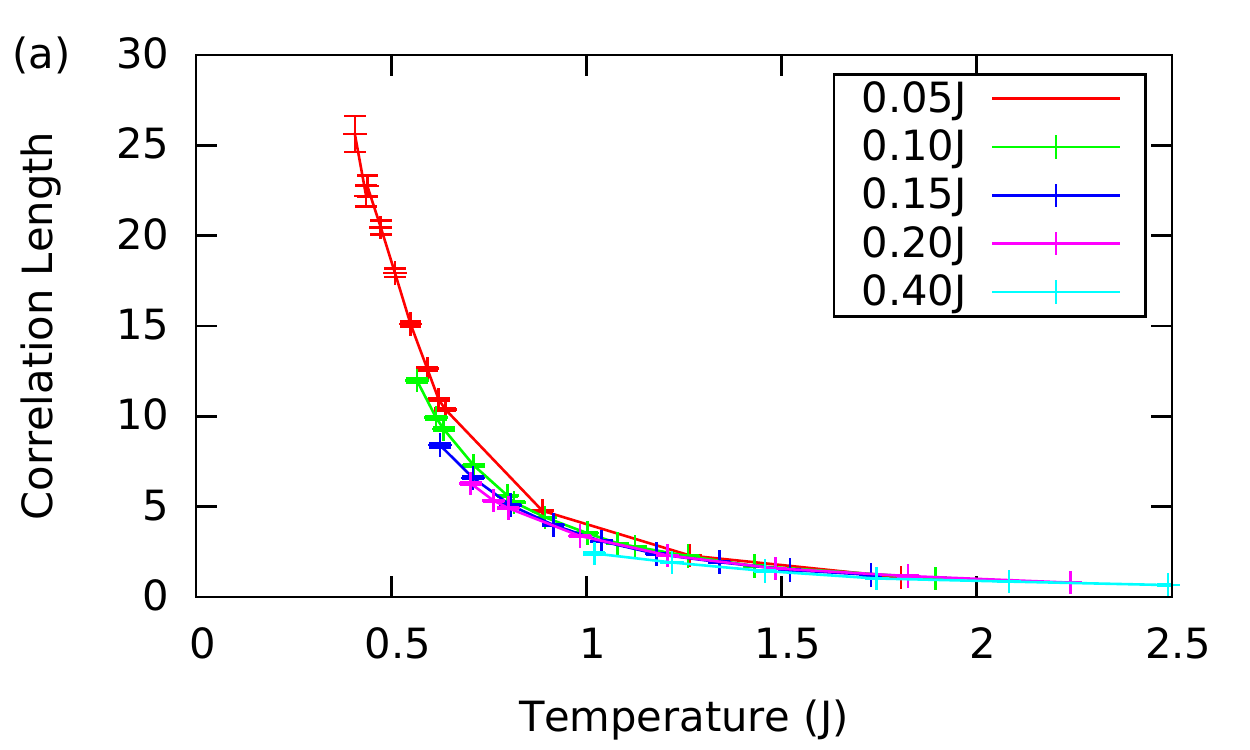}
	\includegraphics[width=0.48\textwidth,page=2]{parameter_plots.pdf}
	\caption{$(a)$ Correlation length $\xi_\perp$ and $(b)$ helix radius $Q$, as a function of temperature for various values of $J_\perp$ in the $abc$ stacking. $\xi_\perp$ is measured in units of lattice spacing, $Q$ in units of inverse lattice spacing. Dashed lines are SCGA predictions for $Q$ from Eq.~(\ref{Qabc}). Results were obtained in a system of size $L=72, L_z = 12$. Data for each value of $J_\perp$ extend to the lowest temperature employed in parallel tempering that was above $T_{\rm c}$.}
	\label{fig:fit_params_v_T}
\end{figure}

Results are shown in Fig.~\ref{fig:fit_params_v_T}. The correlation length $\xi_\perp$ increases rapidly with decreasing $T$ for $T\lesssim J$, as demonstrated in Fig.~\ref{fig:fit_params_v_T}(a). It reaches large values within the paramagnetic phase if $J_\perp/J$ is small. Its dependence on $J_\perp$ at fixed $T$ is very weak, because its value is determined by the density of vortices in the height field [see  Sec.~\ref{behaviour}] and for $J_\perp \ll J$ this in turn is controlled mainly by the value of $T/J$.
The variation of the helix radius $Q$ with $J_\perp$ and $T$ is illustrated in Fig.~\ref{fig:fit_params_v_T}(b). Its value is given quite accurately by the SCGA,  Eq.~(\ref{Qabc}), for $T\gtrsim J$, and shows a small increase with decreasing temperature.

\begin{figure}
	\includegraphics[width=0.48\textwidth]{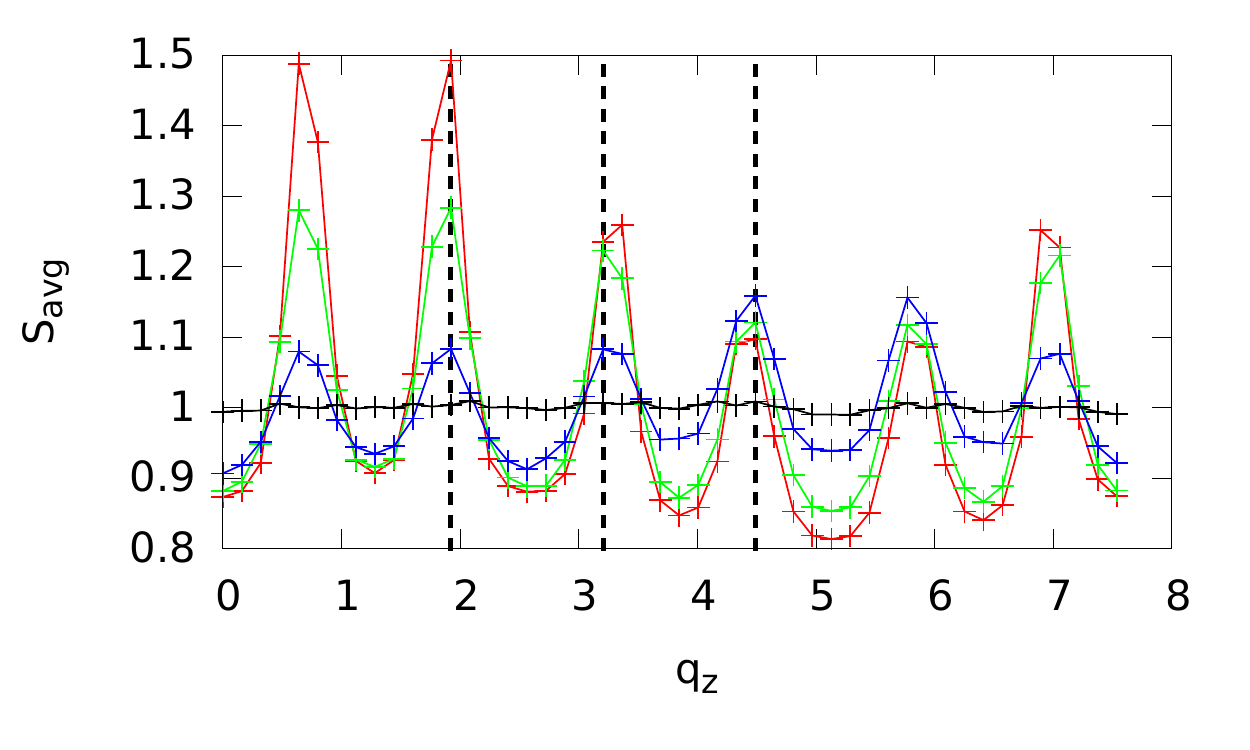}
	\includegraphics[width=0.48\textwidth]{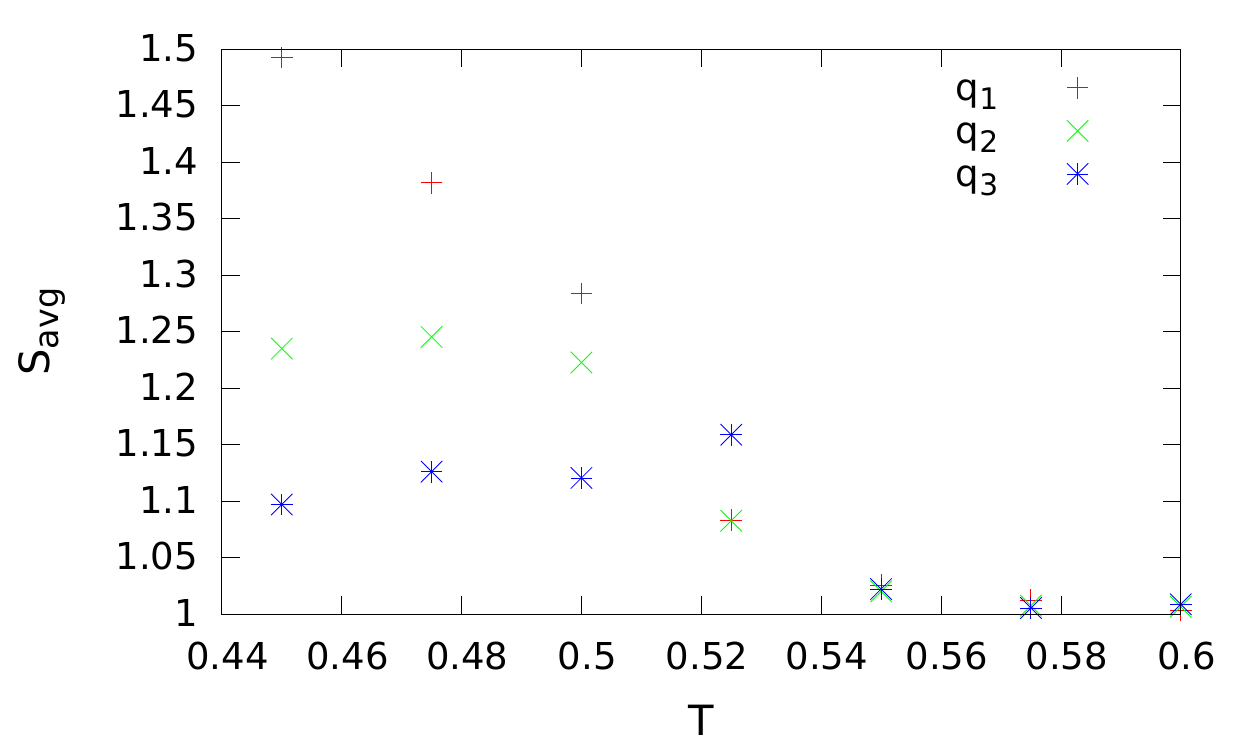}
	\caption{Development of Bragg peaks in the ordered phase for the $abc$ stacking. Top: $S_\text{avg}(q_z)$ [Eq.~(\ref{Savg})] as a function of $q_z$ at four selected temperatures near the transition, in a system with $J_\perp = 0.1J$. Data are for 
		$T =0.45J$, $0.5J$, $0.52J$ and $0.6J$, in order of decreasing peak intensity, and the transition temperature is $T_{\rm c} \approx 0.54J$. Bottom: $S_\text{avg}(q_z)$ as a function of $T$ for the three values of $q_z$ that are marked with vertical dashed lines in the top panel. Results for both panels were obtained in a system of size $L=36$, $L_z = 48$.}
	\label{fig:q_z_modulations}
\end{figure}

In the ordered phase, Bragg peaks are expected in the structure factor, in place of a continuous distribution of weight on helices. We probe the evolution between the two behaviours by computing
\begin{equation}\label{Savg}
S_\text{avg}(q_z) = \frac{1}{L^2}\sum_{q_x,q_y}S(\mathbf{q})\,.
\end{equation}
Results in Fig.~\ref{fig:q_z_modulations} show the rapid development of Bragg peaks as temperature is lowered through the transition. Although we believe  that the transition is first order for the value of $J_\perp/J$ studied here, discontinuities are not apparent in the temperature dependence of $S_\text{avg}(q_z)$, presumably because of finite-size rounding. Indeed, since evaluation of correlation functions is more computationally demanding than calculation of energy distributions, the results presented in Fig.~\ref{fig:q_z_modulations} are for smaller system size than those in Fig.~\ref{fig:output206_equilibration}; we find (data not shown) that the energy distribution at the transition is not bimodal for the smaller size.

\subsubsection{The $abab$ stacking}\label{ab}

Because the $abab$-stacked lattice has two sites in a primitive unit cell, the relation between fluctuations and correlations is less direct than for the $abc$ stacking, in which the unit cell has a single site. More specifically, the form of $S({\bf q})$ is affected by interference between contributions from the two sites. Within the SCGA, this is apparent from Eq.~(\ref{SCGAS(q)}), where contributions involving a given eigenvalue $\epsilon^l_{\bf q}$ of the interaction matrix are weighted by a sum $\sum_{\alpha,\alpha^\prime} u^{l*}_{\mathbf{q}}\left(\alpha\right) u^{l}_{\mathbf{q}} \left(\alpha^\prime\right)$ that includes both site-diagonal ($\alpha=\alpha'$) and interference ($\alpha\not=\alpha'$) terms. In order to eliminate these interference effects and expose fluctuations in the $abab$ stacking in a simple way, we compute the structure factor using contributions only from one of the two sites in each unit cell, by restricting the sum in Eq.~(\ref{fourier_spins}) to this set of sites.

\begin{figure*}
	\includegraphics[width=0.99\textwidth]{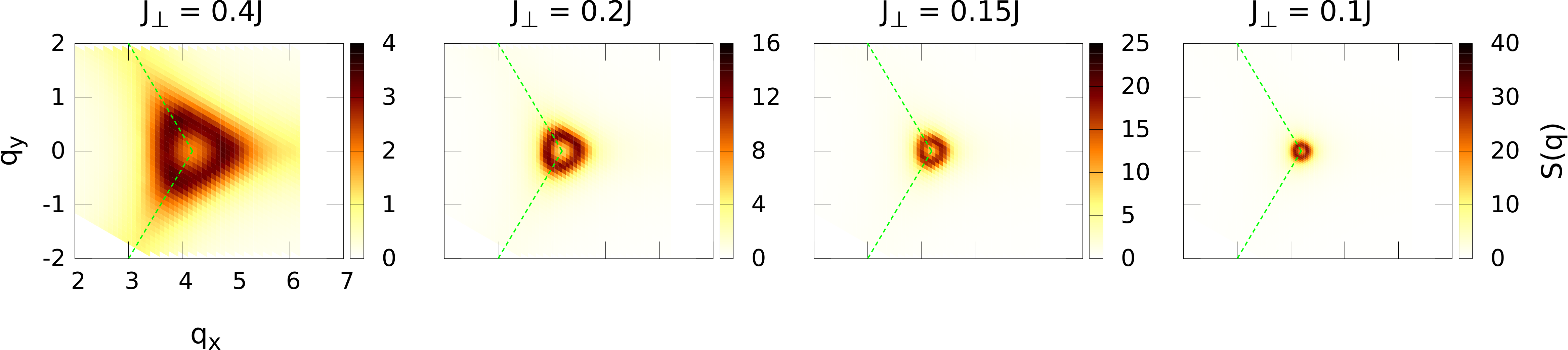}
	\caption{Cross-sections of structure factor at $q_z = 0$ for systems with varying $J_\perp$ in the $abab$ stacking. Intensity is maximum on a closed loop, which is approximately circular for small $J_\perp/J$ but develops triangular distortions with increasing $J_\perp/J$.  Data (from left to right) are for
$T = 1.14J$, $0.71J$, $0.64J$, $0.57J$, obtained in systems of size $L=72$, $90$, $90$, $204$ and $L_z = 12$, $12$, $30$, $6$. Note the changing intensity scale and increasing maximum intensity as $J_\perp$ and $T$ decrease. The ordering temperatures are $T_{\rm c}/J = 0.99\pm 0.008$, $0.680\pm 0.014$, $0.602\pm 0.007$ and $0.502\pm 0.01$.}
	\label{fig:structure_factor_slices_ring}
\end{figure*}

We expect from Eq.~(\ref{Qabab}) that this single-sublattice structure factor will have its maxima lying on closed loops in the $q_z=0$ plane. An overview of our data, illustrating this behaviour, is given in Fig.~\ref{fig:structure_factor_slices_ring}. 

\begin{figure}
	\includegraphics[width=0.45\textwidth]{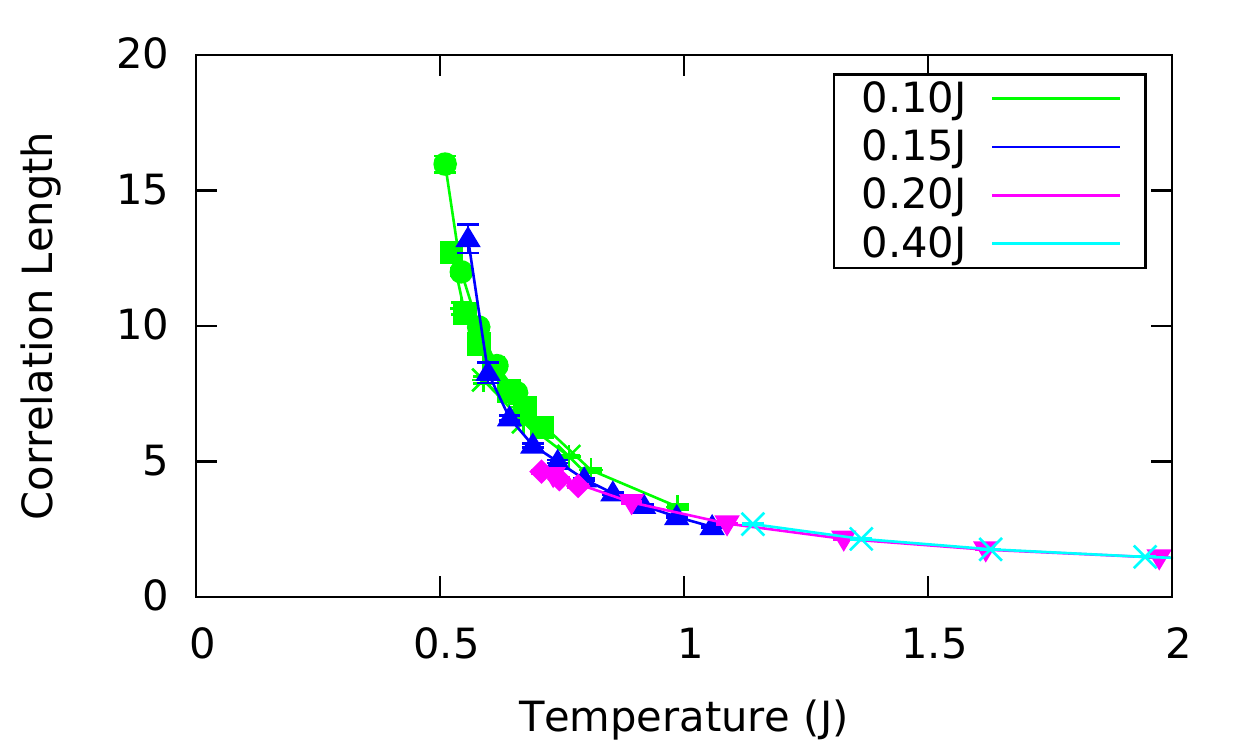}
	\caption{Correlation length, $\xi_\perp$, as a function of temperature for various values of $J_\perp$ in the $abab$ stacking, obtained by fitting to the functional form given in Eq.~(\ref{SQabab}).}
	\label{fig:parameter_ab_layer_fit}
\end{figure}

A simple way to extract a correlation length $\xi_\perp$ is by fitting data for $q_z=0$ and $q_x,q_y$ close to a selected Brillouin zone corner to the functional form
\begin{align}\label{SQabab}
S\left(\mathbf{q}\right) &= \frac{I}{\xi_\perp^{2}\left(Q -\lvert \mathbf{q}_\perp - \mathbf{K}\rvert\right)^2 + 1},
\end{align}
where $\mathbf{K}$ denotes the location of the Brillouin-zone corner and $Q$ specifies the reciprocal-space radius of the ring of intensity.
This fitting function provides a good description of the data for small values of $J_\perp/J$, where the maximum in the structure factor lies on a circle, but it does not capture the triangular distortions for larger $J_\perp/J$ that are apparent in the left-most panel of Fig.~\ref{fig:structure_factor_slices_ring}. As shown in Fig.~\ref{fig:parameter_ab_layer_fit}, and as for the $abc$ stacking, the resulting values of $\xi_\perp$ increase rapidly with decreasing temperature but vary little with $J_\perp$. 

\subsubsection{Self-consistent Gaussian Approximation}\label{scga}
\begin{figure}
	\includegraphics[width=0.35\textwidth]{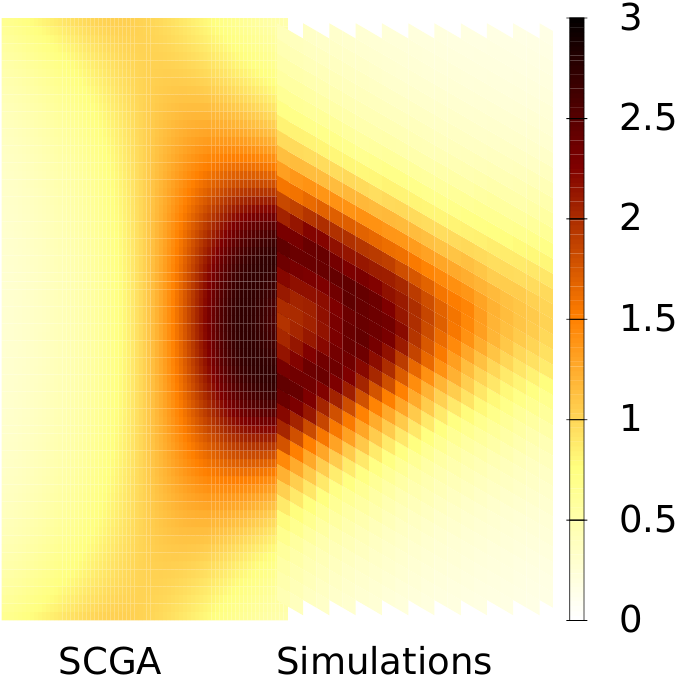}
	\caption{Comparison of SCGA and simulation results for $S(\mathbf{q})$ in the $abab$ stacking. $J_\perp = 0.4J, T = 1.36J$}
	\label{fig:side_by_side}
\end{figure}

\begin{figure}
	\includegraphics[width=0.45\textwidth]{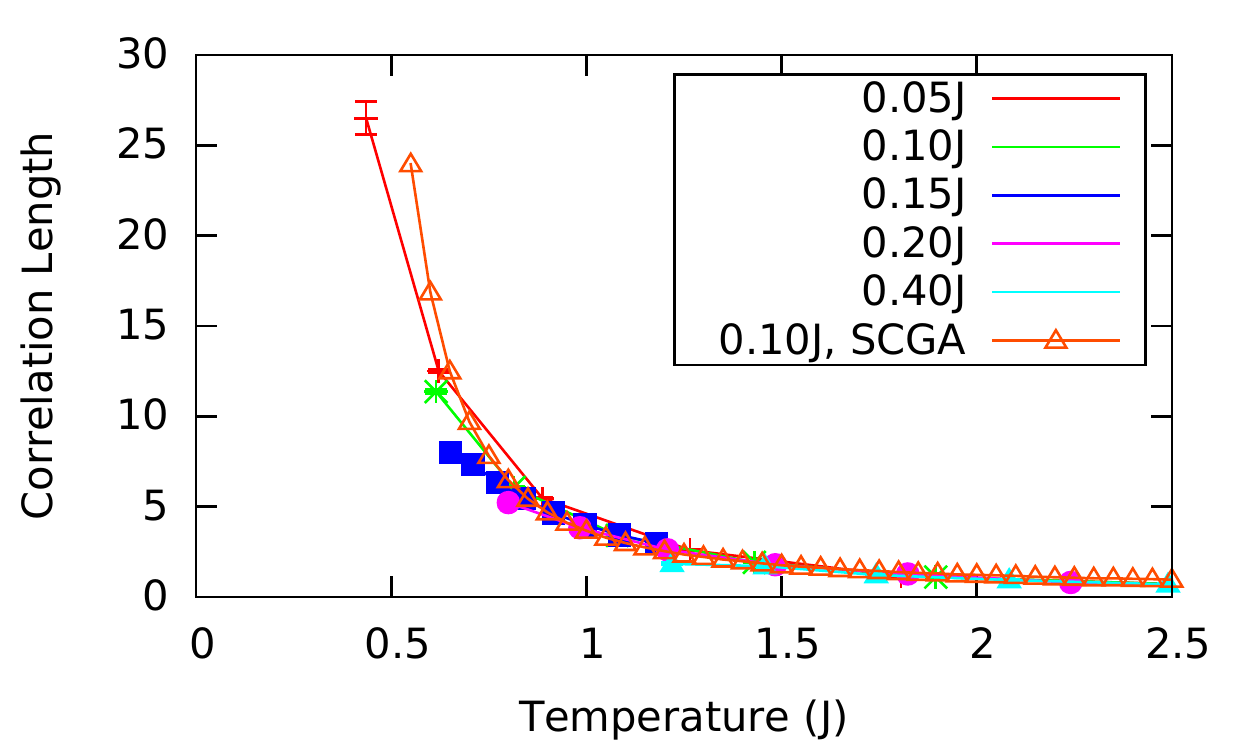}
         \includegraphics[width=0.45\textwidth]{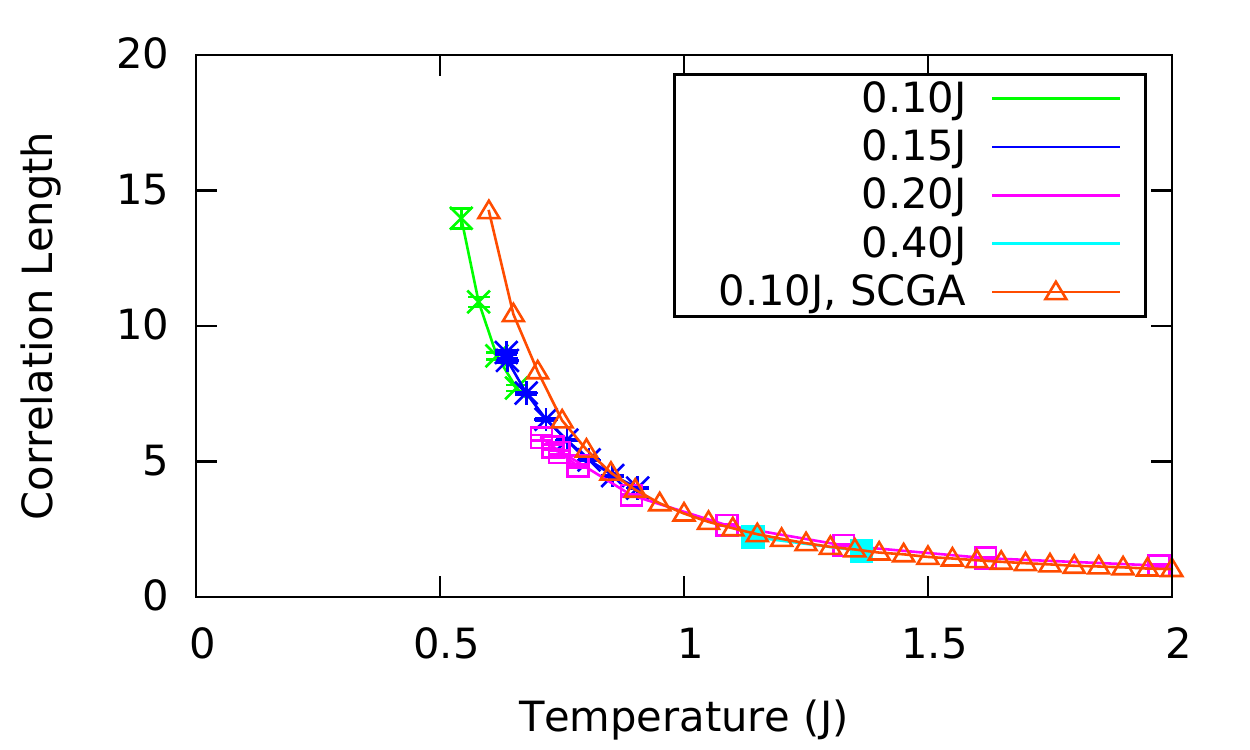}
	\caption{
Correlation length $\xi_\perp$ as a function of temperature for various values of $J_\perp$ as obtained from the SCGA. Top: abc stacking. Bottom: abab stacking. The data labeled `SCGA' have been derived by imposing the condition $\langle|\sigma|^2\rangle=1$ while the variable $\lambda$ is used as a fitting parameter in the other curves. For clarity, results from the first of these approaches are shown only at one value of $J_\perp$; agreement is similar at other values of $J_\perp$.}
	\label{fig:scga_parameter_plot}
\end{figure}

As discussed in section ~\ref{overview}, 
the SCGA provides a useful description of frustrated magnets in the strongly correlated regime. In particular, it offers a simple theoretical prediction for $S({\bf q})$, which we now show to be a good representation of our simulation data. We use the functional form of Eq.~(\ref{SCGAS(q)}) in two ways, which are distinct in principle but yield very similar results. One of these treats the variable $\lambda$ as a fitting parameter with respect to simulations; the other fixes its value using the SCGA condition $\langle |\sigma_i|^2\rangle=1$.

The SCGA form for $S({\bf q})$ is especially helpful at larger values of $J_\perp/J$, when detailed lattice effects are important. The results of these lattice effects for the $abc$ stacking include a dependence of the helix radius [${\bf q}_\perp^0(q_z)$ in Eq.~(\ref{abcS})] on $q_z$. For the $abab$ stacking they generate correlations that are not represented using the circular maximum in $S({\bf q})$ implied by the fitting function given in Eq.~(\ref{SQabab}). The SCGA gives a good description of this physics. 
Most notably, for the $abab$ stacking the SCGA fits are effective in capturing the triangular distortion of the rings, as demonstrated in Fig.~\ref{fig:side_by_side}.

Once the value of $\lambda$ is obtained from the fit, the correlation length can be extracted from the model. The results for $\xi_\perp$ are shown in Fig.~\ref{fig:scga_parameter_plot}. They agree to $\sim 10\%$ with those obtained by fitting the functional forms given in Eq.~(\ref{abcS}) and Eq.~(\ref{SQabab}) for the $abc$ and $abab$ cases respectively (see Figs.~\ref{fig:fit_params_v_T}a and \ref{fig:parameter_ab_layer_fit}). Alternatively, the value of $\lambda$ can be determined without reference to simulations, using the SCGA condition, yielding a theoretical prediction for $\xi_\perp$. From Fig.~\ref{fig:scga_parameter_plot}, it is apparent that both approaches to determining $\lambda$ yield very similar results.

\section{Height model}
\label{height}
\begin{table}[htb]
\begin{tabular}{|c|ccc|ccc|}
\hline
$h= \frac{1}{3} \left(h_A+h_B+h_C\right)$ & $h_A$ & $h_B$ & $h_C$& $\sigma_A$ & $\sigma_B$ & $\sigma_C$ \\
\hline
$0$ & $0$ & $1$ & $5$ & $+$ & $-$ & $-$ \\
$1$ & $0$ & $1$ & $2$ & $+$ & $-$ & $+$ \\
$2$ & $3$ & $1$ & $2$ & $-$ & $-$ & $+$ \\
$3$ & $3$ & $4$ & $2$ & $-$ & $+$ & $+$ \\
$4$ & $3$ & $4$ & $5$ & $-$ & $+$ & $-$ \\
$5$ & $0$ & $4$ & $5$ & $+$ & $+$ & $-$ \\
\hline
\end{tabular}
\caption{\label{3-sublattice-heights} Heights (all modulo 6) defined at triangle centres (column 1) and at triangle corners (columns 2-4), for each ground state spin configuration (columns 5-7) of the triangle. The spin configuration determines the height configuration up to a global shift. The sublattice labelling is illustrated in Fig.~\ref{HeightsMappings}b.\label{table}}  
\end{table}

We now turn to an analytical treatment of stacked triangular lattice Ising antiferromagnets. Although the SCGA, as demonstrated, provides a good approximate description, it is formally correct only for $n$-component spins in the large-$n$ limit. It is therefore not a natural starting point for a systematic approach. By contrast,  the height model provides a representation of a single-layer TLIAFM that is known to capture exactly the physics at low temperatures and long distances. Here we use the height model to construct a description of the multilayer system that allows for a controlled treatment of weak interlayer interactions.

Following Bl\"ote {\it et al.} \cite{heightmodel} and Zeng and Henley \cite{zeng}, we map ground states of a single layer Ising model onto states of a height model in such a way that spin configurations with long-range three-sublattice order correspond to flat height configurations. Because of frustration, domain walls can be introduced without energy cost between regions with different types of three-sublattice order. These domain walls correspond to steps in the height field. In a coarse-grained description, steps are represented by a gradient in the height field, and a large value for this gradient carries an entropy penalty. 

The mapping is conveniently described in two stages. First we define heights at the sites of the triangular lattice, as in Ref. \onlinecite{heightmodel}.  Second, following Ref. \onlinecite{zeng}, we average these site heights to define heights at the centres of triangles, obtaining a height model that is easily coarse-grained.  

\begin{widetext}

\begin{figure}
\begin{center}
\includegraphics[width=.165\linewidth]{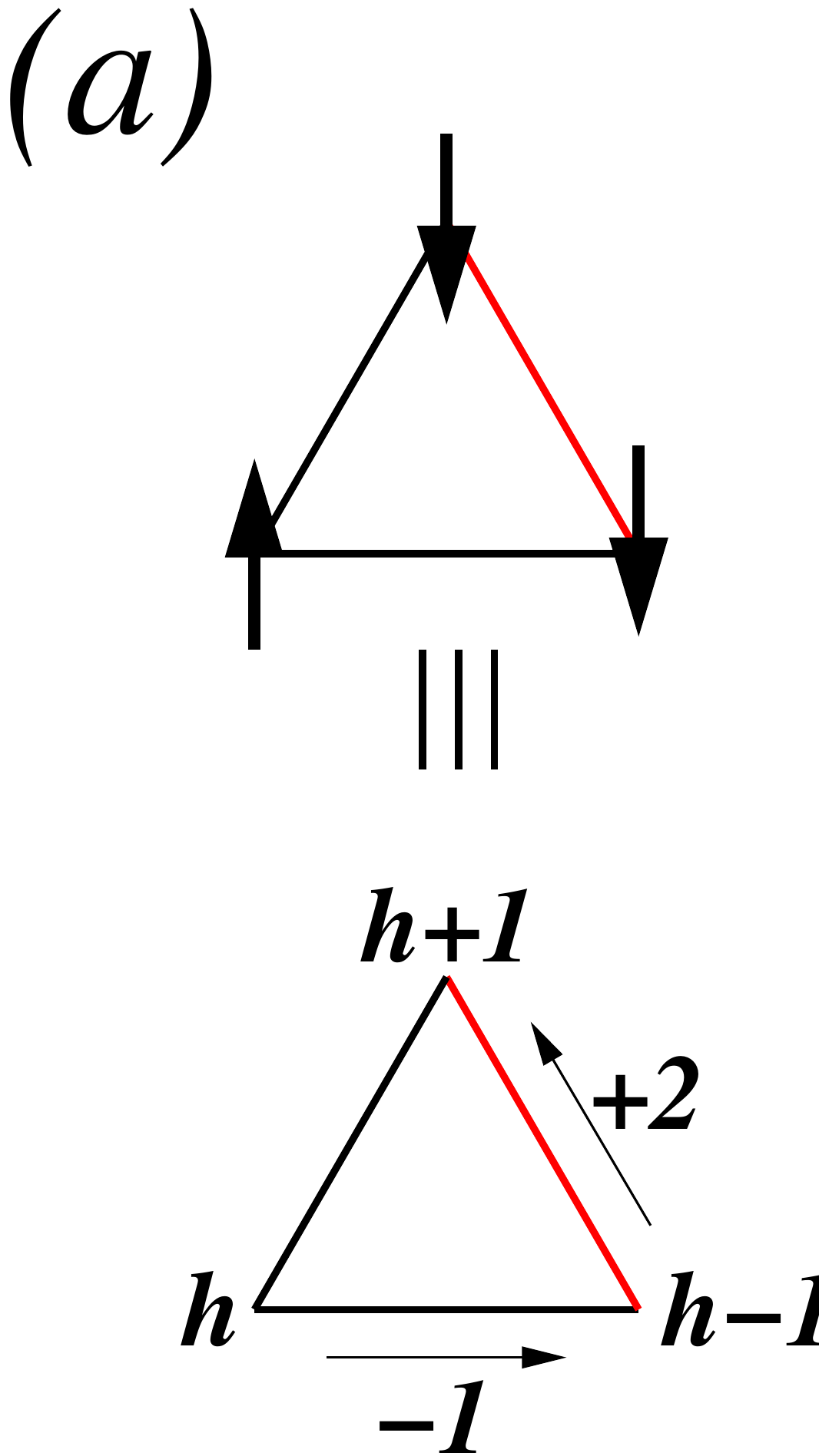}
\includegraphics[width=.7\linewidth]{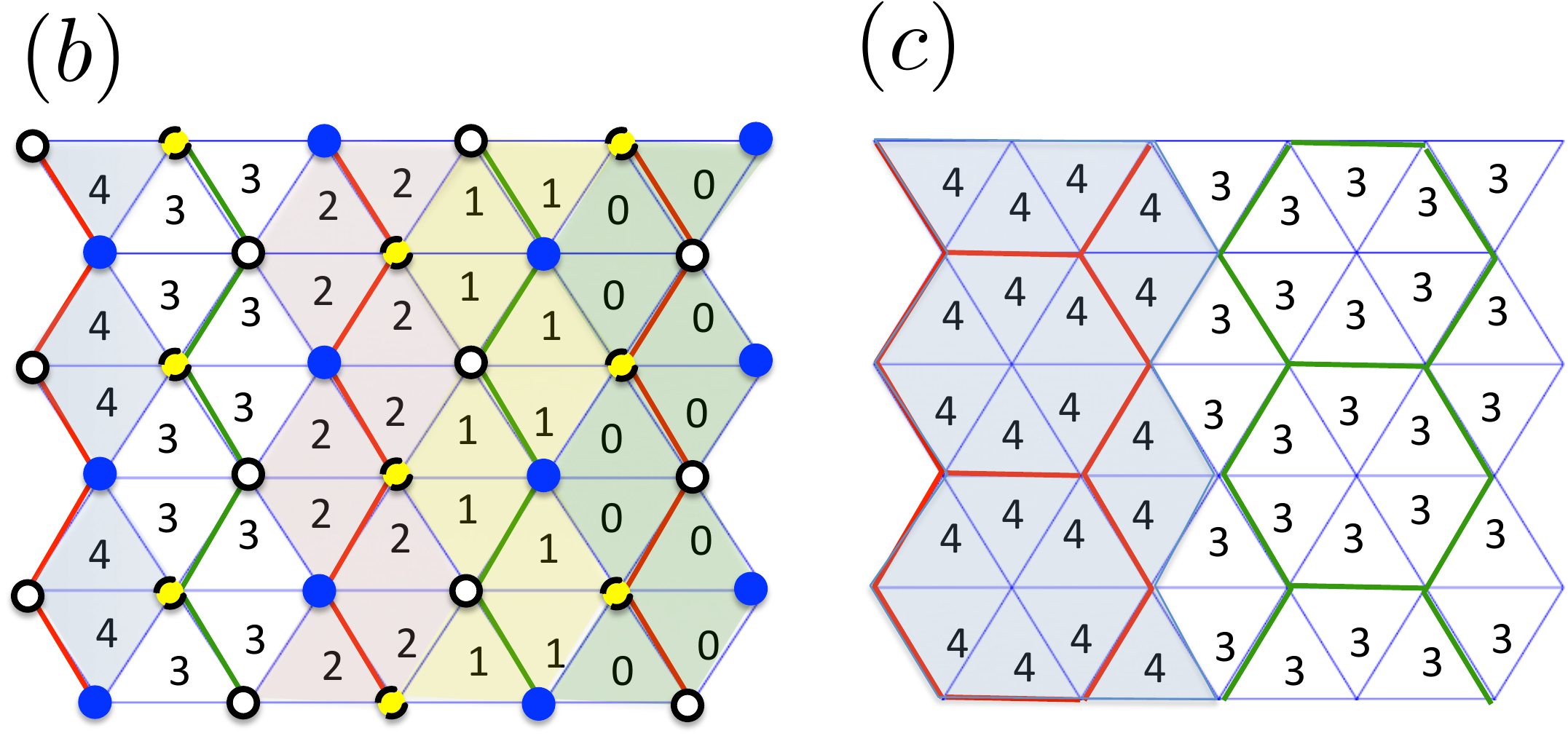}
\caption{Mapping from Ising spins to heights on the triangular lattice.  (a) The height field decreases by 1 (increases by 2) along an unfrustrated (frustrated) bond as an upward-facing triangle is traversed in the counter-clockwise direction.  This ensures that the net change in height field around each triangle is zero provided the triangle is in one of its ground states.  (b) and (c): Sample patterns of frustrated bonds and height fields. Green (red) edges on the triangular lattice represent frustrated bonds between pairs of up (down) spins; blue edges correspond to unfrustrated bonds.  The number at the centre of each triangle indicates the value of the corresponding height variable; the different shades highlight regions with different heights.  (b) shows a maximally tilted configuration (height variables at triangle centres decrease as rapidly as possible from left to right), corresponding to the true ground state for the $abc$ and $abab$ stackings; (c) shows a flat, three-sublattice ordered configuration with a single domain wall (height variables differ only along the domain wall). Our convention for the three sublattices of Table \ref{table}}  is indicated by the coloured circles: $A$= solid blue; $B$= yellow with dashed border; $C$= open white.
\label{HeightsMappings}
\end{center}
\end{figure}

\end{widetext}

To map from a spin configuration to heights at lattice sites, we first assign height zero to a reference site. The heights on all other sites of the lattice are then fixed by the requirement that the height difference between the neighbouring sites $i$ and $j$ is +2 if $\sigma_i = \sigma_j$, and $-1$ if $\sigma_i = - \sigma_j$ going anticlockwise around an up-triangle (or clockwise around a down triangle): see Fig.~\ref{HeightsMappings}a.  Heights at triangle centres are defined as the averages of site heights at vertices. The advantage of this locally-averaged height field is that ground states with three-sublattice order are exactly flat in these variables: see Fig.~\ref{HeightsMappings}c. In the following, we use the term `height field' exclusively for the locally-averaged quantity.

This mapping is summarised for a single triangle in Table \ref{3-sublattice-heights}. Here the sites of the triangular lattice are divided into three sublattices, labelled $A$, $B$, $C$ and indicated by the three colours of dots at the vertices in Fig.~\ref{HeightsMappings}b.  With this convention, the six ground states of each triangle are specified by the orientation of the spin on sublattice $A$ and the location of the frustrated bond. The ground-state spin configuration of a triangle fixes the value of the height $h$ at its centre modulo $6$.

The mapping is unique up to labelling conventions. Permuting the choice of  $A$, $B$ and $C$ sublattices (which results from lattice translations or rotations by $2 \pi/3$ about the centre of a triangle) corresponds to a global shift $h \rightarrow h +2$.   (By contrast, rotations about an axis passing through a site leave the labelling and hence the height field invariant.)   Shifting $h \rightarrow h+3$ corresponds to a global spin flip operation.  The remaining possibilities (shifting $h$ by $1$ or $5$) correspond to a combination of the global spin-flip and re-assignment of the three sublattices.  

The inverse mapping, from a height configuration to a spin configuration, can be expressed in terms of a function $f(h)$ and a constant $s_\alpha$. The function $f(h)\equiv f(h+6)$ takes the values $f(h)=+1$ for $h=-1,0,1$ and $f(h)=-1$ for $h=2,3,4$. The constant $s_\alpha$ takes values $s_A=0$, $s_B=2$ and $s_C=-2$ on sublattices $\alpha = A,B$ or $C$.  The spin orientation is then given by 
\ba  \label{Spin2H}
\sigma_{\alpha}  = f(h+ s_\alpha)\equiv f_\alpha(h).
\ea
For integer $h$ we can represent this function as $f (h) = \frac{4 }{3} \cos \frac{ \pi h}{3} - \frac{1}{3} \cos \pi h $. Note that since each spin is part of six triangles, to fully specify the mapping we must choose which triangle's height dictates which spin.  Reassuringly, one can verify that this choice is unimportant: when the height configurations are integers, and can change by at most $1$ between any pair of adjacent triangles, every convention yields the same spin configuration. 

Excitations of the spin model consist of triangles in which all spins are up, or all are down. They are represented by vortices in the height field, which is multi-valued in their presence: it increases by $6$ on going anticlockwise around an upward-facing excited triangle, and decreases by 6 around a down-facing triangle. An excited state produced from a ground state by reversing a single spin necessarily contains a vortex-antivortex pair, which may be separated by additional spin flips without further energy cost.

\subsection{Height-model analysis for a single layer} \label{TriHeightSec}

Before discussing stacked TLIAFMs, it is instructive to review how the height model captures the physics of a single triangular layer. 
The relative entropic weights of different height configurations are represented by the effective Hamiltonian\cite{heightmodel} 
\begin{equation} \label{1LayerH}
{\cal H} = \frac{K}{2} \int {\rm d}^2 {\bf r}\, |\nabla h({\bf r})|^2 +  \int {\rm d}^2 {\bf r}\, \tilde{V}(h)\,.
\end{equation}
We can determine the value of $K$ (and verify that (\ref{1LayerH}) captures the correct physics) by comparing the correlation functions of this model with $\tilde{V}(h)=0$ to those of the exact solution for the 2D TLIAFM.  Stephenson\cite{stephenson} has shown that at long distances
\be
\langle \sigma_{\alpha} ({\bf r}) \sigma_{\beta} ({\bf r'}) \rangle \sim \frac{\omega^s } {\sqrt{ |{\bf r} - {\bf r'} | } } + {\rm c.\ c.}\,,
\ee
where $s=(s_\alpha-s_\beta)/2$ and $\omega=e^{ i 2\pi/3 }$.
The dominant terms in the expression for the intra-sublattice spin-spin correlation function in terms of the height fields are
\ba \label{TheCors}
 \langle \sigma_\alpha ({\bf r}) \sigma_\beta ({\bf r'}) \rangle &\sim& \langle e^{ i \frac{\pi}{3} [h({\bf r}) - h ({\bf r'}) ]} \rangle \omega^s + {\rm c.\ c.} \n
 &\sim& \text{exp}\left[ - \frac{ 2 \pi}{ 36 K} \ln |{\bf r} - {\bf r'} | \right] (\omega^s + \omega^{-s}) \n
 &\sim&  |{\bf r} - {\bf r'} |^{ - \frac{ 2 \pi}{ 36 K} } (\omega^s + \omega^{-s})
\ea
Hence at zero temperature, to reproduce the long-wavelength properties of the exact solution, we take $K = \pi/9 $.

What about the potential term, which we ignored in the above calculation?    Microscopically the heights are integers; we can account for this by 
including the potential $\tilde{V}(h) = - v \cos (2 \pi h)$.  At short distances $v$ is large and positive.  At longer length scales the effective value of $v$ is determined by  the {scaling dimension} of the operator $\cos 2 \pi h$,  which can be deduced from the 2-point function
\ba
\langle \cos ( 2 \pi h {\bf r}) \cos( 2 \pi h {\bf r'} ) \rangle &\sim& |{\bf r} - {\bf r'} |^{ - \frac{ 2 \pi}{ K} }\n 
{\rm implying}\ \ \ 
\int d^2 r \cos( 2 \pi h {\bf r'} ) &\sim& L^{2 - \frac{\pi}{K}} \ \ .
\ea
This yields the scaling dimension $2 - \frac{\pi}{K} =  -7$ at $T=0$; hence the effective value of the coefficient $v$ decreases rapidly as we probe the system at longer length-scales, and its effect on the long-wavelength correlations is negligible.  

Finally, we can ask about behaviour at finite temperature.  To describe the system at finite temperature we must include the possibility of vortices in the height field.  Dropping $ \tilde{V}(h)$ in Eq.~(\ref{1LayerH}) but including vortices, we recover the physics of the 2D $xy$ model at an effective temperature that is set by the value of $K$.  The scaling dimension of the vortex can be computed by estimating its free energy: for $v=0$ the entropic cost of the gradients in the height field required to insert a single vortex into a triangular layer of side length  $L$ is $\delta {\cal H} =\frac{9 K}{\pi}  \ln L/a$, where $a$ is the lattice constant.   The number of ways to place the vortex in the system is $L^2/a^2$.  Together, these contributions to the free energy of a single vortex are
\be
\delta F = \left(\frac{9K}{\pi} - 2\right)\ln \left(\frac{L}{a}\right)\,.
\ee
For $K=\pi/9$ this grows more negative with increasing $L$. We are therefore in the high-temperature phase of the $xy$ model, where vortices are unbound. The vortex density, determined by the fugacity associated with the vortex excitation energy $4J$, sets the correlation length. This reflects the fact that the triangular layer, which is critical at $T=0$, is a paramagnet at any finite temperature.   

Hence the height model (\ref{1LayerH}) correctly reproduces the phase diagram and correlations of an isolated triangular layer.  The potential $\tilde{V}(h)$ is an irrelevant operator and can be dropped from the long-wavelength analysis; however the vortices arising at finite temperature are relevant, making the system paramagnetic for any $T>0$.

\subsection{Coupled layers in the height model description}
\label{derivation}

We now turn to the situation of interest, in which spins in triangular layers are coupled to their nearest neighbours in the planes directly above and below.  We will derive expressions for these couplings in the height language, and discuss their effect on the physics of the system.

\begin{figure}[h]
\includegraphics[width=0.4\textwidth]{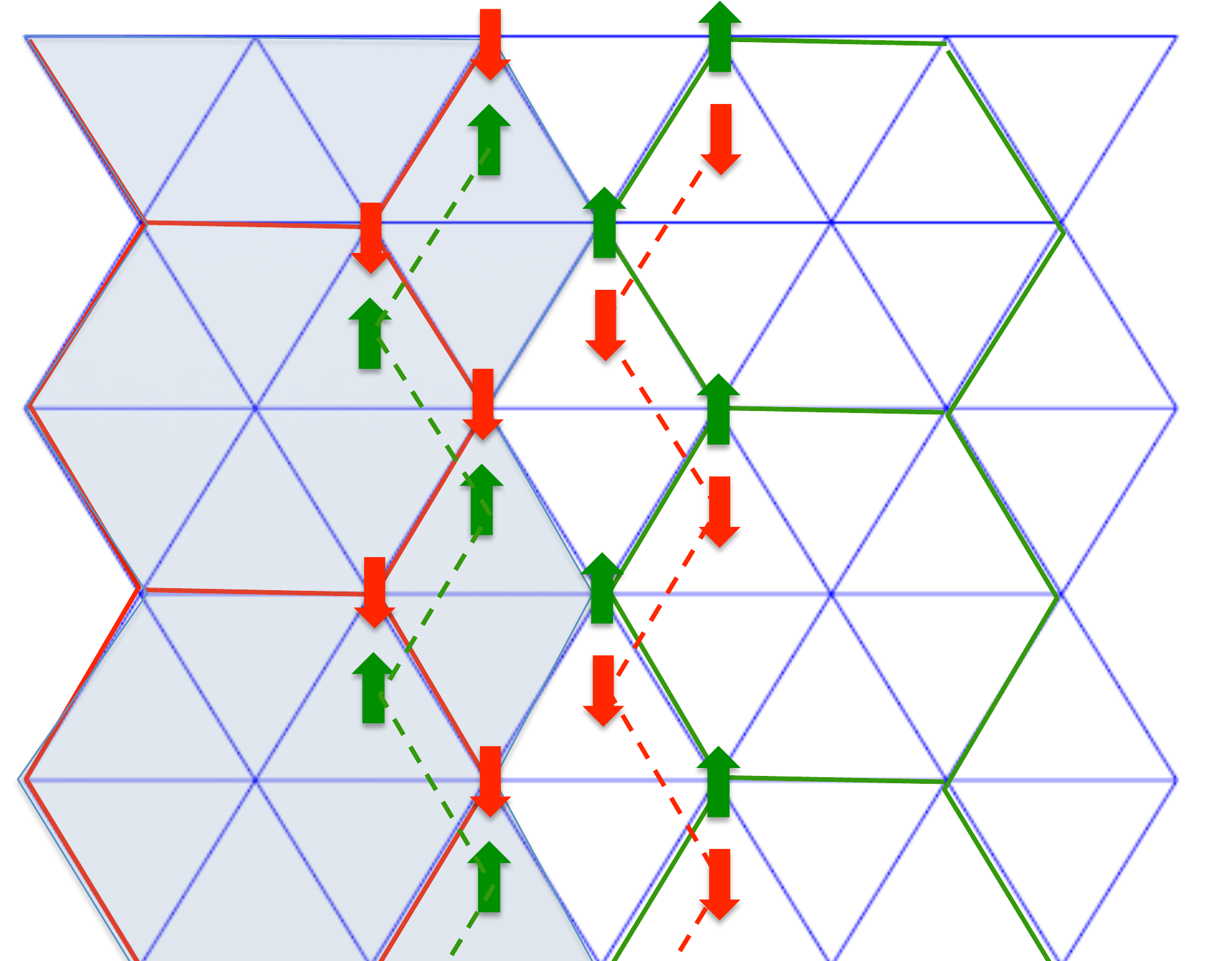}
\caption{\label{DWStackFig} Energetically preferred domain wall stacking. 
Arrows at sites of a triangular lattice represent the spin configuration in one layer.  The height in this layer increases by $1$  moving from the blue region to the white region.  The dashed parallel green and red lines indicate the energetically favourable domain walls in a neighbouring layer, with spin orientations as illustrated.  The height difference between adjacent layers determines the orientation of the domain walls.}
\end{figure}

Frustrated interlayer coupling favours domain walls in the three-sublattice order that is represented by flat configurations of the height field. To minimise the interlayer exchange energy, these domain walls should stack in such a way that a domain wall consisting of up spins sits in the adjacent layer to a domain wall consisting of down spins, as shown in Fig. \ref{DWStackFig}.

To find the functional form of the interlayer coupling in height language, we use Eq. (\ref{Spin2H}) to  express it in terms of the height fields. We then find the scaling dimensions of the various contributions to determine which of these play an important role in the long-wavelength physics.  We will show that, as in the SCGA treatment, for frustrated stackings the relevant terms in the nearest-neighbour model lead to one-parameter sets of degenerate ground states in the height models, whose symmetry can be broken by including further-neighbour couplings. 

\subsubsection{Unfrustrated stacking}

It is instructive to begin by studying the unfrustrated stacking.  For the $aaa$ stacking, the interlayer coupling is
\ba  \label{Stri}
J_\perp ( &\sigma_{A,z}& \sigma_{A,z+1}  + \left. \sigma_{B,z} \sigma_{B,z+1} + \sigma_{C,z} \sigma_{C, z+1} \right ) \n
&=& \frac{8 J_\perp }{3} \cos \frac{ \pi}{3}(h_{z+1}-h_z )  \n &+& \frac{J_{\perp}}{3} \cos \pi h_z \cos \pi h_{z+1} + \ldots
\ea
where $ \ldots $ represents terms of quadratic and higher order in the derivatives, which we drop as they are irrelevant in the scaling sense.  The most relevant term is $\cos \frac{ \pi}{3}(h_{z+1}-h_z )$, which has a scaling dimension of $3/2$ for $K = \pi/9$. 
The term $\cos \pi h_{z} \cos \pi h_{z+1}$ has scaling dimension $-5/2$ and can be neglected.
Hence the effective Hamiltonian of the height model for the $aaa$ stacking is
\ba \label{FSSTI}
{\cal H}^{(aaa)} &=&\frac{K}{2} \sum_z \int d^2 r \left \{  | \nabla h_z({\bf r})|^2  \right . \n
& & \left. + \kappa_3
 \cos \frac{ \pi}{3} (h_{z+1} - h_z )   \right \} \ ,
\ea
with $\kappa_3 = 16 \beta J_\perp/3K$. 
The ground states
\be\label{aaaGS}
h_z({\bf r}) = \gamma
\ee 
of this effective model have a $U(1)$ symmetry under changes of  the constant $\gamma$. This symmetry is broken down to a six-fold discrete symmetry by the interaction $\tilde{V}(h)$, which is irrelevant in the scaling sense at the fixed point describing uncoupled layers, and dangerously irrelevant at the three-dimensional ordering transition \cite{ma}. 

\subsubsection{Frustrated stackings}

\begin{figure}
\begin{center}
\includegraphics[width=0.7\linewidth]{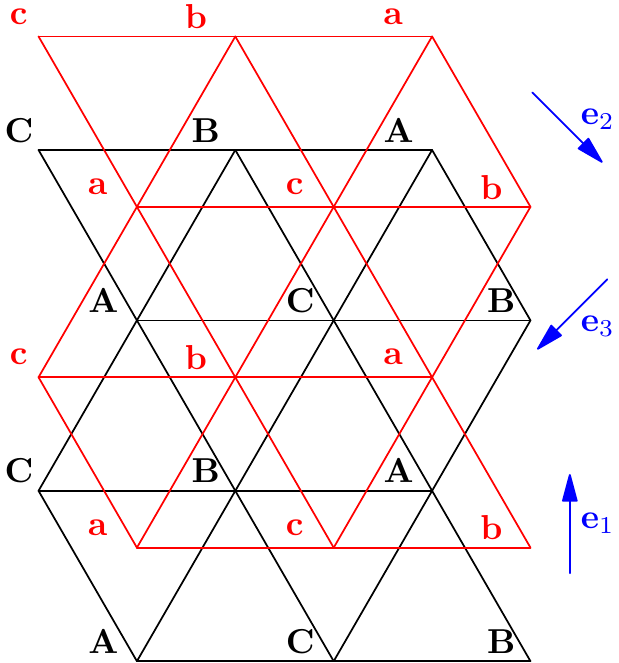}
\caption{Two stacked layers, with sublattice labels and definitions of the vectors ${\bf e}_1$, ${\bf e}_2$, and ${\bf e}_3$.}
\label{stacked}
\end{center}
\end{figure}

For both the $abc$ and the $abab$ stackings, we consider two neighbouring layers as shown in Fig.~\ref{stacked}. There is a coupling between each site on the black lattice and the three sites around it from an up-triangle on the red lattice, or equivalently between each site on the red lattice and the three sites around it from a down triangle on the black lattice. We denote heights on the black lattice by $h_{z+1}({\bf r})$, and ones on the red lattice by $h_{z}({\bf r})$.  
 The coupling is
 
\begin{widetext}

\begin{eqnarray}
{\cal H}_{\perp} &=&J_\perp \sum_{{\bf r} \in A} \sigma_A({\bf r}) [ \sigma_a({\bf r} + {\bf e}_1) + \sigma_b({\bf r} + {\bf e}_2) + \sigma_c({\bf r} + {\bf e}_3)] + \text{symmetry-related terms}\nonumber \\
&=& J_\perp \sum_{{\bf r} \in A} f_A (h_{n+1}({\bf r})) [ f_a(h_{n}({\bf r} + {\bf e}_1 )) + f_b(h_{n}({\bf r} + {\bf e}_2) ) + f_c(h_{n}({\bf r} + {\bf e}_3) )] + \text{symmetry-related terms},
\end{eqnarray}
where `symmetry-related terms' have $B$ or $C$ in place of $A$, and a corresponding permutation of the vectors ${\bf e}_i$. These are defined in terms of the lattice vectors [Eq.~(\ref{latticevectors})] by ${\bf e}_1 =\frac{2}{3}{\bf a}_2 - \frac{1}{3} {\bf a}_1$, ${\bf e}_2 = \frac{2}{3}{\bf a}_1-\frac{1}{3}{\bf a}_2$ and ${\bf e}_3= - \frac{1}{3}{\bf a}_1 -\frac{1}{3} {\bf a}_2$, and are illustrated in Fig.~\ref{stacked} . Expanding  $h({\bf r})$ in a Taylor series, we obtain
\ba \label{HeightHFCC}
{\cal H}_\perp& =& 
- \frac{ 4 \pi J_{\perp} }{9 \sqrt{3} }\sum_{\bf r} \left( \cos \frac{ \pi}{3} (h_{z+1}({\bf r}) - h_z({\bf r}) ) \partial_x  h_z({\bf r}) 
-  \sin \frac{ \pi}{3} (h_{z+1}({\bf r}) - h_z({\bf r}) ) \partial_y  h_z({\bf r}) \right ) + \ldots
\ea 
where $\ldots$ indicates RG-irrelevant terms. Thus keeping only the relevant inter-layer couplings leads to the effective Hamiltonian for the $abc$ stacking
\be\label{FCCFE}
{\cal H}^{(abc)} =
 \frac{K}{2} \sum_z \int {\rm d}^2 r \left \{  \left(\partial_x h_z - \kappa_\perp \cos \frac{ \pi}{3} (h_{z+1} - h_z )  \right )^2 
 +\left( \partial_y h_z + \kappa_\perp \sin \frac{ \pi}{3} (h_{z+1} - h_z )\right )^2
 - \left( \frac{ \kappa_{\perp} }{ K }  \right )^2  \right \} 
\ee
with
$\kappa_\perp \propto \beta J_\perp$.

For the $abab$ stacking, the derivation is identical except that the vertical unit cell contains two layers, with the layers above and below offset in opposite directions.  We use integer $z$ to label unit cells in the vertical direction and $\mu=1,2$ to label layers within each unit cell. The effective Hamiltonian is 
\ba \label{HCPFE}
{\cal H}^{(abab)} &= & \frac{K}{2} \sum_{z}  \int {\rm d}^2 r \bigg \{ \sum_\mu | \nabla h_{z, \mu}|^2 - \kappa_\perp \left \{  \partial_x (  h_{z,1}+  h_{z,2})  \cos  \frac{ \pi}{3}  (h_{z,2} - h_{z,1}) -   \partial_y( h_{z,1}+ \ h_{z,2} ) \sin  \frac{ \pi}{3} (h_{z,2} - h_{z,1})   \right. 
\n 
 &&  \left. + \partial_x ( h_{z+1,1} +   h_{z,2} )  \cos   \frac{ \pi}{3} (h_{z,2} - h_{z+1,1})
 - \partial_y( h_{z+1,1} +h_{z,2} ) \sin  \frac{ \pi}{3} (h_{z,2} - h_{z+1,1})   \right \} \bigg\} \,.
\ea

\end{widetext}

\subsection{Symmetries and further-neighbour couplings}

For both frustrated stackings, emergent continuous symmetries not present in the lattice models are displayed by the effective Hamiltonian of Eqns.~(\ref{FCCFE}) and (\ref{HCPFE}) if terms irrelevant at the $J_\perp=0$ fixed point are omitted. Both models have a  $U(1) \times U(1)$ symmetry.  One $U(1)$ symmetry is associated with global shifts in the height field. It results from the discrete symmetry of the microscopic model related to global shifts in $h$, which -- as for the single-layer height model -- is enhanced to become a continuous symmetry because the pinning potential $\tilde{V}(h)$ is RG-irrelevant and has been omitted. As in the unfrustrated case [see Eq.~(\ref{aaaGS})] we parameterise it with $\gamma$. The second $U(1)$ symmetry is associated with real-space rotations and is reduced to the discrete rotational symmetry of the lattice by irrelevant terms. We parameterise it with $\theta$. 

In detail, these symmetries take the following form. Let $R_\theta$ denote a rotation in the $xy$ plane through the angle $\theta$ and write ${\bf r}^\prime = R_\theta({\bf r})$. Then ${\cal H}^{(abc)}$ is invariant under the transformation
\be\label{u1xu1}
h_z({\bf r}) \to h^\prime_z({\bf r}) = h_z({\bf r}^\prime) + \frac{3z\theta}{\pi} + \gamma\,.
\ee
Similarly ${\cal H}^{(abab)}$ is invariant under $h_{z,\mu}({\bf r}) \to h^\prime_{z,\mu}({\bf r})$ with
\ba
h^\prime_{z,1}({\bf r}) &=& h_{z,1}({\bf r}^\prime) - \frac{3\theta}{2\pi} + \gamma  \n
\mbox{and} \quad h^\prime_{z,2}({\bf r}) &=& h_{z,2}({\bf r}^\prime) + \frac{3\theta}{2\pi} + \gamma\,.
\ea
Ground state configurations of the height model for the $abc$ stacking have the form
\be\label{spiral}
h_z({\bf r}) = \kappa_\perp(x\cos\theta-y\sin\theta) + \frac{3z\theta}{\pi} + \gamma\,.
\ee
For the $abab$ stacking the ground states are 
\ba\label{circle}
h_{z,1}({\bf r}) &=& \kappa_\perp(x\cos\theta-y\sin\theta)- \frac{3\theta}{2\pi} + \gamma  \n
h_{z,2}({\bf r}) &=&  \kappa_\perp(x\cos\theta-y\sin\theta)+ \frac{3\theta}{2\pi} + \gamma\,,
\ea
together with a second symmetry-related set.

The symmetry under continuous changes of $\theta$ is not a feature of the microscopic model: it is broken by the leading irrelevant terms in Eq. (\ref{HeightHFCC}). For the $abc$ stacking these have the form
 \ba\label{Hb}
{\cal H}_b&=& \kappa_b\sum_z \int {\rm d}^2 {\bf r}   \bigg\{ \left[ \left( \partial_x h_z({\bf r}) \right)^2 - \left( \partial_y h_z({\bf r}) \right)^2 \right ]  \cos \delta h_z ({\bf r}) \n
&+& 2 \partial_x h_z({\bf r}) \partial_y h_z({\bf r}) \sin \delta h_z ({\bf r}) \bigg\} ,
 \ea
 where we introduce the notation $\delta_p h_z({\bf r}) = \frac{\pi}{3}[h_{z+p}({\bf r}) - h_z({\bf r})]$ and $\delta h_z({\bf r}) \equiv \delta_1 h_z({\bf r})$. (The form for the $abab$ stacking follows the obvious equivalent pattern.)
 
Significantly, it may also be broken by {\it relevant} further-neighbour couplings, if these are present microscopically, or are generated under renormalisation.  
For the $abc$ stacking, some relevant and marginal couplings that are not included in Eq.~(\ref{FCCFE}) are
\begin{eqnarray}\label{abcH3}
{\cal H}_m &=&  \frac{K_m}{2} \sum_z\int {\rm d}^2{\bf r} \,\nabla h_z({\bf r}) \cdot \nabla h_{z+m}({\bf r})\nonumber \\
{\cal H}_{2} &=&  \kappa_2  \sum_z\int {\rm d}^2{\bf r} \, \left \{  \partial_x h_z ({\bf r})  \cos \frac{\pi}{3}(h_{z+2} - h_z) \right . \n 
&& \left.+ \partial_y h_z ({\bf r}) \sin \frac{\pi}{3}(h_{z+2} - h_z) \right \} \n \label{MakeInterEq}
{\cal H}_{3} &=&  \kappa_3\sum_z \int {\rm d}^2{\bf r}\, \cos \frac{\pi}{3}(h_{z+3} - h_z)\,.
\end{eqnarray}
${\cal H}_3$ is the most relevant of these three: it breaks the degeneracy of Eq.~(\ref{spiral}), selecting ground states for which $3 \theta =0$ ($ \pi$) for $\kappa_3<0 $ ($\kappa_3>0 $).   ${\cal H}_2$ has the same scaling dimension as the bare interlayer coupling.  It also breaks the symmetry, again favouring states for which  $3 \theta =0$ ($ \pi$) for $\kappa_2<0 $ ($\kappa_2>0 $). ${\cal H}_m$ is marginal, and does not break the degeneracy between the ground states identified above, all of which have the same in-plane gradients in each layer.

Therefore as well as potentially being broken spontaneously at low temperature, the emergent $U(1)$ spiral symmetry of the $abc$ model can be broken explicitly at a scale set by the coefficients $\kappa_2$ and  $\kappa_3$.  We discuss this scenario in Sec. \ref{behaviour}.  

For the $abab$ stacking, the perturbations of interest are interlayer gradient couplings similar to ${\cal H}_m$,  and also
\begin{eqnarray}\label{ababH3}
{\cal H}_{3} &=&  \kappa_3 \sum_{z,\mu} \int {\rm d}^2{\bf r}\, \cos \frac{\pi}{3}(h_{z+1,\mu} - h_{z,\mu})  ,
\end{eqnarray}
the unfrustrated coupling between spins two layers apart.  In contrast to the $abc$ case, ${\cal H}_3$ is not expected to be important in determining the ordering temperature: the minimum-energy solutions of the $abab$ model have a definite value of $h_{z+1,\mu}- h_{z,\mu}$, and so this term does not lift the ground-state degeneracy. Instead, symmetry is broken by the irrelevant coupling ${\cal H}_b$, Eq.~(\ref{Hb}).

\section{Behaviour of the height model}\label{behaviour}

To understand the phase diagrams of these coupled-layer height models, we take two successive steps. First we make a perturbative renormalisation group (RG) analysis of the behaviour of weakly coupled layers, as described in Sec.~\ref{TreeRGSec}. Depending on the values of $T$ and $J_\perp$, the model under scaling may remain  weakly coupled: this happens in the weakly-correlated paramagnetic regime. Alternatively, it may flow to strong interlayer coupling. In that case a separate analysis is necessary of the influence of vortex pairs, which is presented in Sec.\ref{VortexFlucSec}. We find that the minimal models with exact $U(1)\times U(1)$ symmetry have anomalously soft excitations. For this reason vortex pairs destroy long-range order, establishing instead a paramagnetic regime with strong interlayer correlations. Symmetry-breaking or `locking' interactions act in competition to vortex pairs, and stabilise the ordered phase when they dominate.

\subsection{Perturbative RG} \label{TreeRGSec}

Our perturbative analysis follows the standard renormalisation-group techniques of Refs.~\onlinecite{KT2,KogutReview}.   For small $J_\perp$ and low $T$, this allows us to use arguments similar to those of Sec.~\ref{TriHeightSec} regarding the phase diagram of these models. If unbound vortices proliferate, the inter-layer coupling flows to zero at long distances, while if the coefficient of one of the cosine terms grows large, a strong-coupling analysis is necessary.

The leading-order behaviour of the RG equations is simply determined by the scaling dimensions of the relevant interlayer couplings and vortices.  (The intra and interplane gradient terms flow only at higher order.) For the interlayer couplings, these can be calculated either from the two-point functions as described in Sect. \ref{TriHeightSec}, or (as is more appropriate for operators involving derivatives of the height field) using a standard momentum-shell RG (see Appendix \ref{DimensionApp}). Using $\ell$ to denote the short-distance cut-off and following the notation of Eqns. (\ref{FCCFE}), (\ref{HCPFE}) and (\ref{MakeInterEq}),  this gives
\ba \label{TreeScale}
\frac{\partial \kappa_\perp}{\partial \ln \ell} &=&  \left( 1 -  \beta_1 \right ) \kappa_\perp\,,  \n
 \frac{\partial \kappa_3}{\partial \ln \ell} &=&\left(  2 - \beta_1 \right ) \kappa_3\n
\mbox{and}\quad  \frac{\partial y }{\partial \ln \ell} &=& \left( 2 -  \alpha_1 \right ) y\,.  
\ea
Here, $\kappa_\perp$ is the frustrated interlayer coupling 
that acts between neighbouring layers in the $abc$ and $abab$ stackings, and $\kappa_3$ is the unfrustrated inter-layer coupling, which couples nearest neighbour layers in the $aaa$ stacking, second neighbours in the $abab$ stacking and third neighbours in the $abc$ stacking. Finally, $y$ is the vortex fugacity, which 
dictates the unbound vortex density. For weakly coupled layers we have
\be \label{RGdims}
\beta_1 = \frac{ \pi}{ 18 K} \quad {\rm and} \quad  \alpha_1=  \frac{9K}{\pi}\,.
\ee
For the unfrustrated stacking the bare value of $\kappa_3$ is $\kappa_{3,0} \sim \beta J_\perp$. For the frustrated stackings
the bare value of the interlayer coupling $\kappa_\perp$ is  $\kappa_{\perp,0} \sim \beta J_\perp$. In both cases, the bare value of the vortex fugacity is $y_0\sim e^{-4\beta J}$. The initial value of $\ell$ is the lattice spacing, which we set to unity.

Let us now consider what we learn from these scaling dimensions about behaviour in the three different models, keeping only nearest-neighbour interactions and the intralayer gradient interaction $K$.  Using the value $K = \frac{ \pi}{9}$ appropriate for decoupled triangular layers, we have $\alpha_1 = 1, \beta_1 = 1/2$, and single-layer vortices are more relevant than their multi-layer counterparts.  
Solving the RG equations (\ref{TreeScale}) gives
\ba
y &=& y_0 \ell, \ \ \ \kappa_{\perp}  = \kappa_{\perp,0} \ell^{ 1/2 } \ \ \ 
\mbox{and}\quad\kappa_{3} =\kappa_{3,0} \ell^{ 3/2 }.\nonumber
\ea
The calculation reaches its limit of validity at the scale $\ell$ where the largest coupling is of order unity, and the physical state of the system is signalled by which coupling first crosses this threshold. If $y\sim 1$ with $\kappa_\perp$ and $\kappa_3\ll 1$, the system is a weakly correlated paramagnet. If either $\kappa_\perp \sim 1$ or $\kappa_3 \sim 1$ with $y \ll 1$, layers are strongly coupled. We turn next to this regime.

\subsection{Strongly coupled layers}\label{VortexFlucSec}

To understand behaviour of the height models at large interlayer coupling, we examine the effective Hamiltonian for each type of stacking at quadratic order in an expansion about the ground states given in Eqns.~(\ref{aaaGS}), (\ref{spiral}) and (\ref{circle}). 

For orientation, consider first the $aaa$ stacking. Let $\varphi_z({\bf r})$ denote the deviation of $h_z$ from a ground-state configuration and introduce its Fourier transform via
\be
\varphi_z({\bf r}) = \frac{1}{(2\pi)^3}\int {\rm d}^3{\bf q}\, \varphi({\bf q}) e^{i ({\bf q}_\perp{\bf r} + q_zz)}\,.
\ee
The energy cost at quadratic order of this deviation from a ground state is
\be
\delta{\cal H} = \frac{K}{2(2\pi)^3}\int {\rm d}^3{\bf q} \, {\cal E}({\bf q}) |\varphi({\bf q})|^2\,
\ee
with
\be
{\cal E}({\bf q}) = q_x^2 + q_y^2 + \tilde{\kappa}(1- \cos q_z),
\ee
where $\tilde{\kappa}_\perp = (\pi^2/9) |\kappa_3|$. Thus, for this unfrustrated stacking, excitations have a dispersion ${\cal E}({\bf q})$ that is conventional in the sense that it is quadratic in wavevector for all orientations of ${\bf q}$.

An equivalent calculation for the $abc$ stacking (for fluctuations around the ground state with $\theta=0$) yields the quite different dispersion relation
\begin{equation}
{\cal E}({\bf q}) = q_x^2 + (q_y - \tilde{\kappa}_\perp \sin q_z)^2 + \tilde{\kappa}_\perp^2(1-\cos q_z)^2,
\end{equation}
where $\tilde{\kappa}_\perp = (\pi/3)\kappa_\perp$. This is anomalously soft, being quartic in wavevector along the line $q_y=\tilde{\kappa}_\perp q_z$.
The soft modes do not give rise to divergent harmonic fluctuations, since
\be \label{Ordh2pts}
 \langle [h_{n+1}({\bf r})- h_n({\bf r})]^2\rangle= \frac{1}{K} \int {\rm d}^3{\bf q}\,\, \frac{(1-\cos q_z)^2}{ {\cal E}({\bf q}) }
 \ee
is finite provided ${\kappa}_\perp \not=0$. 

For the $abab$ stacking, since there are two layers within a unit cell, it is necessary to introduce two fields $\varphi_{z,\mu}({\bf r})$, with $\mu=1,2$.
The resulting quadratic Hamiltonian has two eigenvalues, which for $\theta=0$ are
\be
{\cal E}_\pm({\bf q}) = q_x^2 + q_y^2 + 2\tilde{\kappa}_\perp^2 \pm 2\tilde{\kappa}_\perp |\cos(q_z/2)| \sqrt{q_y^2 + \tilde{\kappa}_\perp^2}\,. 
\ee
In this case as well, the dispersion relation is quartic for one direction, since  ${\cal E}_- = q_x^2 +(\tilde{\kappa}_\perp^2 q_z^2 + q^4_y/\tilde{\kappa}_\perp^2)/4$ for small $|{\bf q}|$, but harmonic fluctuations are bounded for ${\kappa}_\perp\not=0$.

\subsection{Destruction of order by defects}\label{defects}

Our discussion of harmonic height-field fluctuations around ground states of the multilayer model accounts for spin fluctuations within the ground-state manifold of each triangular layer, but a separate treatment is required to understand the effect of excitations out of this ground-state manifold. That is the subject of this subsection. 

The excitations are represented by vortices and antivortices. These are unbound in a single layer, as discussed in Sec.~\ref{TriHeightSec}, but acquire a linear confining potential within ordered states of the multilayer systems. More specifically, suppose that the height field in a layer containing a widely separated vortex-antivortex pair  has a step of height 6 and width $w$: its energy cost per unit length is $ \sim Kw(w^{-2} + \kappa_\perp^2)$ and is minimised by the choice $w \sim \kappa_\perp^{-1}$. Pairs are therefore bound with typical separation $w$ when interlayer correlations are strong. Remarkably, although in other settings bound vortex pairs are typically irrelevant at large scales, we find that they exert a controlling influence in multilayer height models with frustrated stackings.

Height fields in the presence of vortices are in general multivalued, but can be taken to be single-valued in a domain that excludes a core around each vortex-antivortex pair. The presence of these pairs influences the height field far from the cores. A convenient alternative to an explicit treatment of multivalued height fields is to impose a potential that couples linearly to the height field and has the same effect on the far field as a votex-antivortex pair. In order to demonstrate the required form of this potential, consider a single layer containing a pair centred at the origin with separation vector ${\bf b}$. 
This pair is described by the height field configuration
\be
h
({\bf r}) = \frac{3}{ \pi} \left[ \arctan \left( \frac{ 2x + {\bf b} \cdot \hat{x}}{2y+{\bf b} \cdot \hat{y}} \right) -  \arctan \left( \frac{2x - {\bf b}  \cdot \hat{x}}{2y- {\bf b} \cdot \hat{y}} \right)   \right].\nonumber
\ee
For $|{\bf r}| \gg |{\bf b}|$ we have
\be
h
(x,y) \approx \frac{3}{ \pi} \frac{\hat{z} \cdot ( {\bf b} \times {\bf r} )}{r^2 }
\ee
or equivalently
\be \label{FarFieldh}
h({\bf q}) \approx 6 i \frac{ \hat{z} \cdot ({\bf q} \times{\bf b}  ) }{q^2 }. 
\ee
The same far-field height configuration can be induced by adding a potential term $v{(\bf q})$ to the effective Hamiltonian for the height field.  Specifically, for an isolated layer, the effective Hamiltonian  $ (K/[2\pi]^2) \int d^2 {\bf q} \left [  \frac{1}{2} {\mathcal E} ({\bf q})  | \varphi({\bf q}) |^2  - \varphi(-{\bf q}) v({\bf q})  \right]$ has the minimum energy configuration
\be\label{defect}
{\varphi}({\bf q})  = \frac{ v({\bf q})}{{\mathcal E}({\bf q}) } 
\ee
with ${\mathcal E}({\bf q}) = q^2$ for a single layer. Thus choosing a potential
\be
v({\bf q}) = 6 i \hat{z} \cdot ({\bf q} \times  {\bf b} ) 
\ee
we recover the desired far-field configuration.  

To examine the effect of many pairs $j$ with locations ${\bf r}_j,z_j$ and separations ${\bf b}_j$ we impose on the multilayer system the potential 
\be
v_{\rm tot} ({\bf q}) = 6i \sum_j \hat{z} \cdot ({\bf q} \times  {\bf b}_j) \,e^{-i({\bf q}_\perp{\bf r}_j + q_z z_j)}  \,.
\ee
The ground state in the presence of these pairs is again given by (\ref{defect}), but now with the multilayer form for ${\mathcal E}({\bf q})$. We compute the mean square amplitude  of the fluctuations these pairs generate, averaged over bound pair positions with a Poisson distribution at  a density  $\rho$, obtaining 
\be
\langle[\varphi_z({\bf r})]^2\rangle = \frac{\rho}{(2\pi)^3}  \int {\rm d}^3 {\bf q} \frac{\langle|v({\bf q})|^2\rangle}{{\cal E}^2({\bf q})} \,
\ee
where $\langle \ldots \rangle$ indicates an average over pair separations $\bf b$. This integral is convergent at small $q$ for the unfrustrated stacking but divergent for the frustrated systems. Moreover, corrections to a Poisson distribution arising from correlations between pairs appear only at higher order in $\rho$.   Vortex-antivortex pairs in the absence of locking interactions therefore destroy long-range order in the frustrated systems.

We can estimate the correlation length in this disordered state by determining the small-wavevector cut-off for which $\langle[\varphi_z({\bf r})]^2\rangle\sim 1$. We write $\langle |{\bf b}|^2\rangle \sim \ell^2$, where $\ell$ is the cut-off scale at which the system reaches the strong-coupling regime with $\kappa_\perp \sim 1$. This scale is $\ell \sim (\beta J_\perp)^{-2}$. Then for the $abc$ stacking the correlations lengths in the in-plane and $z$-directions are
\be
\xi_\perp \sim \kappa_\perp^{-1}(\ell^2 \rho)^{-2} \quad {\rm and} \quad \xi_z \sim (\ell^2\rho)^{-1}\,.
\ee 
For the $abab$ stacking the corresponding expressions are
\be
\xi_\perp \sim \kappa_\perp^{-1}(\ell^2 \rho)^{-1/2} \quad {\rm and} \quad \xi_z \sim (\ell^2\rho)^{-1/2}\,.
\ee 

The phase transition to a long-range ordered state involves a competition between this disordering effect of bound vortex pairs, and the opposite tendency produced by locking interactions. A simple estimate for the location of the phase boundary is obtained demanding that the locking interaction at the scale $\ell$, integrated over the correlation volume, is of order unity.

The most RG-relevant locking interaction for the $abc$ stacking is $\kappa_3$ [see Eq.~(\ref{MakeInterEq})]. As this is a coupling between layers three apart, it is not present in the bare description of a system with only nearest-neighbour interactions. It is however generated under the first steps of RG, so that the initial value can be taken to be $\kappa_{3,0} \sim (\beta J_\perp)^7$ (see Sec.~\ref{beyond}). At the scale $\ell$ the locking interaction is hence $\kappa_3 \sim (\beta J_\perp)^4$. Note that an important role is played by the fact that $\kappa_3$ is generated only at high order: if instead one had $\kappa_{3,0} \sim (\beta J_\perp)^3$ as might naively have been expected for a third-neighbour coupling, then the value of $\kappa_3$ at scale $\ell$ would be ${\cal O}(1)$ and independent of $J_\perp$. This would leave no scope for a regime with strong interlayer correlations but no long-range order.

For the $abab$ stacking, we have not found locking interactions that are RG-relevant. The leading (least irrelevant) locking term in this case is $\kappa_b$, given in Eq.~(\ref{Hb}). At the scale $\ell$ it is of order $\beta J_\perp \ell^{-1/2} \sim (\beta J_\perp)^2$. 

\subsection{Phase diagram}\label{phasediagram}

Combining results from our discussion of RG for weakly coupled layers with our results on the effect of defects in strongly coupled layers, we can determine regimes of behaviour and phase boundaries for systems with each type of stacking, in the limit $J_\perp \ll J$. The phase boundaries determined theoretically in this section are compared with Monte Carlo results in Sec.~\ref{discussion}.

For the unfrustrated stacking, bound vortex pairs have no important effects. The phase boundary is the point at which $y\sim \kappa_3 \sim 1$. From the results of Sec.~\ref{TreeRGSec}, this implies $\ell \sim e^{4\beta J}$ and $\beta J_\perp e^{6\beta J} \sim 1$. Solving approximately in the limit $J_\perp \ll J$, the phase boundary is at $J_\perp \approx J e^{-6\beta J}$. Interlayer correlations are weak for $J_\perp \ll J e^{-6\beta J}$ while the system has long-range order for $J_\perp \gg J e^{-6\beta J}$. Within the minimal model of Eq.~(\ref{FSSTI}), the set of ordered states has a  U(1) symmetry, as displayed in Eq.~(\ref{aaaGS}). This is broken by the (RG-irrelevant) interaction $\tilde{V}(h)$, introduced for a single layer in Eq.~(\ref{1LayerH}). It selects integer values of the height field, corresponding to six possible types of three-sublattice spin order.

In contrast, for both types of frustrated stacking, the condition $y\sim \kappa_\perp \sim 1$ implies $J_\perp \approx Je^{-2\beta J}$. Interlayer correlations in this case are weak for $J_\perp \ll Je^{-2\beta J}$. The paramagnetic regime with only weak interlayer correlations therefore extends to parametrically lower temperatures and larger values of $J_\perp$ in these systems than in the unfrustrated stacking. Moreover, because of the effect of bound vortex pairs in systems with frustrated stacking, long range order appears at still lower temperatures or larger values of $J_\perp$ than strong interlayer correlations. 

In the case of the $abc$ stacking, if long range order is stabilised by generation of the RG-relevant third-neighbour coupling $\kappa_3$ the condition $\kappa_3 \xi_\perp^2 \xi_z \sim 1$ implies order for $J_\perp \gtrsim J e^{-5\beta J/3}$. Alternatively, order may be stabilised by residual contributions from the RG-irrelevant coupling $\kappa_b$. Specifically, RG flow stops on the scale at which $\kappa_\perp \sim 1$. At this scale, interactions (whether RG-relevant or RG-irrelevant) that break the $U(1)\times U(1)$ symmetry of Eq.~(\ref{u1xu1}) down to a discrete one will act coherently over a correlation volume. This ordering tendency competes with the disordering effect of bound vortex-antivortex pairs. Since $\kappa_3$ is generated rather slowly under RG, RG-irrelevant interactions turn out to be the dominant cause of locking if microscopic interactions are just nearest neighbour.\cite{correction} The condition $\kappa_b  \xi_\perp^2 \xi_z \sim 1$ implies order for $J_\perp \gtrsim J e^{-20\beta J/11}$.

For the $abab$ stacking, locking is driven only by irrelevant interactions. Taking into account the dependence of $\xi_\perp$ and $\xi_\parallel$ on $\rho$ for the $abab$ stacking, the condition $(\beta J_\perp)^2 \xi^2_\perp \xi_z \sim 1$ yields a boundary for long range order at $J_\perp \approx J e^{-5\beta J/3}$. 

In summary, with $J_\perp \ll J$, the classical spin liquid regime, in which correlations are strong both within and between layers, extends for both types of frustrated stacking over the interval
\be
e^{-2\beta J} \lesssim J_\perp/J \lesssim e^{-c\beta J}
\ee
with $c=20/11$ for the $abc$ stacking and $c=5/3$ for the $abab$ stacking.

\subsection{Spin correlations from the height model}\label{correlations}

In the classical spin liquid regime, in which interlayer correlations are strong but there is no long-range order, the system is approximately ordered within each correlation volume $\xi_\perp^2 \xi_z$ but different correlation volumes are essentially independent. We can compute correlations approximately in this regime as an average over all ground states. The starting point for this calculation is the expression (\ref{Spin2H}) for spin variables in terms of height fields, and the expressions (\ref{spiral}) and (\ref{circle}) for ground states of the minimal height models in the systems with frustrated stackings.

We require Fourier components of the spin density at wavevectors that are close in-plane to either of the corners ${\bf K}$ and ${\bf K}^\prime$ of the triangular-lattice Brillouin zone. To obtain the leading contribution at long distance it is sufficient to use the approximation $\sigma_{j,z} \sim \cos\frac{\pi}{3}(h_z({\bf r}_j) + s_\alpha)$, omitting higher harmonics in $h_z({\bf r}_j)$. 

Recalling that $s_\alpha=0,\pm 2$ on the three sublattices, we have for the $aaa$ stacking $e^{i{\bf K}\cdot {\bf r}_{j,z}} = e^{i\pi s_\alpha/3}$ and $e^{i{\bf K}^\prime\cdot {\bf r}_{j,z}} = e^{-i\pi s_\alpha/3}$. The same result holds for the $abab$ stacking on one of the two layers in the unit cell, but for the $abc$ stacking it is necessary to take account of the  relative displacement $\bm{e}_1$ of neighbouring sites on the same sublattice in successive layers. 
We have (modulo $2\pi$)
\ba
({\bf K} &+& n_1 {\bf A}_1 + n_2 {\bf A}_2)\cdot {\bf r}_{j,z}\n &=& \frac{\pi}{3}s_\alpha - z({\bf K} + n_1 {\bf A}_1 + n_2 {\bf A}_2)\cdot {\bm{e}_1} 
= \frac{\pi}{3}(2pz+s_\alpha)\nonumber
\ea
with $p=n_1+n_2$, and
\be
({\bf K}^\prime + n_1 {\bf A}_1 + n_2 {\bf A}_2)\cdot {\bf r}_{j,z}= \frac{\pi}{3}(2p^\prime z-s_\alpha)\nonumber
\ee
with $p^\prime = 2 +n_1 + n_2$. 
%
Retaining only smoothly varying contributions, we can then write for ${\bf q}_\perp$ small but $q_z$ arbitrary 
\be
\sum_j \sigma_{j,z} e^{i({\bf K}+{\bf q})\cdot {\bf r}_{j,z}}  \sim  \int {\rm d}^2{\bf r} \,\, e^{-i\frac{\pi}{3}h_z({\bf r})} \,e^{i({\bf q}_\perp\cdot{\bf r}+[q_z +\frac{2\pi}{3}p]z)}\nonumber
\ee
and
\be
\sum_{j} \sigma_{jz} \, e^{i({\bf K}^\prime+{\bf q})\cdot{\bf r}_{jz}} \sim \int {\rm d}^2{\bf r} \,\, e^{i\frac{\pi}{3}h_z({\bf r})} \,e^{i({\bf q}_\perp\cdot{\bf r}+[q_z\frac{2\pi}{3}p^\prime]z)}\nonumber
\ee
where we can include the $aaa$ and the $a$-layers of the $abab$ stacking by setting  $p=p^\prime=0$ in these cases.

We use these expressions to evaluate 
\be
S({\bf K}+{\bf q}) = \sum_{j,z} \langle \sigma_{0,0}\sigma_{j,z} \rangle e^{i({\bf K}+{\bf q})\cdot {\bf r}_{j,z}}
\ee
and the equivalent with ${\bf K}^\prime$ in place of ${\bf K}$, computing the average $\langle \ldots \rangle$ over ground states [Eqns.~(\ref{spiral}) and (\ref{circle})]. For the $abc$ stacking this gives
\ba
S({\bf K}+{\bf q}) &\propto& \delta(q_x - \frac{\pi}{3}\kappa_\perp \cos [q_z+\frac{2\pi}{3}p] )\n
&\times& \delta(q_y + \frac{\pi}{3}\kappa_\perp \sin [q_z+\frac{2\pi}{3}p])
\ea
and
\ba
S({\bf K}^\prime+{\bf q}) &\propto&  \delta(q_x + \frac{\pi}{3}\kappa_\perp \cos [q_z+\frac{2\pi}{3}p^\prime] )\n
&\times&\delta(q_y + \frac{\pi}{3}\kappa_\perp \sin [q_z+\frac{2\pi}{3}p^\prime])\,.
\ea
For the $abab$ stacking, following our discussion in Sec.~\ref{ab}, we focus on the contribution to the structure factor from sites on only one of the two sublattices by restricting $\sum_{z,\mu}$ to the layer $\mu=1$. This gives
\be
S({\bf K}+{\bf q}) = S({\bf K}^\prime+{\bf q}) \propto \delta(q_z) \delta^{(2)}(q_\perp^2 - [\frac{\pi}{3}]^2\kappa_\perp^2)\,.
\ee 
It is reasonable to expect that the main consequence of finite correlation lengths $\xi_\perp$ and $\xi_z$ will be broadening of the delta functions in these expressions for $S({\bf q})$. Making that allowance, we see that the height model calculation produces results similar to the ones from the SCGA and from Monte Carlo simulations.

\section{Renormalisation group flows beyond leading order}\label{beyond}

Our calculation of RG flow is perturbative in interlayer coupling and vortex fugacity. We can improve the estimates of the previous section by including terms to higher order. Qualitatively, this has two potentially important consequences.  First, the in-plane stiffness $K$ becomes scale-dependent and interlayer gradient couplings are generated under the RG flow.  This in turn modifies the dimensions of the various operators discussed above. Second, for the $abc$ stacking, the relevant further-neighbour couplings that break the U(1) symmetry under spatial rotations are generated from the irrelevant contribution to the nearest-neighbour interlayer coupling, Eq. (\ref{Hb}).

\subsection{Simply stacked triangular layers}

To set the stage, it is instructive to consider the case of $aaa$-stacked triangular layers.  The model [Eq.~(\ref{FSSTI})] is simply a 3D $XY$ model, in which the coupling between neighbouring layers is much weaker than the intra-layer coupling.  For small interlayer couplings there is a regime where the RG flows are well-described by  those of a system of coupled 2D $XY$ models \cite{2D3Dxy,SomebodyElse}.  Though this treatment is not adequate to describe the transition between the low-temperature ordered phase and the high-temperature paramagnet, which is in the 3D $XY$ universality class, it represents behaviour well so long as the renormalised interlayer coupling is not strong.

For  uncoupled layers, two different ways exist to derive RG equations.  The original work by Kosterlitz and Thouless\cite{KT1,KT2} on the 2D $XY$ model used a real-space calculation, integrating out vortex-antivortex pairs separated by less than a minimum length scale $\ell$, and this method has been extended to include models analogous to (\ref{FSSTI})  with vortices \cite{Minnehagen}.  Somewhat later, the momentum-shell RG approach was applied to these systems \cite{MomentumShellRG} and  we use this second approach, which is more transparent in the case of the frustrated $abc$ and $abab$ stackings. We review the method and give technical details of our calculations in Appendix \ref{RGApp}; here we discuss the physical implications of the results.

Including the most relevant interlayer couplings, the marginal gradient couplings  introduced in Eq. (\ref{MakeInterEq}), and a new second-layer coupling term $\cos \frac{ \pi}{3} (h_{z+2}(r) - h_z(r) )$ with coefficient $g_2$, the RG equations additional to (\ref{TreeScale}) to quadratic order in $\kappa_3$ and $y$ are
\ba \label{RGSTRI}
\frac{\partial K}{ \partial \ln \ell} &=& c_1 \kappa_3^2 -  y ^2 K^2 \n 
\frac{\partial K_1 }{\partial \ln \ell} & =&-  c_1\kappa_3^2   \n 
\frac{ \partial g_2}{ \partial \ln \ell} &=& g_2( 2 - \frac{ \pi}{ 18 K }  )- c_2 \kappa_3^3.
\ea
Here we have allowed for the effect of fluctuating bound vortex pairs on the stiffness. A deficiency of the momentum-space approach is that this correction cannot be evaluated easily, and so we take the result computed in the real-space RG using Coulomb gas methods \cite{Nienhuis}.
The constants $c_1$ and $c_2$ are given in Eq.~(\ref{RGcoeffs}). 

The RG flow described by Eqs.~(\ref{TreeScale}) and (\ref{RGSTRI}) includes several important effects. First, at this order the stiffness $K$ flows towards smaller values if vortices dominate. As the interlayer coupling $\kappa_3$ is irrelevant if $K$ is sufficiently small, this ensures that the paramagnetic phase is stable to weak interlayer coupling. Second, new interlayer couplings are generated from $\kappa_3$: the marginal gradient coupling $K_1$ and the relevant second-neighbour coupling $g_2$. The latter contributes to stabilising long-range order if vortices are not dominant. 

Interlayer gradient couplings change the scaling dimensions of other interlayer couplings and of the fugacity for multilayer complexes of vortices. The scaling dimensions of Eq. (\ref{RGdims}) become more generally
\ba
\beta_1 &= &  \frac{\pi}{18} \int_{-\pi}^{\pi} \frac{ d k_z}{2 \pi } \left [  \frac{1- \cos k_z}{ K_0 + \sum_p K_p\cos p k_z  } 
\right ] \n
\alpha_1 &=& \frac{9}{\pi}  \sum_{i,j}  \sigma_i \sigma_{j} K_{|i-j|},
\ea
where $\sigma_i$ is the vortex strength in layer $i$. 

A striking consequence of interlayer gradient couplings that follows from these results for scaling dimensions is the possibility of a sliding phase, \cite{sliding} in which for appropriate values of $\{K_p\}$ neither vortices nor interlayer cosine couplings are relevant. The window of stability of this phase is however quite narrow, and it does not seem likely that it would be reached by RG flow starting from stacked TLIAFMs with only nearest-neighbour interactions, whether frustrated or not. 

\subsection{$abc$ stacking} \label{FCCRG}

We now consider the $abc$ stacking.  As for the $aaa$ stacking, under RG at second order the stiffness $K$ flows and further-neighbour interactions are generated. The most important of these are shown in Eq.~(\ref{MakeInterEq}) with coupling constants denoted by $\kappa_2$ and $\kappa_3$. As they break the spatial U(1) symmetry of ${\cal H}^{(abc)}$ [see Eq~(\ref{FCCFE})], their generation involves the RG-irrelevant nearest-neighbour interaction $\kappa_b$ appearing in Eq.~(\ref{Hb}).  
The coupled RG equations
\ba \label{RGeqs}
\frac{\partial K}{\partial  \ln \ell} &=& c_3 \kappa_\perp^2 -y^2K^2 \n
\frac{\partial K_1}{\partial  \ln \ell} &=& - c_4 \kappa_\perp^2 \n
\frac{\partial \kappa_b}{\partial \ln \ell} &=&  -\frac{\pi}{18 K} \kappa_b  \n 
\frac{\partial \kappa_2}{\partial  \ln \ell} &=&  \kappa_2\left( 1 -\frac{\pi}{18 K} \right) + c_5  \kappa_\perp \kappa_b   \n
\frac{\partial \kappa_3}{\partial  \ln \ell} &=& \kappa_3 \left( 2 -\frac{\pi}{18 K} \right)   + c_6  \kappa_\perp \kappa_2 
\ea
and values of the constants $c_3$, $c_4$, $c_4$ and $c_6$ are given in Eq.~(\ref{RGcoeffs}); both $c_5$ and $c_6$ are proportional to $K_1$ for small $K_1$.  For each coupling, we have included the flow due to its scaling dimension, as well as (for those not initially present in the nearest-neighbour model) the leading-order term that generates it. Flow of the vortex fugacity $y$ is given in Eq.~(\ref{TreeScale}). For the in-plane stiffness $K$, we have included the leading-order non-vanishing contributions to its RG flow, demonstrating that this is slow.

The most important physical effect captured by this second-order calculation is the generation of the locking interaction $\kappa_3$ from $\kappa_b$ (which appears microscopically in a nearest-neighbour model) via the coupling $\kappa_2$. Since $\kappa_3$ is more strongly RG-relevant than $\kappa_2$ (which has the same scaling dimension as $\kappa_\perp$), it is the key interaction. It is generated only in the presence of non-zero $K_1$, itself produced from the nearest-neighbour interaction $\kappa_\perp$. Combining these steps, we find for a system with initial values $\kappa_\perp = \kappa_{\perp,0}$, $\kappa_{b} \sim \kappa_{\perp,0}$ and $K_{1} = \kappa_{2}  = \kappa_{3} =0$, that $\kappa_3 \sim (\kappa_{\perp,0})^7$ is generated after an RG scale change of order one. As discussed in Sec.~\ref{phasediagram}, this locking interaction stabilises long-range order if it dominates over the disordering effects of vortex-antivortex pairs.

We have not examined RG for the $abab$ stacking in detail beyond leading order, since we have not identified RG-relevant interactions that break the continuous ground-state symmetry of the minimal model. Symmetry is instead broken by RG-irrelevant nearest-neighbour interactions that are present microscopically, as discussed in Sec.~\ref{phasediagram}.

\section{Discussion}\label{discussion}

The results from the three approaches we have presented -- the self-consistent Gaussian approximation, Monte Carlo simulations, and analysis of height models -- establish a consistent picture. They show that triangular lattice Ising antiferromagnets with frustrated stackings exhibit classical spin liquid behaviour over an extended temperature range if interlayer coupling is weak. In this regime, there are strong correlations within and between layers, but without long-range order. 

The most significant weakness of the SCGA is that it fails to capture the ordering transition, giving instead a finite correlation length at all non-zero temperatures. The SCGA also predicts a temperature-independent value for the helix radius $Q$, while within the height model $Q$ is a function of $\beta J_\perp$. Small increases in $Q$ with decreasing $T$ at fixed $J_\perp$ are apparent in Fig.~\ref{fig:fit_params_v_T}(b), although the anticipated continuum behaviour is not fully-developed.

Some more detailed comparisons between Monte Carlo simulations and height model calculations are possible. The prediction of Sec.~\ref{phasediagram} that the ordering transition is at larger values of $J_\perp$ and smaller temperatures in systems with frustrated stacking compared to the unfrustrated case ($J_\perp \approx J e^{-20\beta J/11}$ or $J_\perp \approx J e^{-5\beta J/3}$ compared with $J_\perp \approx J e^{-6\beta J}$) is clearly consistent with simulation results shown in Fig.~\ref{fig:phase_diagram}.  For a quantitative test, we fit the phase boundaries determined in simulations to the form $J_\perp = A J \exp(-c\beta J)$. We obtain $c=1.90 \pm 0.08$ for the $abc$ stacking, $c= 1.63 \pm 0.11$ for the $abab$ stacking, and $c=5.44 \pm 0.2$ for the unfrustrated case, in striking agreement with analytical results. Values of the other fitting parameter are $A= 2.87 \pm 0.2$, $A=2.16 \pm 0.27$ and $A=6.43 \pm 0.5$, respectively.

\section*{Acknowledgements.}
We thank F. H. L Essler, O. A. Starykh and especially P. G. Radaelli for discussions. FJB is supported by NSF-DMR 1352271 and Sloan FG-2015-65927. JTC is supported in part by EPSRC Grants Nos. EP/I032487/1 and EP/N01930X/1. LDCJ is supported by the Okinawa Institute of Science and Technology 
Graduate University.

\appendix

\section{Reciprocal-space form of interaction}\label{scga_app}

In this appendix we discuss the reciprocal-space form of the interaction. This is input for SCGA calculations and is illustrated in Fig.~\ref{fig:scga_plots}. Definitions of the lattice vectors, reciprocal lattice vectors, and $K$, $K^\prime$ points are given in Eqns.~(\ref{latticevectors}) and (\ref{Kpoints}).

The contribution for all three stackings from in-plane couplings is \be
{\bf J}_{2D}({\mathbf{q}}) = 
J [\cos (q_x ) + 2 \cos\left({q_x}/{2}\right) \cos(  \sqrt{3} q_y/{2}) ]\,.
\ee

For the $aaa$ stacking, the interplane interactions contribute 
${\bf J}_\perp({\bf q}) = \cos q_z$ and  the combined minima of ${\bf J}({\bf q}) \equiv {\bf J}_{2D}({\mathbf{q}}) + {\bf J}_\perp({\bf q})$ are isolated points in reciprocal space, at $(\frac{ 4 \pi}{3},0,\pi)$ and $(\frac{ 2 \pi}{3},\frac{ 2 \pi}{\sqrt{3}},\pi)$.

For the $abc$ stacking, setting $\zeta =1 + e^{ i \mathbf{q} \cdot \mathbf{a}_1} + e^{ i \mathbf{q} \cdot \mathbf{a}_2}$, we can write the interplane coupling as ${\bf J}_\perp({\bf q}) = J_\perp( \zeta e^{-i{\bf q}\cdot\bm{\delta}}+{\rm c.c.})/2$. The in-plane coupling can also be expressed in terms of $\zeta$, as ${\bf J}_{2D}({\mathbf{q}}) = J(|\zeta|^2 - 3)/2$. The combined interaction can hence be put into the form
\begin{align} \label{Jq1}
{\bf J}({\mathbf{q}})  = \frac{J}{2}\left\lvert \zeta e^{-i \mathbf{q} \cdot \bm{\delta} } + {J_\perp}/{J}\right\rvert^2 - \frac{3J}{2} - \frac{J_\perp^2}{2J}.
\end{align}
From this it is clear that the minima of  ${\bf J}({\mathbf{q}})$ lie on the lines $\zeta = -({J_\perp}/{J})e^{i \mathbf{q} \cdot \bm{\delta} }$. If $J_\perp \ll J$, these lines are helixes with axes passing through $K$-points [Eq.~(\ref{Kpoints})] of the triangular-lattice Brillouin zone: for ${\bf J}({\bf k})$ with ${\bf k} = {\bf K} + n_1{\bf A}_1 +n_2{\bf A}_2 + {\bf q}$,  the line is
\begin{align}\label{SpiralMin}
q_x &\approx  \frac{2  J_\perp  }{\sqrt{3} J}   \cos  (q_z +\frac{2\pi}{3}p),\nonumber \\
q_y &\approx - \frac{2  J_\perp }{\sqrt{3} J}  \sin   (q_z +\frac{2\pi}{3}p)
\end{align}
where $p=n_1+n_2$, as in Sec.~\ref{correlations}. For ${\bf k} = {\bf K}^\prime + n_1{\bf A}_1 +n_2{\bf A}_2 + {\bf q}$ the line is
\begin{align}
q_x &\approx  -\frac{2  J_\perp  }{\sqrt{3} J}   \cos(q_z +\frac{2\pi}{3}p^\prime),\nonumber \\
q_y &\approx -\frac{2  J_\perp }{\sqrt{3} J}  \sin (q_z + \frac{2\pi}{3}p^\prime),
\end{align}
where $p^\prime = 2+n_1+n_2$.  At larger values of $J_\perp/J$, the helix is deformed, acquiring triangular projection in the $x-y$ plane, but the degeneracy of the line of minima is not lifted.   

As the $abab$ stacking has two sites per unit cell, the combined interaction in this case is represented by a matrix
\begin{align}
{\bf J}({\mathbf{q}})  =   \begin{pmatrix}  {\bf J}_{2D}({\mathbf{q}}) & {\bf J}_\perp^{ab}({\mathbf{q}}) \\
{\bf J}_\perp^{ba}({\mathbf{q}}) & {\bf J}_{2D}({\mathbf{q}})\end{pmatrix} 
\end{align}
with ${\bf J}_\perp^{ab}({\mathbf{q}}) = \zeta \cos(q_z/2) e^{i{\bf q}\cdot \bm{\delta}}$ and ${\bf J}_\perp^{ba}({\mathbf{q}}) = [{\bf J}_\perp^{ab}({\mathbf{q}})]^{*}$. 
The eigenvalues are
\begin{align}
\epsilon^\pm_{\mathbf{q}}	&= \frac{J}{2}(\lvert\zeta\rvert^2 - 3) \pm  J_\perp \cos\left(q_z/2\right)\lvert\zeta\rvert.
\end{align}
Minima lie on the line $q_z=0$, $\lvert\zeta\rvert = {J_\perp}/{J}$. For $J_\perp \ll J$ they form circles around the $K$-points of the triangular-lattice Brillouin zone, as shown in Fig. \ref{fig:scga_plots}.

\section{Analysis of Monte Carlo results}\label{num_app}

In this appendix we discuss in further detail our Monte Carlo results for $S({\bf q})$ and the fitting procedures used to analyse them. 

\begin{figure}[hb]
	\includegraphics[width = 0.4\textwidth]{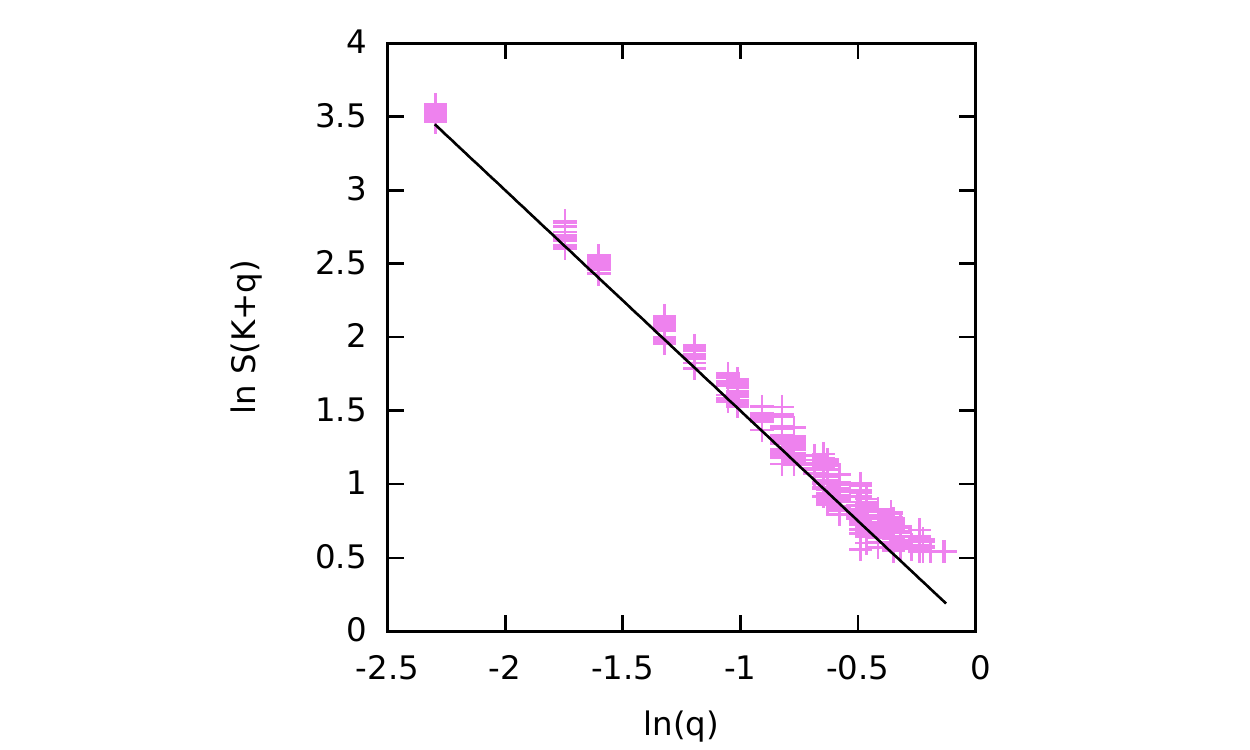}
	\caption{Illustration of power-law behavior without interlayer coupling: line has slope $-\frac{3}{2}$; system parameters are $L=72$ and $T=0.31J$. }
	\label{fig:fitting_diagnostic_00}
\end{figure}

As a simple check, we start by considering uncoupled layers, which are expected to display power-law correlations at low temperature with $S({\bf K} +{\bf q}) \propto q^{-3/2}$. The behaviour illustrated in Fig.~\ref{fig:fitting_diagnostic_00} matches this quite accurately. Interlayer interactions produce significant changes in $S({\bf q})$, and no clear remnant of the $3/2$ power law is identifiable even for the smallest values of $J_\perp/J$ that we have investigated. Instead, we find for non-zero $J_\perp$ that $S({\bf q})$ is well-represented using Lorenztian functions of wavevector.

\subsection{Correlations for the $abc$ stacking}\label{fitting_app}

The data displayed in Fig.~\ref{fig:structure_factor_slices} show helices of high intensity with axes passing through the $K$-points of the triangular-lattice Brillouin zone. In broad terms, we extract the correlation length $\xi_\perp$ and the helix radius $Q$ by analysing simulation results for S({\bf q}) separately at each $q_z$, and fitting data near the maximum to a sum of Lorentzian contributions, one from for each helix that intersects the plane. 

\begin{figure}
	\includegraphics[width=0.35\textwidth]{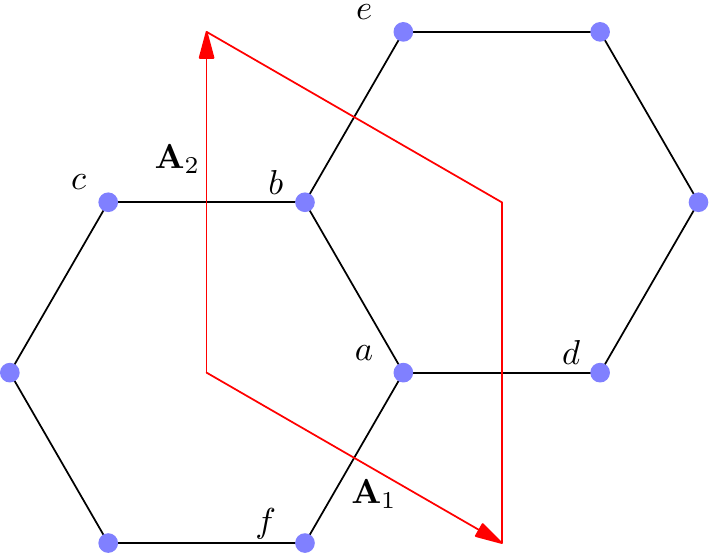}
	\caption{Brillouin zone for the triangular lattice, with $K$-points labelled $a$--$f$.}
	\label{fig:brill_zone}
\end{figure}

\begin{figure}
        \begin{minipage}{0.48\textwidth} 
		\includegraphics[width=0.95\textwidth,page=2]{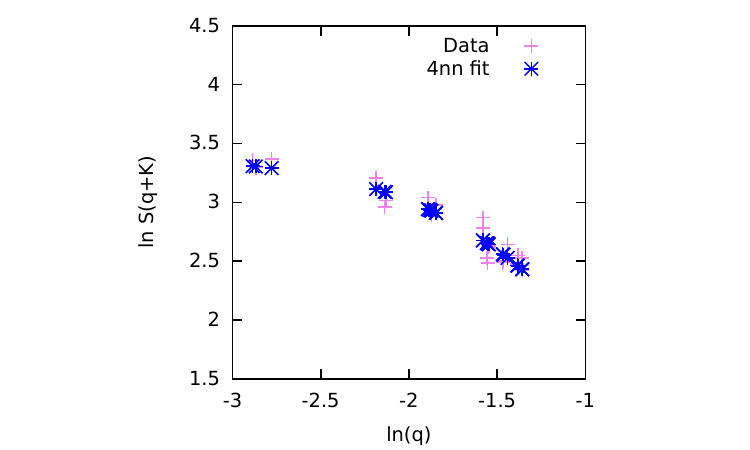}
        \end{minipage} 
	\caption{Comparison of $F_{4nn}$ with data for $L=72,L_z = 12$, $J_\perp = 0.2J$, $T=0.8J$ in the $abc$ stacking.}
	\label{fig:fitting_diagnostic_02}
\end{figure}

In detail, we consider values of $S({\bf q})$ at fixed $q_z$ with $(q_x,q_y)$ spanning one Brillouin zone. To focus on the maxima, we retain the $N$ largest values of $S(\mathbf{q})$ from a total of $L^2$ points within each $q_z$-plane. If  $N$ is too large, some points are included that are too far in reciprocal space from the helix to be well-represented by the fitting function; if $N$ is too small, statistical accuracy is sacrificed. Results are insensitive to the choice of $N$ in the range $20\leq N \leq 200$, and we use $N=50$. Referring to Fig.~\ref{fig:brill_zone}, the form of $S({\bf q})$ near the $K$-points labelled $a$ and $b$ should be dominated by helices with their axes passing through these $K$-points, but may also be influenced by helices with axes passing through the four $K$-points $c$ -- $f$ if the helix radius is large. Our fitting function 
\begin{align}
F_{4nn}(\mathbf{q}_\perp) &= \sum_i\frac{I}{\xi_\perp^{2}\left(\mathbf{q}_\perp - \mathbf{q}_{\perp,i}\right)^2+1} \label{4nn_fit}
\end{align}
therefore includes six terms, labelled by $i$. Since the different values of $ \mathbf{q}_{\perp,i}$ are related by symmetry, it contains four real scalar fitting parameters. 
The quality of fit we obtain in this way is illustrated in Fig.~\ref{fig:fitting_diagnostic_02}. 

\begin{figure}
	\includegraphics[width = 0.48\textwidth]{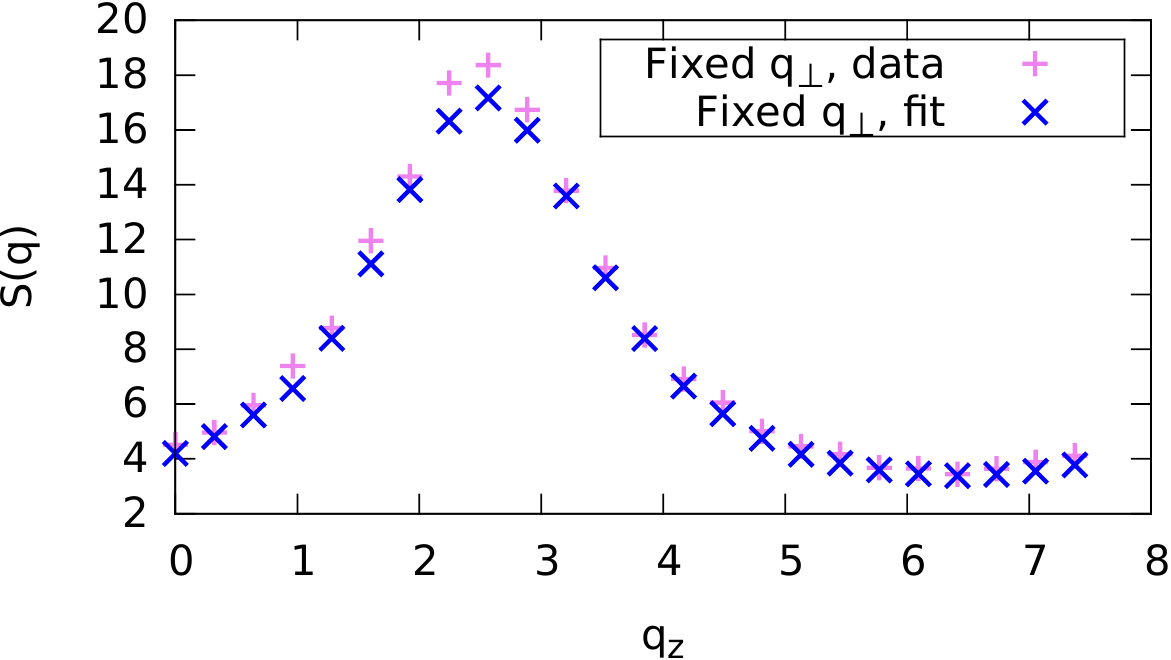}
	\caption{$S(\mathbf{q})$ vs $q_z$  for fixed $q_x$, $q_y$ in the $abc$ stacking, comparing data and fitting function. $J_\perp = 0.1J$, $L=36,L_z = 48$, $T=0.56J$.}
	\label{fig:fixed_q_x_q_y}
\end{figure}

In principle, one expects  $S({\bf q})$ to be characterised by two distinct correlation lengths, $\xi_\perp$ and $\xi_z$, as discussed in Sec.~\ref{behaviour}. In practice, we have been unable to extract a second correlation length from our Monte Carlo data for the $abc$ stacking, for reasons we now discuss. Consider first the ideal form of correlations, reached in the limit of divergent correlation lengths:
\begin{align}
S_\text{ideal}\left( \mathbf{q}\right)     &= \delta\left(q_x - q_x^0\left(q_z\right)\right)\delta\left(q_y - q_y^0\left(q_z\right)\right)\,.
\end{align}
The consequences of finite correlation lengths can be represented by convolving $S_\text{ideal}\left( \mathbf{q}\right)$ with a form factor that is characterised by its width in two directions transverse to the line $q^0_x(q_z)$, $q^0_y(q_z)$. The fitting function $F_{4nn}\left(\mathbf{q}_\perp\right)$ corresponds to a choice for this form factor that has circular contours in the $q_x$--q$_y$ plane. More general possibilities have elliptical contours; we have made fits of this type, but find they do not show significant in-plane anisotropy.  As a demonstration that the form $F_{4nn}\left(\mathbf{q}_\perp\right)$ is an adequate representation of our data, we show in Fig.~\ref{fig:fixed_q_x_q_y} a comparison of it with Monte Carlo data, as a function of $q_z$ at fixed $q_x$, $q_y$, on a line passing through the helix. The close match indicates that the broadening within the $q_x$--$q_y$ plane that is contained in $F_{4nn}\left(\mathbf{q}_\perp\right)$ also accounts for the broadening of the helix along $q_z$. 

\subsection{Correlations for the $abab$ stacking}

\begin{figure}
	\includegraphics[width=0.48\textwidth,page=2]{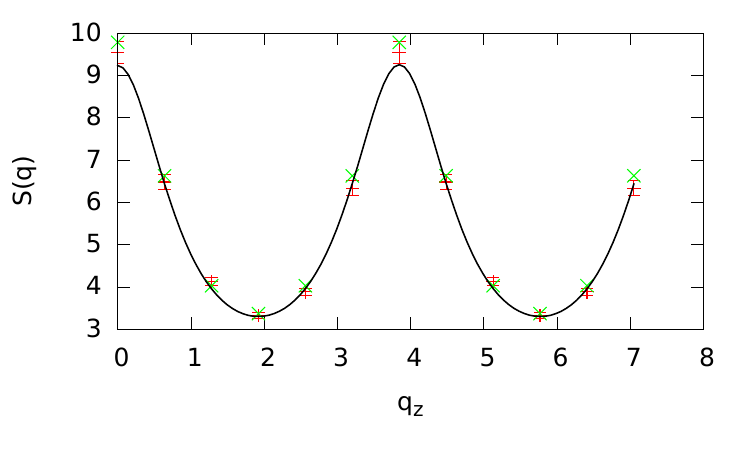}
	\caption{$S(\mathbf{q})$ vs $q_z$, for fixed $q_x,q_y$ passing through the maximum, in the $abab$ stacking: data (red); fit to SCGA (green); sum of Lorentzians (black). $J_\perp = 0.20J, T = 0.73J$.}
	\label{fig:interlayer_ab_fitting}
\end{figure}

\begin{figure}
	\includegraphics[width=0.48\textwidth,page=2]{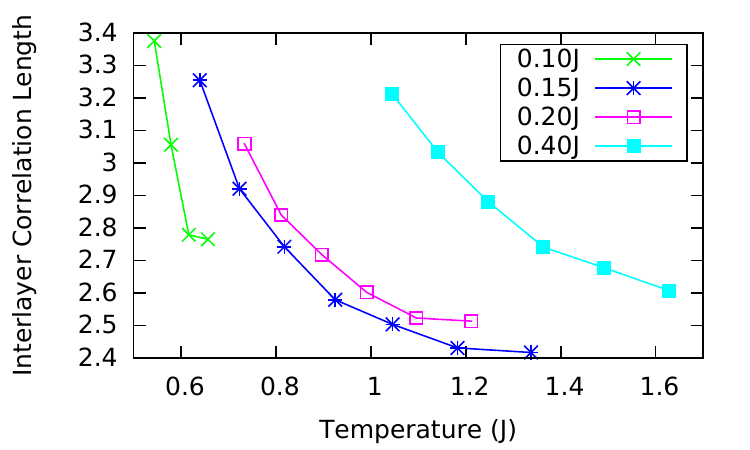}
	\caption{$\xi_z$ vs $T$ for different values of $J_\perp$ in the $abab$ stacking. The unit of length is the spacing between successive $a$-layers.}
	\label{fig:interlayer_correlations}
\end{figure}

For the $abab$ stacking, our fitting of $S({\bf q})$ as a function of $q_x$ and $q_y$  follows similar steps to the ones used for the $abc$ stacking, but analysis of the dependence on $q_z$ has new features. For this stacking the peak width of $S({\bf q})$ as a function of $q_z$ yields directly the interlayer correlation length $\xi_z$. An example of a fit is shown in
Fig.~\ref{fig:interlayer_ab_fitting} and the resulting values of $\xi_z$ are displayed as a function of $J_\perp$ and $T$ in Fig.~\ref{fig:interlayer_correlations}. 

\section{RG calculations}
\label{RGApp} 

Here we present technical aspects of our RG calculations, following a standard momentum-shell approach \cite{KogutReview}.
The general method is as follows.  Our objective is to evaluate correlation functions or the partition function
\be
Z =\int {\cal D}[h]  e^{ - ({\cal H}_0 +  {\cal H}_1 ) }
\ee
with an initial momentum cutoff $\Lambda = 1/\ell$, where $\ell$ is the lattice constant.  Here ${\cal H}_0$ is a quadratic effective Hamiltonian, which may include both in-plane and inter-plane gradient terms:
\ba \label{Hkin}
{\cal H}_{0} = \frac{1}{2}\sum_z  \int& {\rm d}^2& {\bf r} \bigg[  K \left( \nabla h_z({\bf r}) \right)^2  \nonumber \\
&+& \sum_{p>0} K_p \nabla h_z({\bf r}) \cdot \nabla h_{z+p} ({\bf r})  \bigg] \,.
\ea

We divide the height field into short-wavelength and long-wavelength modes by writing
 \ba
 h_z^>({\bf r}) &=& \int_{ \Lambda/s < |q| < \Lambda} {\rm d}^2 {\bf q} \ h_z({\bf q}) e^{i {\bf q} \cdot {\bf r} }  \nonumber \\ 
 {\rm and} \quad h_z^<({\bf r}) &=& \int _{|q|\leq \Lambda/s} {\rm d}^2{\bf q} \ h_z({\bf q}) e^{i {\bf q} \cdot {\bf r} }\,.
 \ea
A new effective Hamiltonian ${\cal H}_{\rm eff}$ with a reduced cutoff $\Lambda/s$ is obtained by integrating out the short-wavelength modes, and then re-scaling all in-plane lengths by $s$.  Note that we retain the layer index $z$ as a discrete variable, and coarse-grain only the in-plane co-ordinates. Expanding in powers of ${\cal H}_1$
\ba \label{CalculateMe}
e^{ -{\cal H}_{\rm eff}}& =&\int \prod_z {\cal D}[h_z^> ]  e^{ - ({\cal H}_0  +   {\cal H}_1 )  } 
\\
&\approx &\int \prod_z {\cal D}[h_z^> ]  e^{ - {\cal H}_0  } \left \{  1 -   {\cal H}_1  + \frac{1}{2}   {\cal H}_1^2  + ... \right \}  \,.
 \nonumber 
\ea
To quadratic order, the effective Hamiltonian with the reduced cutoff $\Lambda/s$ is 
\ba \label{FindSeff}
{\cal H}_{\rm eff} &=&     {\cal H}^\prime_0   +   \langle {\cal H}_1   \rangle_{0} 
- \frac{1}{2}  \langle  {\cal H}_1^2  \rangle_{0} + ...
\ea
where all terms are functions of only the long-wavelength fields $h_z^{<}({\bf r})$, the average $\langle \ldots \rangle_0$ is over short-wavelength fields with weight $e^{ - {\cal H}_0  }$ and ${\cal H}^\prime_0$  is obtained from ${\cal H}_0$ by omitting the short-wavelength fields. As a final step, lengths in ${\cal H}_0^\prime$, and in the expectation values on the right are re-scaled according to ${\bf r} \rightarrow s {\bf r}$.

\subsection{First-order calculation}
\label{DimensionApp}

We derive the first-order RG equations as follows.  Consider the interlayer coupling
\ba
{\cal H}_1& =& \kappa_\perp \int d^2 {\bf r}   \left [ \partial_x h_z({\bf r}) \cos \delta h_z ({\bf r}) -\partial_y h_z({\bf r}) \sin \delta h_z ({\bf r}) \right ]\nonumber\\  & = & \kappa_\perp \text{Im} \left[ \int d^2 {\bf r} \,  \partial_{\overline{\zeta}_{z}} e^{i  \delta h_z ({\bf r})}  \right ] 
\ea
where we have introduced $\zeta_z = x_z + i y_z$, and $z$ denotes the layer with which the coordinates $x,y$ are associated.  We have
\be
\langle {\cal H}_1 \rangle_{0}=  \kappa_\perp \text{Im} \left[   \int d^2 {\bf r} \,  \partial_{\zeta_{z}} e^{i  \delta h_< ({\bf r},z)}  \langle e^{ i \delta h_> ({\bf r},z)} \rangle_{0}  \right ] \,.
\ee
Defining
\ba
 F_n &\equiv& \langle h_>({\bf r},z) h_>({\bf r}, z+n) \rangle \nonumber\\ &=&\frac{ \gamma_n}{4\pi^2 K} \int_{\Lambda > |q|> \Lambda/s} {\rm d}^2 {\bf q}  \frac{1}{q^2}   =\frac{  \gamma_n  }{2 \pi K  } \log s   \n
 \ea
 with
 \ba
 \gamma_n  &=& \frac{1}{2\pi} \int_0^{2\pi} {\rm d}{k_z}   \frac{ K \cos n k_z }{ \left [ K + \sum_{p>0} K_p  \cos p k_z \right ] }  \n 
 {\rm and} \quad \beta_n &=&  \frac{\pi}{18 K}  \left( \gamma_0 - \gamma_n \right ) \,,
 \ea
we obtain
\ba \label{1PtCor}
\langle e^{i  \delta_n h(r,z) } \rangle & = & \text{exp} \left [ - \frac{\pi^2}{18} \langle \delta_n h({\bf r},z) ^2 \rangle  \right ] \n
&=&  \text{exp} \left [ - \frac{\pi^2}{9}\left( F_0 - F_n  \right )  \right ]
= s^{- \beta_n}\,.
 \ea
The re-scaling ${\bf r} \rightarrow s {\bf r}$ gives
\be \label{EqH1p1}
{\cal H}_{\rm eff}   =  \kappa_\perp  \int d^2 {\bf r} \,  \text{Im} \left[   \partial_{\zeta_{z}} e^{i  \delta h_< ({\bf r},z)}  \right ] s^{1-\beta_1}
\ee
In the continuum limit $s\to 1$ we have
\be
\frac{\partial  \kappa_\perp }{\partial \ln \ell} =(1- \beta_1)\kappa_\perp.
\ee
Scaling dimensions of the other operators can be deduced in a similar way.

\subsection{Second-order calculation}

At second order, we must evaluate the quadratic terms in Eq.~(\ref{FindSeff}).  It is useful to introduce some notation. Let ${\cal H}_n(z)$ denote a contribution to interlayer coupling involving the height differences $\delta_p h_z({\bf r}) \equiv \frac{\pi}{3}[h_{z+p}({\bf r}) - h_z({\bf r})]$
and define
\be
\Delta_{m,n, z-z'} =  \langle {\cal H}_m(z) {\cal H}_n (z') \rangle_{0} - \langle {\cal H}_m(z) \rangle_{0}  \langle {\cal H}_n (z') \rangle_{0} \,.\n
\ee
We are primarily interested in two types of such term: those that contribute to the most relevant interlayer couplings, and those that contribute corrections to the marginal gradient couplings. 

\begin{widetext}

\subsubsection{Corrections to gradient couplings}

We first compute corrections to the gradient couplings that are generated by $\Delta_{n,n, 0}$ for various $n$.  An example is
\ba
\Delta_{3,3,0} &=&  \langle {\cal H}_3 (z) {\cal H}_3 (z) \rangle_0 - \langle {\cal H}_3(z) \rangle_0 \langle {\cal H}_3 (z) \rangle_0 \n
&=&\frac{(\kappa_3)^2}{2} \int {\rm d}^2 {\bf r}\, {\rm d}^2{ \bf r'} \left \{ \cos \left( \delta_3 h^<_z({\bf r}) +  \delta_3
 h_z^<({\bf r}')  \right ) \left( \langle e^{i \delta_3 h_z^>({\bf r} ) } e^{  i \delta_3 h_z^>({\bf r}') } \rangle_0 - s^{ - 2 \beta_3} \right )  \right . \n
 && \left. 
 + \cos \left( \delta_3 h_<({\bf r}, z ) -  \delta_3 h_<({\bf r}', z)  \right ) \left( \langle e^{i \delta_3 h_>({\bf r}, z ) } 
  e^{- i \delta_3 h_>({\bf r}', z) } \rangle - s^{ - 2 \beta_3} \right )  \right \} 
\ea
We write
\be
\langle e^{i \delta_3 h_z^>({\bf r} ) } e^{  i \delta_3 h_z^>({\bf r}') } \rangle_0 - s^{ - 2 \beta_3} = s^{ - 2 \beta_3}  \left(e^{ - 4 \pi \beta_3 G({\bf r}  -{\bf r'} )}  -1 \right ),
\ee
where
\be \label{TheGeq}
G({\bf r }) =  \int_{\Lambda /s < |q| < \Lambda} \frac{{\rm d}^2 {\bf q}}{4 \pi^2} 
\frac{e^{ i {\bf q \cdot r} } }{q^2} \ \ .
\ee

Assuming that $\left( e^{  4 \pi \beta_3 G({\bf r}  -{\bf r'} )}  - 1\right ) $ is small unless $|{\bf r}  -{\bf r'}|  \ll 1$, we expand the long-wavelength height fields in ${\bf R} = {\bf r}  -{\bf r'}$ to obtain 
\ba
\Delta_{3,3,0} &=&\frac{(\kappa_3)^2}{2} \int {\rm d}^2 {\bf r}\,{\rm d}^2{ \bf r'} \left [  \cos \left(2 \delta_3 h_z^<({\bf r}) \right )  + ( {\bf r} - {\bf r'} ) \cdot \nabla  \delta_3 h_z^<({\bf r}) \sin \left( 2 \delta_3 h_z^<({\bf r})  \right )  + ... \right ] s^{ - 2 \beta_3}  \left(e^{ - 4 \pi \beta_3 G({\bf r}  -{\bf r'} )}  -1 \right ) \n
 &&
+ \frac{1}{2} \int {\rm d}^2 {\bf r}\, {\rm d}^2{ \bf r'}  \left \{ 1 -  \frac{1}{2}  \left(  ( {\bf r} - {\bf r'} ) \cdot \nabla \delta_3 h_z^<({\bf r}) \right )^2 s^{ - 2 \beta_3}  \left( e^{  4 \pi \beta_3 G({\bf r}  -{\bf r'} )}  - 1\right ) \right \} 
 \ea
 The terms in the first line are new, less relevant couplings between spins 3 layers apart, and can be ignored.  The first term in the second line is a constant, and the second term in the second line is the contribution to the gradient energy that we are interested in.  After performing the angular integration, only terms of the form $(\nabla \delta h^< )^2$ remain and we obtain
 \be \label{Delta2FinalEq}
  \Delta_{3,3,0} =  -  \frac{(\kappa_3 s^{ -  \beta_3}  )^2}{8} B \int d^2 {\bf r}\,  | \nabla (\delta_3 h_z^<({\bf r}))|^2 +  \text{const} + . . . ,
\ee
where the ellipsis represents the less relevant interlayer couplings, and
\be
B=\int d^2 { \bf R} \,    R^2  \left( e^{  4 \pi \beta_3 G({\bf R} )}  - 1\right )\,.
\ee

The other important corrections to the gradient energy come from $\Delta_{1,1,0}$ and $\Delta_{2,2,0}$.
These contain two types of terms, with the forms
\ba
\Delta_{1,1,0}(+) &=& 
-(\kappa_\perp)^2 \int {\rm d}^2 {\bf r} \,{\rm d}^2 {\bf r'} \, \left [  \langle \partial_{\overline{\zeta}_{z}}  \partial_{\overline{\zeta'}_{z}} e^{i  ( \delta h_z ({\bf r})+ \delta h_z ({\bf r'}) )} \rangle_0 -\langle \partial_{\overline{\zeta}_{z}}   e^{i  \delta h_z ({\bf r})} \rangle_0 \langle \partial_{\overline{\zeta'}_{z}}e^{i \delta h_z ({\bf r'}) } \rangle_0  + c.c.\right ]  \n
{\rm and} \quad \Delta_{1,1,0}(-) &=& 
 (\kappa_\perp)^2\int {\rm d}^2 {\bf r}\, {\rm d}^2 {\bf r'} \, \left [  \langle \partial_{\overline{\zeta}_{z}}  \partial_{\zeta'_{z}} e^{i  ( \delta h_z ({\bf r})- \delta h_z ({\bf r'}) )} \rangle_0 -  \langle \partial_{\overline{\zeta}_{z}}  e^{i  \delta h_z ({\bf r})} \rangle_0 \langle \partial_{\zeta'_{z}}  e^{-i \delta h_z ({\bf r'}) } \rangle_0 + c.c. \right ]\,.
\ea
Terms of the first type generate new (but irrelevant) inter-layer couplings that do not lift the helical degeneracy; they are not important for our analysis.  We are interested in terms of the second type, which reduce to
\ba
\Delta_{1,1,0} (-)  &=& \frac{(\kappa_\perp)^2}{2}\int {\rm d}^2 {\bf r}\, {\rm d}^2{ \bf r'} \partial_{\overline{\zeta}_{z}}  \partial_{\zeta'_{z}} \left \{  \cos \left( \delta h_z^<({\bf r} ) -  \delta h_z^<({\bf r}')  \right ) s^{ - 2 \beta_1}  \left( e^{  4 \pi \beta_1 G({\bf r}  -{\bf r'} )}  - 1\right )  \right \} 
\ea
Differentiating both slow and fast fields, and expanding for small ${\bf R}$, we obtain the four terms
\be
\Delta_{1,1,0} (-)  =  \frac{(\kappa_\perp)^2}{2}  s^{ - 2 \beta_1} \int {\rm d}^2 {\bf r}\,\left \{  C_0 | \nabla h_z^< ({\bf r}) |^2  -C_1 \left( \nabla h_z^< ({\bf r}) \right)\cdot \left( \nabla \delta h_z^< ({\bf r}) \right) + C_2 | \nabla \delta h_z^< ({\bf r}) |^2  + C_3\right \}+ \text{ irrel.}
\ee
where
\ba \label{Ceqs}
C_0 &=&   \int {\rm d}^2 {\bf R }  \left( e^{  4 \pi \beta_1 G({\bf R} )}  - 1\right ), \qquad
C_1 =  4 \pi \beta_1 \int {\rm d}^2 {\bf R } \,\left( {\bf R} \cdot \nabla_{\bf R}  G({\bf R} ) \right) e^{  4 \pi \beta_1 G({\bf R} )}  =-  C_0 \n
C_2 &=& - \frac{1}{4}  \int {\rm d}^2 {\bf R }  \, {R}^2 \left[    \nabla_R^2 e^{  4 \pi \beta_1 G({\bf R} )}  -8 \pi^2 \beta_1^2 \left | \nabla_R  G({\bf R} ) \right |^2 e^{  4 \pi \beta_1 G({\bf R} )} \right ]  = - \frac{1}{4} C_0  + \frac{ \pi^2}{9 K} \left( \gamma_1 + \gamma_0 \right )  \int d^2 {\bf R }  \, {R}^2 \nabla_R^2 G({\bf R} )  e^{  4 \pi \beta_1 G({\bf R} )}\n
C_3 &=&  \int {\rm d}^2 {\bf R }  \, \left(  \frac{ \pi^2}{9 K} \gamma_0  \nabla_R^2 G ({\bf R}) 
+ 2 \pi \beta_1^2 | \nabla_{\bf R} G({\bf R}) |^2 \right )e^{  4 \pi \beta_1 G({\bf R} )}
\ea
and we have exploited symmetries in the integration over ${\bf R}$.  

Summing over layers, the contribution to the gradient energy is
 \be
\delta {\cal H}   = \frac{(\kappa_\perp s^{ -  \beta_1})^2}{2}  \sum_z \int {\rm d}^2 {\bf r}\left \{   2 C_2 | \nabla h_z^< ({\bf r}) |^2  
 - \left( 2 C_2 - C_0  \right)\left( \nabla h_z^< ({\bf r}) \right)\cdot \left( \nabla \delta h_{z+1}^< ({\bf r}) \right)  \right \}\,.
\ee
A similar contribution arises from $\Delta_{2,2,0}$.
Although the leading irrelevant terms in the interlayer coupling also renormalise the gradient energy, we will neglect their effect here as it influences only the initial part of the RG flow.

\subsubsection{Generation of new interlayer couplings}

We now turn to the most important part of our RG calculation, which is to determine at what order in $\kappa_\perp$ the relevant inter-layer coupling
${\cal H}_3$ is generated in a microscopic theory with only nearest-layer couplings. (Recall from Eq.~(\ref{abcH3}) that ${\cal H}_3$ couples layers three apart in the $abc$ stacking.) Importantly, we show that though one might expect ${\cal H}_3$ to be produced at order $\kappa_\perp^3$, in fact this is not the case: generating this interaction requires $K_1\neq 0$, and it consequently appears at order $\kappa_\perp^7$.

 If ${\cal H}_3$ is absent, then to generate it we must keep the leading irrelevant term that breaks the U(1) symmetry, which is the interlayer coupling ${\cal H}_b(z)$ of Eq.~(\ref{Hb}). 
Then ${\cal H}_2$ [Eq.~(\ref{abcH3})] is generated by the bilinear
\be
\Delta_{b,1,1} = \langle {\cal H}_1(z) {\cal H}_b (z+1) \rangle  - \langle {\cal H}_1(z) \rangle \langle {\cal H}_b (z+1) \rangle  
\ee
and ${\cal H}_3$ is generated by
\be
 \Delta_{2,1,1} =\langle {\cal H}_1(z) {\cal H}_2(z+1) \rangle - \langle {\cal H}_1(z)\rangle \langle {\cal H}_2(z+1) \rangle  .
\ee
Other cross-terms, such as $\langle {\cal H}_1 (z) {\cal H}_1(z+1) \rangle $ and $\langle {\cal H}_3 (z) {\cal H}_1(z) \rangle $ also generate new inter-layer couplings.  However, for our purposes these can be ignored: they are either less relevant than the terms listed above, or equally relevant but appear at a higher order in $\kappa_\perp$.  

We have
\ba
\Delta_{2,1,1} &=& - \kappa_\perp \kappa_2 \int {\rm d}^2 {\bf r}\, {\rm d}^2 {\bf r'} \, \left [  \langle \partial_{\overline{\zeta}_{z}}  \partial_{\zeta'_{z+1}} e^{i  ( \delta h_z ({\bf r})+ \delta_2 h_{z+1} ({\bf r'}) )} \rangle -\langle \partial_{\overline{\zeta}_{z}}   e^{i  \delta h_{z} ({\bf r})} \rangle \langle \partial_{\zeta'_{z+1}}e^{i \delta_2 h_{z+1} ({\bf r'}) } \rangle + h.c.\right ]   + ...  \n
&\approx & -\left( \kappa_\perp s^{ -  \beta_1} \right ) \left(  \kappa_2 s^{- \beta_2} \right )C_3(2)  \int {\rm d}^2 {\bf r}  \cos \left( \delta_3 h_z ({\bf r}) \right )  + \text{ less relevant terms }
\ea
where $+ ...$ represents a contribution that generates terms of the form $\text{exp} \left[ i \frac{\pi}{3} ( h_{z+3}({\bf r'} ) + h_z({\bf r} ) -  h_{z+1}({\bf r'} ) -h_{z+1}({\bf r} )  ) \right ]$, which produce inter-layer couplings less relevant than the terms of interest, which have been neglected in the second line.  Additionally, in the second line we have kept only terms in which all derivatives are applied to the fast height fields, as these generate the most relevant inter-layer coupling, and neglected all but the leading order term in a derivative expansion of the argument of the cosine term for small ${\bf R} = {\bf r} - {\bf r}'$. The coefficient is 
\ba \label{C3def}
C_3(n) & =&  \frac{ \pi^2}{9 K}\int {\rm d}^2  {\bf R} \, 
  \left \{ -  \gamma_1 \nabla_{\bf R}^2 G({\bf R})   +\frac{\pi^2}{9 K}  \left( \gamma_{1}   - \gamma_{n+1} \right)   \left( \gamma_{1}   - \gamma_0 \right)  \left| \nabla_{\bf R} G({\bf R} ) \right| ^2\right \} e^{ \frac{  \pi^2}{9 K}  \left(\gamma_0 - \gamma_1 + \gamma_{n+1} - \gamma_n \right)  G({\bf R} ) } \,.
\ea

The frustrated second-layer coupling is generated by
\be \label{SecondMake}
\Delta_{b,1,1} = -4 \kappa_\perp \kappa_b  \int {\rm d}^2 {\bf r}\, {\rm d}^2 {\bf r'} \, \left [  \langle \partial_{\overline{\zeta}_{z}} e^{i  \delta h_z ({\bf r}) }( \partial_{\zeta'_{z+1}} e^{i  \delta h_{z+1} ({\bf r'})/2 } )^2 \rangle -\langle \partial_{\overline{\zeta}_{z}}   e^{i  \delta h_{z} ({\bf r})} \rangle \langle ( \partial_{\zeta'_{z+1}} e^{i  \delta h_{z+1} ({\bf r'})/2 } )^2 \rangle + h.c.\right ]  + ...\nonumber
\ee
where again, $+ ...$ generates interlayer couplings of the form $\text{exp}\left[ i \frac{\pi}{3} ( h_{z+2} + h_z - 2 h_{z+1} ) \right ]$, which we omit as they are less relevant. In this case, because ${\cal H}_2$ involves one derivative of the slow height fields, we must calculate two terms. 
First, applying all three of the derivatives to $h^>$ gives the terms
\ba
\kappa_\perp \kappa_b\int &{\rm d}^2 {\bf r}& {\rm d}^2 {\bf r'}e^{ i  \left( \delta h_z^<({\bf r}) + \delta h_{z+1}^{<} ({\bf r'}) \right )} \lim_{{\bf r}'' \rightarrow {\bf r}'} \partial_{\overline{\zeta}_{z}} \partial_{\zeta'_{z+1}}  \partial_{\zeta''_{z+1}} \langle  e^{i  \delta h_z^> ({\bf r}) } e^{i  \delta h_{z+1}^{>} ({\bf r'})/2 } e^{i  \delta h_{z+1}^{>} ({\bf r''})/2 }  \rangle + h.c. \n
&=& \left( \kappa_\perp s^{ - \beta_1 } \right ) \left( \kappa_b s^{ - \beta_1 } \right )  \left( \frac{\pi^2}{9 K} \right)^3 \left[ (\gamma_1 - \gamma_{n+1})( \gamma_1-\gamma_0)^2 \right]   \n
 &\times& \int {\rm d}^2 {\bf r}\, {\rm d}^2 {\bf r'}
\left [ e^{i  \left( \delta h_z^<({\bf r}) + \delta h_{z+1}^{<} ({\bf r'}) \right ) }
\partial_{\overline{\zeta}_z} G({\bf r} - {\bf r'} ) ( \partial_{\zeta'_{z'}} G({\bf r} - {\bf r'} ) )^2 + h.c. \right ] e^{ - \frac{  \pi^2}{9 K}  \left(\gamma_0 + \gamma_2- 2 \gamma_1 \right)  G({\bf R} ) } \,.
 \nonumber
\ea
Next, we Taylor expand for small ${\bf R} = {\bf r} - {\bf r'}$.  After integrating over ${\bf R}$, the leading-order term vanishes, and the most relevant term that we are left with is
\be
\left( \kappa_\perp s^{ - \beta_1 } \right ) \left( \kappa_b s^{ - \beta_1 } \right ) C_4 \int {\rm d}^2 {\bf r} \left[  \partial_x  \delta_2 h_z^<({\bf r}) \cos  \left( \delta_2 h_z^<({\bf r}) \right )  +  \partial_y  \delta_2 h_z^<({\bf r}) \sin  \left( \delta_2 h_z^<({\bf r}) \right )   \right ]
\ee
where 
\ba \label{Eqc4}
C_4 =\left( \frac{\pi^2}{9 K} \right)^3  \left[ (\gamma_1 - \gamma_{n+1})( \gamma_1-\gamma_0)^2 \right] \int {\rm d}^2 {\bf R} \, R_x \partial_x G({\bf R}) |\nabla G({\bf R}) |^2 e^{ - \frac{  \pi^2}{9 K}  \left(\gamma_0 + \gamma_2- 2 \gamma_1 \right)  G({\bf R} ) } 
\ea
Second, applying one derivative to $h^<$ in Eq. (\ref{SecondMake}) generates the contribution to ${\cal H}_2$
\ba
4  \kappa_\perp \kappa_b\int &{\rm d}^2 {\bf r}&\, {\rm d}^2 {\bf r'} \, \left [ e^{i  \delta h_{z+1}^< ({\bf r'})/2 } \partial_{\zeta'_{z}} e^{i  \delta h_{z+2}^< ({\bf r'})/2 } \langle \partial_{\overline{\zeta}_{z}} e^{i  \delta h_z^> ({\bf r}) }( e^{i  \delta h_{z+1}^{>} ({\bf r'})/2 } \partial_{\zeta'_{z}} e^{i  \delta h_{z+1}^{>} ({\bf r'})/2 } ) \rangle 
 + h.c.\right ]  \n
& =& \kappa_\perp \kappa_b\int {\rm d}^2 {\bf r}\, {\rm d}^2 {\bf r'} \, \left [  \partial_{\zeta'_{z}} e^{i  \delta h_{z+1}^{<} ({\bf r'}) } \langle \partial_{\overline{\zeta}_{z}} e^{i  \delta h_z^> ({\bf r}) } \partial_{\zeta'_{z}} e^{i  \delta h_{z+1}^{>} ({\bf r'}) } ) \rangle 
 + h.c.\right ]  \n
&\approx&  \left( \kappa_\perp s^{ - \beta_1 } \right ) \left( \kappa_b s^{ - \beta_1 } \right )C_3 (1)  \int {\rm d}^2 {\bf r}  \left ( \partial_x h_z^<({\bf r}) \cos ( \delta_2 h_z^<({\bf r}) ) + \partial_y h_z^<({\bf r}) \sin (\delta_2 h_z^<({\bf r}) ) \right ) 
\ea
with $C_3(n)$ as defined in Eq. (\ref{C3def}).  
The remaining contributions, which come from applying two or three derivatives to $h^<$, necessarily produce irrelevant couplings, and are safely omitted from our RG calculation.  
In total, we therefore obtain 
\ba
\Delta_{b,1,1} &=&  \left( \kappa_\perp s^{ - \beta_1 } \right ) \left( \kappa_b s^{ - \beta_1 } \right )  \int {\rm d}^2 {\bf r} \left [  \left( C_3(1) - \frac{ \pi}{3} C_4 \right  )\left  \{ \partial_x h_z^< ({\bf r}) \cos \left( \delta_2 h_z^< ({\bf r}) \right )+ \partial_y  h_z^<({\bf r}) \sin  \left( \delta_2 h_z^<({\bf r}) \right )  \right \}  \right . \n
&& \left. + \frac{ \pi}{3} C_4  \left  \{ \partial_x h_{z+2}^< ({\bf r}) \cos \left( \delta_2 h_z^< ({\bf r}) \right )+ \partial_y  h_{z+2}^<({\bf r}) \sin  \left( \delta_2 h_z^<({\bf r}) \right )  \right \}  \right ] + \text{irrelevant terms}
\ea
Though the first terms in the second line are just as relevant as the terms in the first line, they do not contribute to generating ${\cal H}_3$ until inter-layer kinetic terms  $K_n$ are generated for $ n>1$ (see Eq. (\ref{Hkin})).  Hence we have neglected them in our discussion, as  their impact on the other couplings in the RG is very weak.  We will also see presently that the coefficient $C_4$ is negligibly small compared to $C_3(1)$.

We emphasise that  both $C_3(1)$ and $C_4$ are of order at least $\kappa_\perp^2$, since for $K_1 \ll K$ and $K_n =0, n>1$, we have
\be
\gamma_0 = \frac{1}{ \sqrt{1 - \left ( \frac{K_1}{K} \right)^2 }} \approx 1 \ , \ \ \ \ \gamma_1 = \frac{K}{K_1} \left( 1 -  \frac{1}{ \sqrt{1 - \left ( \frac{K_1}{K} \right)^2 }} \right ) \approx \frac{ K_1}{2 K} \ , \ \ \ \gamma_n =0\ ,\  n>1 \ ,
\ee
and $K_1$ is generated only at order $\kappa_\perp^2$.   Therefore, in summary,  $\kappa_3$ is generated not at order $\kappa_\perp^3$, as one might naively have expected, but at order $\kappa_\perp^7$.  A similar effect was noted for frustrated couplings in Ref. \onlinecite{Starykh}.   

\subsection{Evaluation of coefficients} \label{CoefficientSect}

To proceed further, we must evaluate the coefficients $B, C_0, C_1, C_2$, and $C_3(n)$. 
In order to compute the relevant integrals, we expand the exponentials for small $G({\bf r})$.  (We will justify this expansion presently).
To ensure that all integrals are absolutely convergent, we take our system to have a finite size $L_x = L_y = L$, and use periodic boundary conditions.  
In this case the first-order terms vanish after integration, since
\be \label{Gdelts}
\int {\rm d}^2 {\bf r} \, G({\bf r }) =  \frac{1}{L^2} \int {\rm d}^2 {\bf r} \, \sum^\prime_{\Lambda/s< q< \Lambda} \frac{e^{ i {\bf q} \cdot {\bf r} } }{q^2}\n
= \frac{1}{L^2}  \sum^\prime_{\Lambda/s< q< \Lambda}    \frac{ 1 }{q^2}
\int {\bf d}^2 {\bf r}\, e^{ i {\bf q \cdot r} }   = \sum^\prime_{\Lambda/s< q< \Lambda}  \begin{cases} 1 & \text{ if } {\bf q} = 0 \\
0 & \text{ else }
\end{cases}\,.\nonumber
\ee
 A similar derivation applies for derivatives of a single power of $G$, which also vanish.  
The leading-order contributions are therefore quadratic in $G$.  Keeping only these terms, the integrals of interest are
\be
I_0 = 
  \int {\rm d}^2 {\bf r }  \left(  G({\bf r} ) \right )^2, \ \
I_1=   
\int {\rm d}^2 {\bf r } \, { r}^2  \left(  G({\bf r} ) \right )^2,\ \ 
I_2 =   
 \int {\rm d}^2 {\bf r }  \, { r}^2 G({\bf r} ) \nabla^2 G ({\bf r}) \ \
{\rm and} \ \ I_3  =   
\int {\rm d}^2 {\bf r } \, G({\bf r} ) \nabla^2 G ({\bf r})  - \int {\rm d}^2 {\bf r }  | \nabla G({\bf r}) |^2\,. \nonumber
\ee
The two integrals not involving explicit powers of ${\bf r}$ are easily evaluated as
\ba \label{Gdelts2}
I_0 &=& \int d^2 {\bf r} \,  \left(   G({\bf r} )  \right )^2 = \frac{1}{L^4}  \int d^2 {\bf r} \sum_{{\bf q},{\bf k}} \frac{e^{ i  ( {\bf q} + {\bf k} ) \cdot {\bf r} } }{q^2 k^2}
= \frac{1}{L^2}  \sum^\prime_{\bf q}   \frac{ 1 }{q^4} \approx \frac{1}{4  \pi^2} \int_{\Lambda/s}^{\Lambda}  \frac{d^2 {\bf q} }{q^4 } = \frac{1}{2 \pi \Lambda^2} \frac{ s^2 -1 }{2} \approx \frac{{\rm d}s}{ 2 \pi  \Lambda^2}  \n 
I_3 &=&  \int d^2 {\bf r }  \, \left(   G({\bf r} )  \nabla_r^2 G ({\bf r})  \right ) = \frac{1}{L^4}  \int d^2 {\bf r }  \sum^\prime_{{\bf q},{\bf k}} \frac{e^{ i  ( {\bf q} + {\bf k} ) \cdot {\bf r} } }{q^2 }
\approx \frac{1}{(2 \pi)^2} \int_{\Lambda/s}^{\Lambda}  \frac{d^2 {\bf q} }{q^2 } = \frac{\log (s) }{2 \pi} \approx \frac{{\rm d}s}{2 \pi}\,.\nonumber
\ea
We note, somewhat surprisingly, that it is the terms quadratic in $G({\bf r})$ -- rather than the linear terms --  that are proportional to ${\rm d}s$.   
Our Taylor expansion is nevertheless justified: for higher powers of $G$, the $\delta$-function constraint takes the form $\delta ( \sum_{i=1}^n {\bf k }_i )$.  In practice, this means that non-zero contributions to momentum integrals require both that $|{\bf k}_i|$ is within the momentum shell for each $i$ and also that $|\sum_{i=1}^{n-1} {\bf k }_i |$ lies in this shell.  This leads to a strong phase-space suppression of the relevant angular integrals for $n>2$, justifying the quadratic approximation used here.  

Evaluating $I_1$ and $I_2$, we encounter a second difficulty: the resulting integrals retain an explicit dependence not only on the cutoff $\Lambda$ but also on the system size $L$.  
This stems from the factors of ${r}^2$ in the integrands, which arise from Taylor expansions of the type 
\be
f ( h^< ({\bf R} ) - h^< ( {\bf R} + {\bf r} ))  \approx f ( {\bf r} \cdot \nabla h^< ({\bf R} ) ) + ... 
\ee
followed by an
expansion of $f$ for small $r$.  The expansion is justified if $G({\bf r})$ falls off sufficiently quickly in $r$ that only small values of $r$ contribute;
however, in the cases at hand this is not so. 

To circumvent this difficulty, we instead expand the function $f$ to quadratic order in the difference $h^< ({\bf R} ) - h ^<( {\bf R} + {\bf r} )$ of the height fields, without making a Taylor expansion of the height fields in powers of ${\bf r}$.  The approach amounts to the substitution
\ba
\int& {\rm d}^2& {\bf q}  \, q^2 h^<_{\bf q} h^<_{- {\bf q} }  \int {\rm d}^2 {\bf r } \, r^2 F({\bf r} )\approx \int {\rm d}^2 {\bf q}  \,  h^<_{\bf q} h^<_{- {\bf q} }  \int {\rm d}^2 {\bf r }  \, 2( 1 - \cos {\bf q \cdot \bf r} ) F({\bf r} )\,. \nonumber
\ea
Using this, we obtain
\ba
q^2 I_1&=& 2  \int {\rm d}^2 {\bf r }  \,( 1 - \cos {\bf q \cdot \bf r} ) \left( G({\bf r} ) \right)^2  = \frac{1}{L^4} \sum_{ {\bf k}_1, {\bf k}_2 }  \int {\rm d}^2 {\bf r } \left (2  \frac{e^{ i  ( {\bf k}_1 + {\bf k}_2 ) \cdot {\bf r} } }{k_1^2 k_2^2 } -  \frac{e^{ i  ( {\bf k}_1 + {\bf k}_2 + {\bf q}  ) \cdot {\bf r} } }{k_1^2 k_2^2 } - \frac{e^{ i  ( {\bf k}_1 + {\bf k}_2 - {\bf q}  ) \cdot {\bf r} } }{k_1^2 k_2^2 }
\right )  \n
& \approx  & \frac{1}{4 \pi^2} \int \frac{{\rm d}^2{\bf k}  }{k^2} \left( \frac{2}{ k^2} - \frac{1}{ | {\bf k} + {\bf q} |^2} - \frac{1}{ | {\bf k} - {\bf q} |^2} \right ) ,
\approx   -q^2  \frac{{\rm d}s}{ \pi \Lambda^4}
\ea
where in the last line we have kept terms only to quadratic order in $q$, as higher-order terms are RG-irrelevant.  
Similarly, we may evaluate
\ba
q^2 I_2 &=& 2  \int {\rm d}^2 {\bf r }  \,( 1 - \cos {\bf q \cdot \bf r} )  G({\bf r} ) \nabla_{\bf r}^2 G({\bf r} )   = \frac{1}{L^4} \sum_{ {\bf k}_1, {\bf k}_2 }  \int {\rm d}^2 {\bf r } \left (2  \frac{e^{ i  ( {\bf k}_1 + {\bf k}_2 ) \cdot {\bf r} } }{k_1^2 } -  \frac{e^{ i  ( {\bf k}_1 + {\bf k}_2 + {\bf q}  ) \cdot {\bf r} } }{k_1^2 } - \frac{e^{ i  ( {\bf k}_1 + {\bf k}_2 - {\bf q}  ) \cdot {\bf r} } }{k_1^2 }
\right )  \n
& \approx  & \frac{1}{4 \pi^2} \int {\rm d}^2{\bf k} \left( \frac{2}{ k^2} - \frac{1}{ | {\bf k} + {\bf q} |^2} - \frac{1}{ | {\bf k} - {\bf q} |^2} \right ) 
\approx  -q^2  \frac{{\rm d}s}{ \pi \Lambda^2}.
\ea
Note that in both of these evaluations, we have ignored an important constraint, which is that for the terms involving ${\bf q}$, we must have $\Lambda/s \leq | {\bf k} + {\bf q} | \leq \Lambda$, in addition to $\Lambda/s \leq k \leq \Lambda$.  However this constraint, if included, will modify the result by a factor of order unity, provided that $q$ is not large compared to $\Lambda - \Lambda/s$.  
The final results are
\ba
B &=& -  16 \pi  \beta_3 ^2\frac{ {\rm d}s }{\Lambda^4}, \qquad
C_0 =   \frac{1}{2} \left(4 \pi \beta_1 \right) ^2 I_0 = 4 \pi \beta_1^2  \frac{ {\rm d}s}{\Lambda^2},  \qquad C_4 = 0\\
C_2 &=&   - \frac{C_0}{4}  + \frac{\pi^2}{9 K} (4 \pi \beta_1) (\gamma_1 + \gamma_0 ) I_2= -   \pi\left( \beta_1^2  +   8 \left( \frac{ \pi}{18 K} \right)^2 \left( \gamma_0^2- \gamma_1^2 \right ) \right ) \frac{ {\rm d}s}{\Lambda^2} \n
C_3 (n) &=&  - \left( \frac{ \pi^2}{9 K } \right)^2  \left[   \gamma_1 (\gamma_0 - \gamma_1 + \gamma_{n+1} - \gamma_n )  +  (\gamma_1 - \gamma_{n+1} ) ( \gamma_1 - \gamma_0 )  \right ]  I_3  = \frac{{\rm d}s}{ 2 \pi}  \left( \frac{ \pi^2}{9 K } \right)^2  \left[   \gamma_1 \gamma_n - \gamma_0 \gamma_{n+1} \right ] . \nonumber
\ea
\end{widetext}
Here the factors of $\Lambda$ in each coefficient reflect the total engineering dimension of the couplings involved; these factors can be eliminated by defining appropriate dimensionless couplings. From these expressions, we can extract values for the constants appearing in Eqns.~(\ref{RGSTRI}) and (\ref{RGeqs}), obtaining
\be\label{RGcoeffs}
\begin{array}{rclrcl}
c_1 {\rm d}s &=& -\frac{1}{8}  B \qquad &
c_2  {\rm d}s &=& 2 \pi \beta_3^2 \frac{ds}{\Lambda^2}\\
c_3  {\rm d}s &=&  C_2 \qquad &
c_4  {\rm d}s &=& \frac{C_0}{2} - C_2 \\
c_5  {\rm d}s &=& -\frac{1}{2}C_3(1) \qquad \qquad &
c_6  {\rm d}s &=&\frac{1}{2} C_3(2)\,.
\end{array}
\ee


\begin{thebibliography}{99}
\bibitem{tlafm} G. Wannier, Phys. Rev. {\bf 79}, 357 (1950); R. M. F. Houtappel, Physica {\bf 16} 425 (1950).
\bibitem{review} For reviews, see: J. T. Chalker, in {\it Highly Frustrated Magnetism}, edited by C. Lacriox, P. Mendels, and F. Mila (Springer, 2010); 
L. Balents, Nature {\bf 464}, 199 (2010).
\bibitem{kagome} J. T. Chalker, P. C. W. Holdsworth and E. F. Shender, Phys. Rev. Lett. {\bf 68}, 855 (1992).
\bibitem{pyrochlore} J. N. Reimers, A. J. Berlinsky, and A.-C. Shi, Phys. Rev. B {\bf 43}, 865 (1991).
\bibitem{bergman} D. Bergman, J. Alicea, E. Gull, S. Trebst and L. Balents, Nature Phys. {\bf 3}, 487 (2007).
\bibitem{rastelli} E. Rastelli, A. Reatto and A. Tassi, J. Phys C {\bf 16}, L331 (1983). 
\bibitem{heightmodel} H. W. J. Bl\"ote and H. J. Hilhorst, J. Phys. A {\bf 15}, L631 (1982);
B. Nienhuis, H. J. Hilhorst, and H. W. Bl\"ote, {\it ibid.} {\bf 17}, 3559 (1984); B. Nienhuis, Phys. Rev. Lett. {\bf 49}, 1062 (1982).
\bibitem{zeng} C. Zeng and C. L. Henley, Phys. Rev, B {\bf 55}, 14935 (1997).
\bibitem{yamada} Y. Yamada, K. Kitsuda, S. Nohdo, and N. Ikeda, Phys. Rev. B {\bf 62}, 12 167 (2000), and J. Phys. Soc. Jpn. {\bf 66},
3733 (1997). 
\bibitem{radaelli} A. J. Hearmon, D. Prabhakaran, H. Nowell, F. Fabrizi, M. J. Gutmann, and P. G. Radaelli, Phys. Rev. B {\bf 85}, 014115 (2012).
\bibitem{harris} A. B. Harris, and T. Yildirim, Phys. Rev. B {\bf 81}, 134417 (2010). See also erratum: Phys. Rev. B {\bf 82}, 029902(E) (2010).
\bibitem{fe-review} For a recent review, see: N. Ikeda, T. Nagata, J. Kano,
and S. Mori, J. Phys Cond. Matt. {\bf 27}, 053201 (2015).
\bibitem{nagano} A. Nagano, M. Naka, J. Nasu, and S. Ishihara, Phys. Rev. Lett. {\bf 99}, 217202 (2007); A. Nagano and S. Ishihara,
J. Phys. Cond. Matt. {\bf 19} (2007) 145263.
\bibitem{domany} D. Auerbach, E. Domany and J. E. Gubernatis, Phys. Rev. B {\bf 37}, 1719 (1988).
\bibitem{diep} D.-T Hoang and H. T. Diep, Phys. Rev. E {\bf 85}, 041107 (2012).
\bibitem{kallin} D. S. Zimmerman, C. Kallin, and A. J. Berlinsky, Phys. Rev. B {\bf 37}, 7766 (1988).
\bibitem{kallin-aaa} A. Bunker, B. D. Gaulin, and C. Kallin, Phys. Rev. B {\bf 48}, 15861 (1993).
\bibitem{coppersmith} S. N. Coppersmith, Phys. Rev. B {\bf 32}, 1584 (1985). 
\bibitem{ma} D. Blankschtein, M. Ma, A. N. Berker, G. S. Grest, and C. M. Soukoulis, Phys. Rev. B {\bf 29}, 5250 (1984).
\bibitem{moessner2001} R. Moessner and S. L. Sondhi, Phys. Rev. B {\bf 63}, 224401 (2001).
\bibitem{previous} F. J. Burnell and J. T. Chalker, Phys. Rev. {\bf 92}, 220417 (2015).
\bibitem{SCGA} B. Canals and D. A. Garanin, Can. J. Phys, {\bf 79}, 1323 (2001).
\bibitem{isakov} S.V. Isakov, K. Gregor, R. Moessner and S. L. Sondhi, Phys. Rev. Lett. {\bf 93}, 167204 (2004).
\bibitem{Swendsen1986} R. H. Swendsen, and J.-S. Wang, Phys. Rev. Lett. 57, 2607 (1986).
\bibitem{Parisi1992} E. Marinari and G. Parisi, Europhys. Lett. {\bf 19}, 451 (1992).
\bibitem{FCC1980} M. K. Phani, J. L. Lebowitz, and M. H. Kalos, Phys. Rev. B {\bf 21}, 4027 (1980).
\bibitem{FCC2006} A. D. Beath and D. H. Ryan, Phys. Rev. B, {\bf 73}, 174416 (2006).  
\bibitem{stephenson} J. Stephenson, J. Math. Phys. {\bf 5}, 1009 (1964); {\it ibid.} {\bf 11}, 413 (1970).
\bibitem{correction} In Ref.~\onlinecite{previous} only the effects of $\kappa_3$ were considered, and not those of $\kappa_b$. 
\bibitem{KT1} J. M. Kosterlitz and D. J. Thouless, J. Phys. C {\bf 6}, 1181 (1973).
\bibitem{KT2} J. M. Kosterlitz, J. Phys. C {\bf 7}, 1046 (1974).
\bibitem{Nienhuis} B. Nienhuis in {\it Phase Transitions and Critical Phenomena} vol. 11, edited by C. Domb and J. L. Lebowitz (Academic Press, London, 1987).
\bibitem{Minnehagen} P. Minnhagen, Rev. Mod. Phys. {\bf 59}, 1001 (1987).
\bibitem{MomentumShellRG} P.B Wiegmann, J. Phys. C{\bf 11}, 1583 (1978).
\bibitem{2D3Dxy} S. R. Shenoy and  B. Chattopadhyay, Phys. Rev. B {\bf 51}, 9129 (1995).
\bibitem{SomebodyElse} S. Hikami and T. Tsuneto, Prog. Theor. Phys. {\bf 63}, 387 (1980).
\bibitem{sliding} C. S. O'Hern, T. C. Lubensky, and J. Toner, Phys. Rev. Lett {\bf 83}, 2745 (1999).
\bibitem{KogutReview} K. G. Wilson and J. Kogut, Phys. Rep. {\bf 12C}, 75 (1974).  
\bibitem{Starykh}  O. A. Starykh and L. Balents, Phys. Rev. Lett. {\bf 98}, 077205 (2007). 


\end{thebibliography}
\end{document}